\documentclass[12pt]{article}

\usepackage{amsmath,amssymb,amsthm,amscd,fullpage,cite}

\title{Hopf Algebras in Noncommutative Geometry}

\author{Joseph C. V\'arilly%
  \thanks{Regular Associate of the Abdus Salam ICTP.\quad
          Email: \texttt{varilly@cariari.ucr.ac.cr}}\\
        The Abdus Salam International Centre
	for Theoretical Physics, Trieste\\
        \textit{and}\\
	Depto.~de Matem\'atica,
        Universidad de Costa Rica, \\
        2060 San Jos\'e, Costa Rica}

\date{October 23, 2001}

\makeatletter
\def\section{\@startsection{section}{1}{\z@}{-3.5ex plus -1ex minus
 -.2ex}{2.3ex plus .2ex}{\large\bf}}
\def\subsection{\@startsection{subsection}{2}{\z@}{-3.25ex plus -1ex
 minus -.2ex}{1.5ex plus .2ex}{\normalsize\bf}}
\makeatother

\numberwithin{equation}{section}    

\theoremstyle{plain}
\newtheorem{thm}{Theorem}[section]  
\newtheorem{prop}[thm]{Proposition} 

\theoremstyle{definition}
\newtheorem{defn}{Definition}[section] 
\newtheorem{exer}{Exercise}[section]   

\theoremstyle{remark}
\newtheorem{exmp}{Example}[section] 

\newcommand{\marker}{\vspace{6pt}\noindent{$\rtri$}\enspace}

\newbox\ncintdbox \newbox\ncinttbox 
\setbox0=\hbox{$-$}
\setbox2=\hbox{$\displaystyle\int$}
\setbox\ncintdbox=\hbox{\rlap{\hbox
    to \wd2{\hskip-.125em \box2\relax\hfil}}\box0\kern.1em}
\setbox0=\hbox{$\vcenter{\hrule width 4pt}$}
\setbox2=\hbox{$\textstyle\int$}
\setbox\ncinttbox=\hbox{\rlap{\hbox
    to \wd2{\hskip-.175em \box2\relax\hfil}}\box0\kern.1em}

\newcommand{\hideqed}{\renewcommand{\qed}{}} 

\DeclareRobustCommand{\qef}{
  \ifmmode 
  \else \leavevmode\unskip\penalty9999 \hbox{}\nobreak\hfill
  \fi
  \quad\hbox{\qefsymbol}}
\newcommand{\qefsymbol}{$\lozenge$} 

\newcommand{\hideqef}{\renewcommand{\qef}{}} 

\newcommand{\ldbrack}{[\mkern-2mu[}    
\newcommand{\rdbrack}{]\mkern-2mu]}    
\newcommand{\stroke}{\mathbin|}        
\newcommand{\tribar}{|\mkern-2mu|\mkern-2mu|} 

\renewcommand{\a}{\alpha}           
\renewcommand{\b}{\beta}            
\newcommand{\dl}{\delta}            
\newcommand{\Dl}{\Delta}            
\newcommand{\eps}{\varepsilon}      
\newcommand{\Ga}{\Gamma}            
\newcommand{\ga}{\gamma}            
\newcommand{\id}{\iota}             
\newcommand{\La}{\Lambda}           
\newcommand{\la}{\lambda}           
\newcommand{\Om}{\Omega}            
\newcommand{\om}{\omega}            
\newcommand{\sg}{\sigma}            
\renewcommand{\th}{\theta}          
\newcommand{\Ups}{\Upsilon}         
\newcommand{\vf}{\varphi}           

\newcommand{\A}{\mathcal{A}}        
\newcommand{\D}{\mathcal{D}}        
\newcommand{\E}{\mathcal{E}}        
\newcommand{\F}{\mathcal{F}}        
\newcommand{\G}{\mathcal{G}}        
\renewcommand{\H}{\mathcal{H}}      
\renewcommand{\L}{\mathcal{L}}      
\newcommand{\Rr}{\mathcal{R}}       
\newcommand{\Tee}{\mathcal{T}}      
\newcommand{\U}{\mathcal{U}}        

\newcommand{\C}{\mathbb{C}}         
\newcommand{\FF}{\mathbb{F}}        
\newcommand{\N}{\mathbb{N}}         
\newcommand{\Q}{\mathbb{Q}}         
\newcommand{\R}{\mathbb{R}}         
\newcommand{\Sf}{\mathbb{S}}        
\newcommand{\T}{\mathbb{T}}         
\newcommand{\Z}{\mathbb{Z}}         

\newcommand{\cc}{\mathbf{c}}        

\newcommand{\g}{\mathfrak{g}}       
\newcommand{\hl}{\mathfrak{h}}      

\DeclareMathOperator{\Aut}{Aut}     
\DeclareMathOperator{\ch}{ch}       
\DeclareMathOperator{\Cl}{Cl}       
\DeclareMathOperator{\CCl}{\C l}    
\DeclareMathOperator{\Der}{Der}     
\renewcommand{\det}{\operatorname{det}} 
\DeclareMathOperator{\Dom}{Dom}     
\DeclareMathOperator{\Exp}{Exp}     
\DeclareMathOperator{\grad}{grad}   
\DeclareMathOperator{\Hom}{Hom}     
\DeclareMathOperator{\im}{im}       
\DeclareMathOperator{\Isom}{Isom}   
\DeclareMathOperator{\lin}{span}    
\DeclareMathOperator{\Prim}{Prim}   
\DeclareMathOperator{\Ran}{Ran}     
\DeclareMathOperator{\sign}{sign}   
\DeclareMathOperator{\Spin}{Spin}   
\DeclareMathOperator{\supp}{supp}   
\DeclareMathOperator{\Tr}{Tr}       
\DeclareMathOperator{\tr}{tr}       

\renewcommand{\d}{\mathrm{d}}       
\newcommand{\dR}{\mathrm{dR}}       
\newcommand{\even}{\mathrm{even}}   
\newcommand{\odd}{\mathrm{odd}}     

\newcommand{\ssJ}{{\scriptscriptstyle J}} 

\newcommand{\del}{\partial}         
\newcommand{\downto}{\downarrow}    
\newcommand{\hookto}{\hookrightarrow} 
\newcommand{\op}{\oplus}            
\newcommand{\opp}{\circ}            
\newcommand{\ox}{\otimes}           
\newcommand{\rtri}{\blacktriangleright} 
\newcommand{\semi}{\rtimes}         
\newcommand{\w}{\wedge}             
\newcommand{\x}{\times}             
\newcommand{\7}{\dagger}            
\newcommand{\8}{\bullet}            
\renewcommand{\.}{\cdot}            
\renewcommand{\:}{\colon}           

\newcommand{\Coo}{C^\infty}         
\newcommand{\dm}{\dot\mu}           
\newcommand{\Gaoo}{\Gamma^\infty}   
\newcommand{\Hoo}{\H^\infty}        
\newcommand{\Rbar}{\overline{R}}    
\newcommand{\Uhat}{{\widehat U}}    

\newcommand{\Dslash}{{D\mkern-12.5mu/\,}} 

\newcommand{\aspa}{\mathbin{\#}}        

\newcommand{\hatox}{\mathrel{\mathchoice{\widehat\otimes}
                    {\widehat\otimes}{\hat\otimes}{\hat\otimes}}}

\newcommand{\opyop}{\oplus\cdots\oplus}    
\newcommand{\oxyox}{\otimes\cdots\otimes}  
\newcommand{\wyw}{\wedge\cdots\wedge}      

\newcommand{\longto}{\mathop{\longrightarrow}\limits}

\renewcommand{\injlim}{\varinjlim}
\newcommand{\ncint}{\mathop{\mathchoice{\copy\ncintdbox}%
           {\copy\ncinttbox}{\copy\ncinttbox}%
		   {\copy\ncinttbox}}\nolimits}  
\newcommand{\PVint}{\,\mathrm{P}\!\!\int} 
\DeclareMathOperator{\tsum}{{\textstyle\sum}} 

\newcommand{\shalf}{{\scriptstyle\frac{1}{2}}}  
\newcommand{\teighth}{\tfrac{1}{8}}   
\newcommand{\thalf}{\tfrac{1}{2}}     
\newcommand{\tihalf}{\tfrac{i}{2}}    
\newcommand{\tquarter}{\tfrac{1}{4}}  
\newcommand{\tthird}{\tfrac{1}{3}}    

\renewcommand{\Bar}[1]{\overline{#1}} 
\newcommand{\Onda}[1]{\widetilde{#1}} 

\newcommand{\sepword}[1]{\quad\text{#1}\quad} 

\newcommand{\bra}[1]{\langle{#1}\rvert} 
\newcommand{\ket}[1]{\lvert{#1}\rangle} 
\newcommand{\set}[1]{\{\,#1\,\}}     
\newcommand{\snorm}[1]{\mathopen{\tribar}{#1}\mathclose{\tribar}}

\newcommand{\braCket}[3]{\langle#1\stroke#2\stroke#3\rangle} 
\newcommand{\dst}[2]{\langle#1,#2\rangle} 

\def\<#1,#2>{\langle#1\stroke#2\rangle} 
\def\(#1,#2){(#1\stroke#2)}
\def\?#1,#2?{\{#1\stroke#2\}}     

\newcommand{\dd}[1]{\frac{\partial}{\partial#1}} 
\newcommand{\ddto}[1]{\frac{d}{d#1}\biggr|_{#1=0}} 

\newcommand{\twobytwo}[4]{\begin{pmatrix}#1 & #2 \\
                                #3 & #4\end{pmatrix}} 
\newcommand{\twobytwoeven}[2]{\begin{pmatrix}#1 & 0 \\
                            0 & #2\end{pmatrix}} 
\newcommand{\twobytwoodd}[2]{\begin{pmatrix}0 & #1 \\
                            #2 & 0\end{pmatrix}} 

\newcommand{\miss}[1]{\widehat{#1}}  

\newcommand{\row}[3]{{#1}_{#2},\dots,{#1}_{#3}}   

\newcommand{\punto}{\;\begin{picture}(2,2)(0,-2)
\put(0,0){\circle{4}}
\end{picture}\;}

\newcommand{\match}{\;\begin{picture}(5,10)(0,5)
\put(0,10){\circle{4}}
\put(0,10){\line(0,-1){10}}
\put(0,0){\circle*{3}}
\end{picture}\;}

\newcommand{\baton}{\;\begin{picture}(5,20)(0,10)
\put(0,20){\circle{4}}
\put(0,20){\line(0,-1){10}}
\put(0,10){\circle*{3}}
\put(0,10){\line(0,-1){10}}
\put(0,0){\circle*{3}}
\end{picture}\;}

\newcommand{\legs}{\;\begin{picture}(20,10)(0,5)
\put(10,10){\circle{4}}
\put(10,10){\line(-1,-1){10}}
\put(0,0){\circle*{3}}
\put(10,10){\line(1,-1){10}}
\put(20,0){\circle*{3}}
\end{picture}\;}

\newcommand{\stick}{\;\begin{picture}(5,30)(0,15)
\put(0,30){\circle{4}}
\put(0,30){\line(0,-1){10}}
\put(0,20){\circle*{3}}
\put(0,20){\line(0,-1){10}}
\put(0,10){\circle*{3}}
\put(0,10){\line(0,-1){10}}
\put(0,0){\circle*{3}}
\end{picture}\;}

\newcommand{\crook}{\;\begin{picture}(20,20)(0,10)
\put(10,20){\circle{4}}
\put(10,20){\line(-1,-1){10}}
\put(0,10){\circle*{3}}
\put(10,20){\line(1,-1){10}}
\put(20,10){\circle*{3}}
\put(0,10){\line(0,-1){10}}
\put(0,0){\circle*{3}}
\end{picture}\;}

\newcommand{\claw}{\;\begin{picture}(20,10)(0,5)
\put(10,10){\circle{4}}
\put(10,10){\line(-1,-1){10}}
\put(0,0){\circle*{3}}
\put(10,10){\line(1,-1){10}}
\put(20,0){\circle*{3}}
\put(10,10){\line(0,-1){10}}
\put(10,0){\circle*{3}}
\end{picture}\;}

\newcommand{\biped}{\;\begin{picture}(20,20)(0,10)
\put(10,20){\circle{4}}
\put(10,20){\line(0,-1){10}}
\put(10,10){\circle*{3}}
\put(10,10){\line(-1,-1){10}}
\put(0,0){\circle*{3}}
\put(10,10){\line(1,-1){10}}
\put(20,0){\circle*{3}}
\end{picture}\;}

\newcommand{\arch}{\;\begin{picture}(20,20)(0,10)
\put(10,20){\circle{4}}
\put(10,20){\line(-1,-1){10}}
\put(0,10){\circle*{3}}
\put(10,20){\line(1,-1){10}}
\put(20,10){\circle*{3}}
\put(0,10){\line(0,-1){10}}
\put(0,0){\circle*{3}}
\put(20,10){\line(0,-1){10}}
\put(20,0){\circle*{3}}
\end{picture}\;}

\scrollmode 

\begin{document}

\maketitle

\begin{abstract}
We give an introductory survey to the use of Hopf algebras in several
problems of noncommutative geometry. The main example, the Hopf
algebra of rooted trees, is a graded, connected Hopf algebra arising
from a universal construction. We show its relation to the algebra of
transverse differential operators introduced by Connes and Moscovici
in order to compute a local index formula in cyclic cohomology, and to
the several Hopf algebras defined by Connes and Kreimer to simplify
the combinatorics of perturbative renormalization. We explain how
characteristic classes for a Hopf module algebra can be obtained from
the cyclic cohomology of the Hopf algebra which acts on it. Finally,
we discuss the theory of noncommutative spherical manifolds and show
how they arise as homogeneous spaces of certain compact quantum
groups.
\end{abstract}

\subsection*{Introduction}

These are lecture notes for a course given at the Summer School on
Geometric and Topological Methods for Quantum Field Theory, sponsored
by the Centre International de Math\'ematiques Pures et Appliqu\'ees
(CIMPA) and the Universidad de Los Andes, at Villa de Leyva, Colombia,
from the 9th to the 27th of July, 2001.

These notes explore some recent developments which place Hopf
algebras at the heart of the noncommutative approach to geometry and
physics. Many examples of Hopf algebras are known from the literature
on ``quantum groups'', some of which provide algebraic deformations of
the classical transformation groups. The main emphasis here, however,
is on certain other Hopf algebras which have recently appeared in two
seemingly unrelated contexts: in the combinatorics of perturbative
renormalization in quantum field theories, and in connection with
local index formulas in noncommutative geometry.

These Hopf algebras act on ``noncommutative spaces'', and certain
characteristic classes for these spaces can be obtained, by a
canonical procedure, from corresponding invariants of the Hopf
algebras. This comes about by pulling back the cyclic cohomology of
the algebra representing the noncommutative space, which is the
receptacle of Chern characters, to another cohomology of the Hopf
algebra.

Recently, some interesting spaces have been discovered, the
noncommutative spheres, which are completely specified by certain
algebraic relations. They turn out to be homogeneous spaces under the
action of certain Hopf algebras: in this way, these Hopf algebras
appear as ``quantum symmetry groups''. We shall show how these
symmetries arise from a class of quantum groups built from Moyal
products on group manifolds.

\vspace{6pt}

Section 1 is introductory: it offers a snapshot of noncommutative
geometry and the basic theory of Hopf algebras; as an example of how
both theories interact, we exhibit the Connes--Moscovici Hopf algebra
of differential operators in the one-dimensional case. Section~2
concerns the Hopf algebras which have been found useful in the
perturbative approach to renormalization. We develop at length a
universal construction, the Connes--Kreimer algebra of rooted trees,
which is a graded, commutative, but highly noncocommutative Hopf
algebra. Particular quantum field theories give rise to related Hopf
algebras of Feynman graphs; we discuss briefly how these give a
conceptual approach to the renormalization problem.

The third section gives an overview of cyclic cohomology for both
associative and Hopf algebras, indicating how the latter provide
characteristic classes for associative algebras on which they act. The
final Section~4 explains how cyclic-homology Chern characters lead to
new examples of noncommutative spin geometries, whose symmetry groups
are compact quantum groups obtained from the Moyal approach to
prequantization.

\vspace{12pt}

I am grateful to Jos\'e M. Gracia-Bond\'{\i}a and Chryssomalis
Chryssomalakos for several remarks on an earlier version, and to Jean
Bellissard for helpful comments at the time of the lectures. I wish to
thank Sergio Adarve, Hern\'an Ocampo, Marta Kovacsics and especially
Sylvie Paycha for affording me the opportunity to talk about these
matters in a beautiful setting in the Colombian highlands.

\newpage

\tableofcontents

\newpage


\section{Noncommutative Geometry and Hopf Algebras}

Noncommutative geometry, in the broadest sense, is the study of
geometrical properties of singular spaces, by means of suitable
``coordinate algebras'' which need not be commutative. If the space in
question is a differential manifold, its coordinates form a
commutative algebra of smooth functions; but even in this case, adding
a metric structure may involve operators which do not commute with the
coordinates. One learns to replace the usual calculus of points,
paths, integration domains, etc., by an alternative language involving
the algebra of coordinates; by focusing only on those features which
do not require that the coordinates commute, one arrives at an
algebraic (or operatorial) approach which is applicable to many
singular spaces also.

\subsection{The algebraic tools of noncommutative geometry}
\label{sec:brief-NCG} 

The first step is to replace a topological space $X$ by its algebra of
complex-valued continuous functions $C(X)$. If $X$ is a compact
(Hausdorff) space, then $C(X)$ is a commutative $C^*$-algebra with
unit~$1$ and its norm $\|f\| := \sup_{x\in X} |f(x)|$ satisfies the
$C^*$-property $\|f\|^2 = \|f^*f\|$. The first Gelfand--Na\u{\i}mark
theorem~\cite{GelfandN} says that any commutative unital $C^*$-algebra
$A$ is of this form: $A = C(X)$ where $X = M(A)$ is the space of
\textit{characters} (nonzero homomorphisms) $\mu\: A \to \C$, which is
compact in the weak* topology determined by the maps
$\mu \mapsto \mu(a)$, for $a \in \C$. Indeed, the characters of $C(X)$
are precisely the evaluation maps $\eps_x\: f \mapsto f(x)$ at each
point $x \in X$.

We shall mainly deal with the compact case in what follows. A locally
compact, but noncompact, space $Y$ can be handled by passing to a
compactification (that is, a compact space in which $Y$ can be densely
embedded). For instance, we can adjoin one ``point at infinity'': if
$X = Y \uplus \{\infty\}$, then $\set{f \in C(X) : f(\infty) = 0}$ is
isomorphic to $C_0(Y)$, the commutative $C^*$-algebra of continuous
functions on~$Y$ ``vanishing at infinity''; thus, by dropping the
constant functions from $C(X)$, we get the commutative nonunital
$C^*$-algebra $C_0(Y)$ as a stand-in for the locally compact space
$Y$. There is also a maximal compactification $\beta Y := M(C_b(Y))$,
called the Stone--\v{C}ech compactification, namely, the character
space of the (unital) $C^*$-algebra of bounded continuous functions
on~$Y$.

This construction $X \mapsto C(X)$ yields a contravariant functor: to
each continuous map $h\: X_1 \to X_2$ between compact spaces there is
a morphism%
\footnote{By a morphism of unital $C^*$-algebras we mean a
$*$-homomorphism preserving the units.}
$\vf_h\: C(X_2) \to C(X_1)$ given by $\vf_h(f) := f \circ h$.

By relaxing the commutativity requirement, we can regard
\textit{noncommutative} $C^*$-algebras (unital or not) as proxies for
``noncommutative locally compact spaces''. The characters, if any, of
such an algebra may be said to label ``classical points'' of the
corresponding noncommutative space. However, noncommutative
$C^*$-algebras generally have few characters, so these putative spaces
will have correspondingly few points. The recommended course of
action, then, is to leave these pointless spaces behind and to adopt
the language and techniques of algebras instead.

There is a second Gelfand--Na\u{\i}mark theorem~\cite{GelfandN}, which
states that any $C^*$-algebra, commutative or not, can be faithfully
represented as a (norm-closed) algebra of bounded operators on a
Hilbert space. The data for a ``noncommutative topology'' consist,
then, of a pair $(A,\H)$ where $\H$ is a Hilbert space and $A$ is a
closed subalgebra of $\L(\H)$.

\marker
Vector bundles over a compact space also have algebraic counterparts.
If $X$ is compact and $E \longto^\pi X$ is a complex vector bundle,
the space $\Ga(X,E)$ of continuous sections is naturally a module over
$C(X)$, which is necessarily of the form $eC(X)^m$, where
$e = e^2 \in M_m(C(X))$ is an idempotent matrix of elements of $C(X)$.
More generally, if $A$ is any algebra over~$\C$, a right $A$-module of
the form $eA^m$ with $e = e^2 \in M_m(A)$ is called a \textit{finitely
generated projective module} over $A$. The Serre--Swan
theorem~\cite{Swan} matches vector bundles over $X$ with finitely
generated projective modules over $C(X)$. The idempotent~$e$ may be
constructed from the transition functions of the vector bundle by
pulling back a standard idempotent from a Grassmannian bundle: see
\cite[\S 1.1]{FedosovBook} or \cite[\S 2.1]{Polaris} for details.

A more concrete example is that of the tangent bundle over a compact
Riemannian manifold $M$: by the Nash embedding theorem
\cite[Thm~14.5.1]{TaylorPDE}, one can embed $M$ in some $\R^m$ so that
the metric on $TM$ is obtained from the ambient Euclidean metric; if
$e(x)$ is the orthogonal projector on~$\R^n$ with range $T_xM$, then
$e = e^2 \in M_m(C(M))$ and the module $\Ga(M,TM)$ of vector fields
on~$M$ may be identified with the range of~$e$.

In the noncompact case, one can use Rennie's nonunital version of the 
Serre--Swan theorem~\cite{RennieSwan}: $C_0(Y)$-modules of the form 
$eC(X)^m$, where $X$ is some compactification of~$Y$ and
$e = e^2 \in M_m(C(X))$, consist of sections vanishing at infinity
(i.e., outside of~$Y$) of vector bundles $E \to X$. One can take $X$
to be the \textit{one-point} compactification of~$Y$ only if $E$ is
\textit{trivial} at infinity; as a rule, the compactification to be
used depends on the problem at hand.

If $A$ is a $C^*$-algebra, we may replace $e$ by an orthogonal
projector (i.e., a selfadjoint idempotent) $p = p^* = p^2$ so that
$eA^m \simeq pA^m$ as right $A$-modules. If $A$ is faithfully
represented by bounded operators on a Hilbert space $\H$, then
$M_m(A)$ is an algebra of bounded operators on $\H^m = \H \opyop \H$
($m$~times), so we can schematically write $e = \twobytwo{1}{x}{0}{0}$
as an operator on $e\H^m \op (1-e)\H^m$; then
$p := \twobytwoeven{1}{0}$ is the range projector on~$e\H^m$.

The correspondence $E \mapsto \Ga(X,E)$ is a covariant functor which
carries topological invariants of~$X$ to algebraic invariants
of~$C(X)$. In particular, it identifies the $K$-theory group $K^0(X)$,
formed by stable equivalence classes of vector bundles where
$[E] + [F] := [E \op F]$ ---here $\op$ denotes Whitney sum of vector
bundles over~$X$--- with the group $K_0(C(X))$ formed by stable
isomorphism classes of matrix projectors over $C(X)$ where
$[p] + [q] := [p \op q]$ and $\op$ now denotes direct sum of
projectors. The $K$-theory of $C^*$-algebras may be developed in an
operator-theoretic way, see \cite{Blackadar,Murphy,Wegge} and
\cite[Chap.~3]{Polaris}, for instance; or purely algebraically, and
the group $K_0(A)$ turns out to be the same in both approaches.
(However, the group $K_1(A)$, formed by classes of unitaries in
$M_m(A)$, does not coincide with the algebraic $K_1$-group in general:
see, for instance, \cite{RosenbergAlg} or \cite[p.~131]{Polaris}.) The
salient feature of both topological and $C^*$-algebraic $K$-theories
is \textit{Bott periodicity}, which says that two $K$-groups are
enough: although one can define $K_j(A)$ is a systematic way for any
$j \in \N$, it turns out that $K_{j+2}(A) \simeq K_j(A)$ by natural
isomorphisms (in marked contrast to the case of purely algebraic
$K$-theory).

\marker
To deal with a (compact) differential manifold $M$ (in these notes, we
only treat differential manifolds without boundary), we replace the
continuous functions in $C(M)$ by the dense subalgebra of
\textit{smooth} functions $\A = \Coo(M)$. This is no longer a
$C^*$-algebra, but it is complete in its natural topology (that of
uniform convergence of functions, together with their derivatives of
all orders), so it is a Fr\'echet algebra with a $C^*$-completion.
Likewise, given a vector bundle $E \longto M$, we replace the
continuous sections in $\Ga(M,E)$ by the $\A$-module of smooth
sections $\Gaoo(M,E)$. The Serre--Swan theorem continues to hold,
\textit{mutatis mutandis}, in the smooth category.

In the noncommutative case, with no differential structure \textit{a
priori}, we need to replace the $C^*$-algebra $A$ by a subalgebra $\A$
which should (a)~be dense in~$\A$; (b)~be a Fr\'echet algebra, that
is, it should be complete under some countable family of seminorms
including the original $C^*$-norm of~$A$; and~(c) satisfy
$K_0(\A) \simeq K_0(A)$. This last condition is not automatic: it is
necessary that $\A$ be a \textit{pre-$C^*$-algebra}, that is to say,
it should be stable under the holomorphic functional calculus (which
is defined in the larger algebra~$A$). The proof of~(c) for
pre-$C^*$-algebras is given in \cite{Bost}; see also
\cite[\S 3.8]{Polaris}.

\marker
The next step is to find an algebraic description of a Riemannian
metric on a smooth manifold. This can be done in a principled way
through a theory of ``noncommutative metric spaces'' at present under
construction by
Rieffel~\cite{RieffelMetricI,RieffelMetricII,RieffelDist,RieffelCoad}.
But here we shall take a short cut, by defining metrics only over
\textit{spin} manifolds, using the Dirac operator as our instrument;
this was, indeed, the original insight of Connes~\cite{ConnesMetric}.

A metric $g = [g_{ij}]$ on the tangent bundle $TM$ of a (compact)
manifold $M$ yields a contragredient metric $g^{-1} = [g^{rs}]$ on the
cotangent bundle $T^*M$; so we can build a Clifford algebra bundle
$\CCl(M) \longto M$, whose fibre at~$x$ is
$\Cl((T_x^*M)^\C,g_x^{-1})$, by imposing a suitable product structure
on the complexified exterior bundle $(\La^{\!\8} T^*M)^\C$. We assume
that $M$ supports a spinor bundle $S \longto M$, on which $\CCl(M)$
acts fibrewise and irreducibly; on passing to smooth sections, we may
write $c(\a)$ for the Clifford action of a $1$-form $\a$ on spinors.
The spinor bundle comes equipped with a Hermitian metric, so the
squared norm
$\|\psi\|^2 := \int_M |\psi(x)|^2 \,\sqrt{\det g} \,dx$ makes sense;
the completion of $\Gaoo(M,S)$ in this norm is the Hilbert space
$\H = L^2(M,S)$ of square-integrable spinors. Locally, we may write
the Clifford action of $1$-forms as $c(dx^r) := h_\a^r \,\ga^\a$,
where the ``gamma matrices'' $\ga^a$ satisfy
$\ga^\a \ga^\b + \ga^\b \ga^\a = 2\,\dl^{\a\b}$ and the coefficients
$h_\a^r$ are real and obey $h_\a^r \dl^{\a\b} h_\b^s = g^{rs}$. The
Dirac operator is locally defined as
\begin{equation}
\Dslash := -i\,c(dx^r) \,\Bigl( \dd{x^r} - \om_r \Bigr),
\label{eq:Dirac-op-local} 
\end{equation}
where $\om_r = \tquarter \Onda\Ga_{r\a}^\b \,\ga^\a \ga^\b$ are
components of the \textit{spin connection}, obtained from the
Christoffel symbols $\Onda\Ga_{r\a}^\b$ (in an orthogonal basis) of
the Levi-Civita connection. The manifold $M$ is spin whenever these
local formulae patch together to give a well-defined spinor bundle.
There is a well-known topological condition for this to happen (the
second Stiefel-Whitney class $w_2(TM) \in H^2(M,\Z_2)$ must
vanish~\cite{LawsonM}), and when it is fulfilled, $\Dslash$ extends to
a selfadjoint operator on~$\H$ with compact
resolvent~\cite{Polaris,LawsonM}.

Apart from these local formulae, the Dirac operator has a fundamental
algebraic property. If $\psi$ is a spinor and $a \in \Coo(M)$ is
regarded as a multiplication operator on spinors, it can be checked
that
$$
\Dslash(a\psi) = -i\,c(da)\,\psi + a\,\Dslash\psi,
$$
or, more simply,
\begin{equation}
[\Dslash, a] = -i\,c(da).
\label{eq:Dirac-comm-reln} 
\end{equation}

Following~\cite{BerlineGV}, we call a ``generalized Dirac operator''
any selfadjoint operator $D$ on~$\H$ satisfying $[D,a] = -i\,c(da)$
for $a \in \Coo(M)$. Now $c(da)$ is a bounded operator on $L^2(M,S)$
whenever $a$ is smooth, and its norm is that of the gradient of~$a$,
i.e., the vector field determined by $g(\grad a, X) := da(X) = X(a)$.
A continuous function $a \in C(M)$ is called Lipschitz (with respect
to the metric~$g$) if its gradient is defined, almost everywhere, as
an essentially bounded measurable vector field, i.e.,
$\|\grad a\|_\infty$ is finite. Now the Riemannian distance $d_g(p,q)$
between two points $p,q \in M$ is usually defined as the infimum of
the lengths of (piecewise smooth) paths from $p$ to~$q$; but it is not
hard to show (see \cite[\S 9.3]{Polaris}, for instance) that the
distance can also be defined as a supremum:
\begin{equation}
d_g(p,q)
 = \sup\set{|a(p) - a(q)| : a \in C(M),\ \|\grad a\|_\infty \leq 1}.
\label{eq:metr-dist} 
\end{equation}
The basic equation \eqref{eq:Dirac-comm-reln} allows to replace the
gradient by a commutator with the Dirac operator:
\begin{equation}
d_g(p,q)
 = \sup\set{|a(p) - a(q)| : a \in C(M),\ \|[\Dslash,a]\| \leq 1}.
\label{eq:Connes-dist} 
\end{equation}
Thus, the Riemannian distance function $d_g$ is entirely determined by
$\Dslash$. Moreover, the metric $g$ is in turn determined by $d_g$,
according to the Myers--Steenrod theorem~\cite{MyersS}. From the
noncommutative point of view, then, the Dirac operator assumes the
role of the metric. This leads to the following basic concept.

\begin{defn}
\label{df:spec-tri} 
A \textbf{spectral triple} is a triple $(\A,\H,D)$, where $\A$ is a
pre-$C^*$-algebra, $\H$ is a Hilbert space carrying a representation
of~$\A$ by bounded operators, and $D$ is a selfadjoint operator
on~$\A$, with compact resolvent, such that the commutator $[D,a]$ is a
bounded operator on~$\H$, for each $a \in \A$.

Spectral triples comes in two parities, odd and even. In the odd
case, there is nothing new; in the even case, there is a grading
operator $\chi$ on $\H$ (a bounded selfadjoint operator satisfying
$\chi^2 = 1$, making a splitting $\H = \H_+ \op \H_-$), such that the
representation of $\A$ is even ($\chi a = a \chi$ for all $a \in \A$)
and the operator $D$ is odd, i.e., $\chi D = - D\chi$; thus each
$[D,a]$ is a bounded odd operator on~$\H$.
\end{defn}

A \textbf{noncommutative spin geometry} is a spectral triple
satisfying several extra conditions, which were first laid out by
Connes in the seminal paper \cite{ConnesGrav}. These conditions (or
``axioms'', as they are sometimes called) arise from a careful
consideration of the algebraic properties of ordinary metric geometry.
Seven such properties are put forward in \cite{ConnesGrav}; here, we
shall just outline the list. Some of the terminology will be clarified
later on; a more complete account, with all prerequisites, is given in
\cite[\S 10.5]{Polaris}.

\begin{enumerate}
\item \textit{Classical dimension\/}:
There is a unique nonnegative integer $n$, the ``classical dimension''
of the geometry, for which the eigenvalue sums
$\sg_N := \sum_{0\leq k<N} \mu_k$ of the compact positive operator
$|D|^{-n}$ satisfy $\sg_N \sim C \log N$ as $N \to \infty$, with
$0 < C < \infty$; the coefficient is written $C = \ncint |D|^{-n}$,
where $\ncint$ denotes the ``Dixmier trace'' if $n \geq 1$. This $n$
is even if and only if the spectral triple is even. (When
$\A = \Coo(M)$ and $D$ is a Dirac operator, $n$ equals the ordinary
dimension of the spin manifold~$M$).

\item \textit{Regularity\/}:
Not only are the operators $a$ and $[D,a]$ bounded, but they lie in
the smooth domain of the derivation $\dl(T) := [|D|,T]$. (When $\A$
is an algebra of functions and $D$ is a Dirac operator, this smooth
domain consists exactly of the $\Coo$ functions.)

\item \textit{Finiteness\/}:
The algebra $\A$ is a pre-$C^*$-algebra, and the space of smooth
vectors $\Hoo := \bigcap_k \Dom(D^k)$ is a finitely generated
projective left $\A$-module. (In the commutative case, this yields
the smooth spinors.)

\item \textit{Reality\/}:
There is an antiunitary operator $C$ on~$\H$, such that
$[a, Cb^*C^{-1}] = 0$ for all $a,b \in \A$ (thus
$b \mapsto Cb^*C^{-1}$ is a commuting representation on $\H$ of the
``opposite algebra'' $\A^\opp$, with the product reversed); and
moreover, $C^2 = \pm 1$, $CD = \pm DC$, and $C\chi = \pm\chi C$ in the
even case, where the signs depend only on $n \bmod 8$. (In the
commutative case, $C$ is the charge conjugation operator on spinors.)

\item \textit{First order\/}:
The bounded operators $[D,a]$ commute with the opposite algebra
representation: $[[D,a], C b^* C^{-1}] = 0$ for all $a,b \in \A$.

\item \textit{Orientation\/}:
There is a \textit{Hochschild $n$-cycle} $\cc$ on~$\A$ whose natural
representative is $\pi_D(\cc) = \chi$ (even case) or $\pi_D(\cc) = 1$
(odd case). More on this later: such an $n$-cycle is usually a finite
sum of terms like $a_0 \ox a_1 \oxyox a_n$ which map to operators
$$
\pi_D(a_0 \ox a_1 \oxyox a_n) := a_0 \,[D,a_1] \dots [D,a_n],
$$
and $\cc$ is the algebraic expression of the \textit{volume form} for
the metric determined by~$D$.

\item \textit{Poincar\'e duality\/}:
The index map of~$D$ determines a nondegenerate pairing on the
$K$-theory of the algebra~$\A$. (We shall not go into details, except
to mention that in the commutative case, the Chern homomorphism
matches this nondegeneracy with Poincar\'e duality in de~Rham
co/homology.)

\end{enumerate}

It is very important to know that when $A = \Coo(M)$ the usual
apparatus of geometry on spin manifolds (spin structure, metric, Dirac
operator) can be fully recovered from these seven conditions: for the
full proof of this theorem, see \cite[chap.~11]{Polaris}. Another
proof, assuming only that $\A$ is commutative, is developed by Rennie
in~\cite{RennieGeom}.

\subsection{Hopf algebras: introduction}
\label{sec:Hopf-basics} 

The general scheme of replacing point spaces by function algebras and
then moving on to noncommutative algebras also works for symmetry
groups. Now, however, the interplay of algebra and topology is much
more delicate. There are at least two ways of handling this issue. One
is to leave topology aside and develop a purely algebraic theory of
symmetry-bearing algebras: these are the Hopf algebras, sometimes
called ``quantum groups'', about which there is already a vast
literature. At the other extreme, one may insist on using
$C^*$-algebras with special properties; in the unital case, there has
emerged a useful theory of ``compact quantum groups''
\cite{WoronowiczComp}, which only very recently has been extended to
the locally compact case also~\cite{KustermansV}.

We begin with the more algebraic treatment, keeping to the compact
case, i.e., all algebras will be unital unless indicated otherwise.
The field of scalars may be taken as $\C$, $\R$ or~$\Q$, according to
convenience; to cover all cases, we shall denote it by~$\FF$. In this
section, $\ox$ always means the algebraic tensor product.

\begin{defn}
\label{df:bi-alg} 
A \textit{bialgebra} is a vector space $A$ over~$\FF$ which is both an
algebra and a coalgebra in a compatible way. The \textit{algebra}
structure is given by $\FF$-linear maps $m\: A \ox A \to A$ (the
product) and $\eta\: \FF \to A$ (the unit map) where $xy := m(x,y)$
and $\eta(1) = 1_A$. The \textit{coalgebra} structure is likewise
given by linear maps $\Dl\: A \to A \ox A$ (the coproduct) and
$\eps\: A \to \FF$ (the counit map). We write $\id\: A \to A$, or
sometimes $\id_A$, to denote the identity map on~$A$. The required
properties are:
\begin{enumerate}
\item Associativity:
$m(m \ox \id) = m(\id \ox m) : A \ox A \ox A \to A$;
\item Unity: $m(\eta \ox \id) = m(\id \ox \eta) = \id : A \to A$;
\item Coassociativity:
$(\Dl \ox \id) \Dl = (\id \ox \Dl) \Dl : A \to A \ox A \ox A$;
\item Counity: $(\eps\ox\id) \Dl = (\id\ox\eps)\Dl = \id : A \to A$;
\item Compatibility:
$\Dl$ and $\eps$ are unital algebra homomorphisms.
\end{enumerate}
\end{defn}

The first two conditions, expressed in terms of elements $x,y,z$
of~$A$, say that $(xy)z = x(yz)$ and $1_A x = x 1_A = x$. The next two
properties are obtained by ``reversing the arrows''. Commutativity may
be formulated by using the ``flip map''
$\sg\: A \ox A \to A \ox A : x \ox y \mapsto y \ox x$: the bialgebra
is \textit{commutative} if $m\sg = m : A \ox A \to A$. Likewise, the
bialgebra is called \textit{cocommutative} if
$\sg\Dl = \Dl : A \to A \ox A$.

The (co)associativity rules suggest the abbreviations
$$
m^2 := m(m \ox \id) = m(\id \ox m),  \qquad
\Dl^2 := (\Dl \ox \id) \Dl = (\id \ox \Dl) \Dl,
$$
with obvious iterations $m^3\: A^{\ox 4} \to A$,
$\Dl^3\: A \to A^{\ox 4}$; $m^r\: A^{\ox(r+1)} \to A$,
$\Dl^r\: A \to A^{\ox(r+1)}$.

\begin{exer}
\label{xr:coalg-morf} 
If $(C,\Dl,\eps)$ and $(C',\Dl',\eps')$ are coalgebras, a counital
coalgebra morphism between them is an $\FF$-linear map
$\ell: C \to C'$ such that $\Dl' \ell = (\ell \ox \ell) \Dl$ and
$\eps' \ell = \eps$. Show that the compatibility condition is
equivalent to the condition that $m$ and $u$ are
counital coalgebra morphisms.
\end{exer}

\begin{defn}
\label{df:bialg-convl} 
The vector space $\Hom(C,A)$ of $\FF$-linear maps from a coalgebra
$(C,\Dl,\eps)$ to an algebra $(A,m,\eta)$ has an operation of
\textit{convolution}: given two elements $f,g$ of this space, the map
$f * g \in \Hom(C,A)$ is defined as
$$
f * g := m (f \ox g) \Dl : C \to A.
$$
Convolution is associative because
\begin{align*}
(f * g) * h
&= m ((f * g) \ox h) \Dl = m(m \ox \id)(f \ox g \ox h)(\Dl \ox \id)\Dl
\\
&= m(\id \ox m)(f \ox g \ox h)(\id \ox \Dl)\Dl = m (f \ox (g * h)) \Dl
 = f * (g * h).
\end{align*}
This makes $\Hom(C,A)$ an algebra, whose unit is $\eta_A\eps_C$:
\begin{align*}
f * \eta_A\eps_C &= m (f \ox \eta_A\eps_C) \Dl
 = m (\id_A \ox \eta_A)(f \ox \id_\FF)(\id_C \ox \eps_C) \Dl
 = \id_A f \id_C = f,
\\
\eta_A\eps_C * f &= m (\eta_A\eps_C \ox f) \Dl
 = m (\eta_A \ox \id_A)(\id_\FF \ox f)(\eps_C \ox \id_C) \Dl
 = \id_A f \id_C = f.
\end{align*}
\end{defn}

A \textit{bialgebra morphism} is a linear map $\ell \: H \to H'$
between two bialgebras, which is both a unital algebra homomorphism
and a counital coalgebra homomorphism; that is, $\ell$ satisfies the
four identities
$$
\ell m = m'(\ell \ox \ell),  \qquad  \ell\eta = \eta',  \qquad
\Dl'\ell = (\ell \ox \ell)\Dl,  \qquad  \eps'\ell = \eps,
$$
where the primes indicate coalgebra operations for~$H'$.

A bialgebra morphism respects convolution, in the following ways; if
$f,g \in \Hom(C,H)$ and $h,k \in \Hom(H',A)$ for some coalgebra $C$
and some algebra $A$, then
\begin{align*}
\ell(f * g) &= \ell m(f \ox g)\Dl_C = m'(\ell \ox \ell)(f \ox g)\Dl_C
 = m'(\ell f \ox \ell g)\Dl_C = \ell f * \ell g,
\\
(h * k)\ell &= m_A(h \ox k)\Dl'\ell = m_A(h \ox k)(\ell \ox \ell)\Dl
 = m_A(h\ell \ox k\ell)\Dl = h\ell * k\ell.
\end{align*}

\begin{defn}
\label{df:Hopf-alg} 
A \textbf{Hopf algebra} is a bialgebra $H$ together with a
(necessarily unique) convolution inverse $S$ for the identity map
$\id = \id_H$; the map $S$ is called the \textbf{antipode} of~$H$.
Thus,
$$
\id * S = m (\id \ox S) \Dl = \eta\eps,  \qquad
S * \id = m (S \ox \id) \Dl = \eta\eps.
$$
\end{defn}

A bialgebra morphism between Hopf algebras is automatically a Hopf
algebra morphism, i.e., it exchanges the antipodes:
$\ell S = S'\ell$. For that, it suffices to prove that these maps
provide a left inverse and a right inverse for $\ell$ in $\Hom(H,H')$.
Indeed, since the identity in $\Hom(H,H')$ is $\eta'\eps$, it is
enough to notice that
$$
\ell S * \ell = \ell(S * \id) = \ell\eta\eps = \eta'\eps
 = \eta'\eps'\ell = (\id' * S')\ell = \ell * S'\ell,
$$
and associativity of convolution then yields
$$
S'\ell = \eta'\eps * S'\ell = \ell S * \ell * S'\ell
 = \ell S * \eta'\eps = \ell S.
$$

The antipode has an important pair of algebraic properties: it is an
\textit{antihomomorphism} for both the algebra and the coalgebra
structures. Formally, this means
\begin{equation}
Sm = m\sg(S \ox S)  \sepword{and}   \Dl S = (S \ox S) \sg\Dl.
\label{eq:anti-antp} 
\end{equation}
The first relation, evaluated on $a \ox b$, becomes the familiar
antihomomorphism property $S(ab) = S(b)S(a)$. We postpone its proof
until a little later.

\begin{exmp}
\label{eg:group-alg} 
The simplest example of a Hopf algebra is the ``group algebra''
$\FF G$ of a finite group~$G$. This is just the vector space over
$\FF$ with a basis labelled by the elements of~$G$; the necessary
linear maps are specified on this basis. The product is given by
$m(x \ox y) := xy$, linearly extending the group multiplication, and
$\eta(1) := 1_G$ gives the unit map. The coproduct, counit and
antipode satisfy $\Dl(x) := x \ox x$, $\eps(x) := 1$ and
$S(x) := x^{-1}$, for each $x \in G$.
\end{exmp}

\begin{exer}
\label{xr:group-like} 
In a general Hopf algebra $H$, a nonzero element $g$ is called
\textit{grouplike} if $\Dl(g) := g \ox g$. Show that this condition
entails that $g$ is invertible and that $\eps(g) = 1$ and
$S(g) = g^{-1}$.
\end{exer}

There are two main ``classical'' examples of Hopf algebras:
representative functions on a compact group and the enveloping algebra
of a Lie algebra.

\begin{exmp}
\label{eg:rep-fns} 
Now let $G$ be a compact topological group (most often, a Lie group),
and let the scalar field $\FF$ be either $\R$ or~$\C$. The Peter--Weyl
theorem \cite[III.3]{BroeckerD} shows that any unitary irreducible
representation $\pi$ of~$G$ is finite-dimensional, any matrix element
$f(x) := \<u,\pi(x)v>$ is a continuous function on~$G$, and the vector
space $\Rr(G)$ generated by these matrix elements is a dense
subalgebra ($*$-subalgebra in the complex case) of $C(G)$.
Elements of $\Rr(G)$ can be characterized as those continuous
functions $f\: G \to \FF$ whose translates
$f_t : x \mapsto f(t^{-1}x)$ generate a finite-dimensional subspace of
$C(G)$; they are called \textit{representative functions} on~$G$.

The algebra $\Rr(G)$ is a $G$-bimodule in the sense of 
Wildberger~\cite{Wildberger} under left and right translation; indeed, 
it is the algebraic direct sum of the finite-dimensional irreducible 
\mbox{$G$-subbimodules} of~$C(G)$.

The group structure of $G$ makes $\Rr(G)$ a coalgebra. Indeed, we can
identify the \textit{algebraic} tensor product $\Rr(G) \ox \Rr(G)$ with
$\Rr(G \x G)$ in the obvious way ---here is where the
finite-dimensionality of the translates is used
\cite[Lemma~1.27]{Polaris}--- by $(f \ox g)(x,y) := f(x)g(y)$, and then
\begin{equation}
\Dl f(x,y) := f(xy)
\label{eq:coprod-repfns} 
\end{equation}
defines a coproduct on $\Rr(G)$. The counit is $\eps(f) := f(1)$, and
the antipode is given by $Sf(x) := f(x^{-1})$.
\end{exmp}

\begin{exmp}
\label{eg:univ-env-alg} 
The \textit{universal enveloping algebra} $\U(\g)$ of a Lie algebra
$\g$ is the quotient of the tensor algebra $\Tee(\g)$ by the two sided
ideal $I$ generated by the elements $XY - YX - [X,Y]$, for all
$X,Y \in \g$. (Here we write $XY$ instead of $X \ox Y$, to 
distinguish products within $\Tee(\g)$ from elements of 
$\Tee(\g) \ox \Tee(\g)$.) The coproduct and counit are defined on $\g$
by
\begin{equation}
\Dl(X) := X \ox 1 + 1 \ox X,
\label{eq:coprod-tensalg} 
\end{equation}
and $\eps(X) := 0$. These linear maps on $\g$ extend to homomorphisms
of the tensor algebra; for instance,
$$
\Dl(XY) = \Dl(X)\Dl(Y) = XY \ox 1 + X \ox Y + Y \ox X + 1 \ox XY,
$$
and thus
$$
\Dl(XY - YX - [X,Y])
 = (XY - YX - [X,Y]) \ox 1 + 1 \ox (XY - YX - [X,Y]),
$$
so $\Dl(I) \subseteq I \ox \U(\g) + \U(\g) \ox I$. Clearly,
$\eps(I) = 0$, too. Therefore, $I$ is both an ideal and a ``coideal''
in the full tensor algebra, so the quotient $\U(\g)$ is a bialgebra,
in fact a Hopf algebra: the antipode is given by $S(X) := -X$.

{}From \eqref{eq:coprod-tensalg}, the Hopf algebra $\U(\g)$ is clearly
\textit{cocommutative}. The word ``universal'' is appropriate because
any Lie algebra homomorphism $\psi \: \g \to A$, where $A$ is an
unital associative algebra, extends uniquely (in the obvious way) to a
unital algebra homomorphism $\Psi \: \U(\g) \to A$.
\end{exmp}

\begin{exmp}
\label{eg:SUq2} 
Historically, an important example of a Hopf algebra is Woronowicz'
$q$-defor\-ma\-tion of $SU(2)$. The compact group $SU(2)$ consists of
complex matrices $g = \twobytwo{a}{-c^*}{c}{a^*}$, subject to the
unimodularity condition $a^*a + c^*c = 1$. The matrix elements $a$
and~$c$, regarded as functions of~$g$, generate the $*$-algebra
$\Rr(SU(2))$: that is, any matrix element of a unitary irreducible
(hence finite-dimensional) representation of $SU(2)$ is a polynomial
in $a,a^*,c,c^*$.

Woronowicz found \cite{WoronowiczTwist} a noncommutative $*$-algebra
with two generators $a$ and $c$, subject to the relations
$$
ac = qca, \quad  ac^* = qc^*a, \quad  cc^* = c^*c, \quad
a^*a + c^*c = 1, \quad  aa^* + q^2 cc^* = 1,
$$
where $q$ is a \textit{real} number, which can be taken in the range
$0 < q \leq 1$. For the coalgebra structure, take $\Dl$ and $\eps$ be
$*$-homomorphisms determined by
$$
\Dl a := a \ox a - qc^* \ox c,  \qquad  \Dl c := c \ox a + a^* \ox c,
$$
and $\eps(a) := 1$, $\eps(c) := 0$. One can check that, by applying
$\Dl$ elementwise, the matrix $g := \twobytwo{a}{-qc^*}{c}{a^*}$
satisfies $\Dl(g) = g \ox g$. The antipode $S$ is the linear
antihomomorphism determined by
$$
S(a) := a^*, \quad  S(a^*) := a, \quad  S(c) := -qc, \quad
S(c^*) := -q^{-1}c^*,
$$
so that $x \mapsto S(x^*)$ is an antilinear homomorphism, indeed an
involution: $S(S(x^*)^*) = x$ for all~$x$. This last relation is a
general property of Hopf algebras with an involution.

The initial interest of this example was that it could be
represented by a $*$-algebra of bounded operators on a Hilbert space,
whose closure was a $C^*$-algebra which could legitimately be called a
\textit{deformation} of $C(SU(2))$; it has become known as
$C(SU_q(2))$. In this way, the ``quantum group'' $SU_q(2)$ was born.
Nowadays, many $q$-deformations of the classical groups are known,
although $q$ may not always be real: for example, to define
$SL_q(2,\R)$, one needs selfadjoint generators $a$ and $c$ satisfying
$ac = qca$, which is only possible if $q$ is a complex number of
modulus~$1$.
\end{exmp}

\marker
If $u_{ij}(x) := \<e_i,\pi(x)e_j>$, for $i,j = 1,\dots,n$, are the
matrix elements of an $n$-dimensional irreducible representation of a
compact group $G$ with respect to an orthonormal basis $\{\row e1n\}$,
then \eqref{eq:coprod-repfns} and $\pi(xy) = \pi(x)\pi(y)$ show that
\begin{subequations}
\label{eq:matr-coprod} 
\begin{equation}
\Dl u_{ij} = \tsum_{k=1}^n u_{ik} \ox u_{kj},
\label{eq:matr-coprod-one} 
\end{equation}
and the coassociativity of $\Dl$ is manifested as
\begin{equation}
\Dl^2 u_{ij} = \tsum_{k,l} u_{ik} \ox u_{kl} \ox u_{lj},
\label{eq:matr-coprod-two} 
\end{equation}
\end{subequations}
reflecting the associativity of matrix multiplication. This may be
generalized by a notational trick due to Sweedler~\cite{Sweedler}:
if $a$ is an element of any Hopf algebra, we write
$$
\Dl a =: \tsum a_{:1} \ox a_{:2} \quad\text{(finite sum)}.
$$
(The prevalent custom is to write $\Dl a = \tsum a_{(1)} \ox a_{(2)}$,
leading to a surfeit of parentheses.) The equality of
$(\Dl \ox \id)(\Dl a) = \tsum a_{:1:1} \ox a_{:1:2} \ox a_{:2}$ and
$(\id \ox \Dl)(\Dl a) = \tsum a_{:1} \ox a_{:2:1} \ox a_{:2:2}$ is
expressed by rewriting both sums as
$$
\Dl^2 a = \tsum a_{:1} \ox a_{:2} \ox a_{:3}.
$$
The matricial coproduct \eqref{eq:matr-coprod-two} is a particular
instance of this notation. The counit and antipode properties can now
be rewritten as
\begin{subequations}
\label{eq:Swee-notn} 
\begin{align}
\tsum \eps(a_{:1})\,a_{:2} &= \tsum a_{:1}\,\eps(a_{:2}) = a,
\label{eq:Swee-notn-counit} 
\\
\tsum S(a_{:1})\,a_{:2} &= \tsum a_{:1}\,S(a_{:2}) = \eps(a)\,1.
\label{eq:Swee-notn-antp} 
\end{align}
\end{subequations}
The coalgebra antihomomorphism property of $S$ is expressed as
\begin{equation}
\Dl(S(a)) = \tsum S(a_{:2}) \ox S(a_{:1}).
\label{coanti-antp} 
\end{equation}

We can now prove the antipode properties \eqref{eq:anti-antp}.
We show that $Sm \: a \ox b \mapsto S(ab)$ and
$m\sg(S \ox S) : a \ox b \mapsto S(b)S(a)$ are one-sided convolution
inverses for~$m$ in $\Hom(H\ox H, H)$, so they must coincide. The 
coproduct in $H \ox H$ is $(\id\ox\sg\ox\id)(\Dl\ox\Dl)
 : a \ox b \mapsto \tsum a_{:1}\ox b_{:1}\ox a_{:2}\ox b_{:2}$, and so
\begin{align*}
(Sm * m)(a \ox b)
&= m(Sm \ox m)\bigl(\tsum a_{:1}\ox b_{:1}\ox a_{:2}\ox b_{:2}\bigr)
 = \tsum S(a_{:1} b_{:1}) a_{:2} b_{:2}
\\
&= (S * \id)(ab) = \eta\eps(ab) = \eta\eps_{H\ox H}(a \ox b).
\end{align*}
On the other hand, writing $\tau := m\sg(S \ox S)$,
\begin{align*}
(m * \tau)(a \ox b)
&= m(m \ox \tau)\bigl(\tsum a_{:1}\ox b_{:1}\ox a_{:2}\ox b_{:2}\bigr)
 = \tsum a_{:1} b_{:1} S(b_{:2}) S(a_{:2})
\\
&= \eps(b) \tsum a_{:1} S(a_{:2}) = \eps(a)\eps(b)\,1_H
 = \eta\eps(ab) = \eta \eps_{H\ox H}(a \ox b).
\end{align*}
Thus, $Sm * m = \eta_H\eps_{H\ox H} = m * \tau$, as claimed. In like
fashion, one can verify \eqref{coanti-antp} by showing that
$\Dl S * \Dl = \eta_{H\ox H}\eps = \Dl * ((S \ox S)\sg\Dl)$ in
$\Hom(H, H \ox H)$; we leave the details to the reader.

\begin{exer}
\label{xr:coanti-antp} 
Carry out the verification of $\Dl S = (S \ox S)\sg\Dl$.
\end{exer}

Notice that in the examples $H = \Rr(G)$ and $H = \U(\g)$, the 
antipode satisfies $S^2 = \id_H$, but this does not hold in the 
$SU_q(2)$ case. We owe the following remark to Matthias 
Mertens~\cite[Satz~2.4.2]{Mertens}: $S^2 = \id_H$ if and only if
\begin{equation}
\tsum S(a_{:2})\,a_{:1} = \tsum a_{:2}\,S(a_{:1}) = \eps(a)\,1
  \sepword{for all} a \in H.
\label{eq:antp-invol} 
\end{equation}
Indeed, if $S^2 = \id_H$, then 
$$
\tsum S(a_{:2})\,a_{:1} = \tsum S(a_{:2})\,S^2(a_{:1})
 = S\bigl(\tsum S(a_{:1})\,a_{:2}\bigr) = S(\eps(a)\,1) = \eps(a)\,1,
$$
while the relation $\tsum S(a_{:2})\,a_{:1} = \eps(a)\,1$ implies that
$$
(S * S^2)(a) = \tsum S(a_{:1})\,S^2(a_{:2})
 = S\bigl(\tsum S(a_{:2})\,a_{:1}\bigr) = S(\eps(a)\,1) = \eps(a)\,1,
$$
so that \eqref{eq:antp-invol} entails $S * S^2 = S^2 * S = \eta\eps$,
hence $S^2 = \id_H$ is the (unique) convolution inverse for~$S$. Now, 
the relations \eqref{eq:antp-invol} clearly follow from
\eqref{eq:Swee-notn-antp} if $H$ is either commutative or
cocommutative (in the latter case, 
$\Dl a = \tsum a_{:1} \ox a_{:2} = \tsum a_{:2} \ox a_{:1}$). It 
follows that $S^2 = \id_H$ if $H$ is either commutative or
cocommutative.

\marker
Just as locally compact but noncompact spaces are described by 
nonunital function algebras, one may expect that locally compact
but noncompact groups correspond to some sort of ``nonunital Hopf 
algebras''. The lack of a unit requires substantial changes in the 
formalism. At the purely algebraic level, an attractive alternative is 
the concept of ``multiplier Hopf algebra'' due to van~Daele
\cite{VanDaeleMult,VanDaeleDual}.

If $A$ is an algebra whose product is nondegenerate, that is, $ab = 0$
for all $b$ only if $a = 0$, and $ab = 0$ for all $a$ only if $b = 0$,
then there is a unital algebra $M(A)$ such that $A \subseteq M(A)$,
called the \textit{multiplier algebra} of~$A$, characterized by the
property that $xa \in A$ and $ax \in A$ whenever $x \in M(A)$ and
$a \in A$. Here, $M(A) = A$ if and only if $A$ is unital. A coproduct
on~$A$ is defined as a homomorphism $\Dl\: A \to M(A \ox A)$ such
that, for all $a,b,c \in A$,
$$
(\Dl a)(1 \ox b) \in A \ox A,  \sepword{and}
(a \ox 1)(\Dl b) \in A \ox A,
$$
and the following coassociativity property holds: 
$$
(a \ox 1 \ox 1)\,(\Dl\ox\id)\,((\Dl b)(1 \ox c))
 = (\id\ox\Dl)\,((a \ox 1)(\Dl b))\,(1 \ox 1 \ox c).
$$
There are then two well-defined linear maps from $A \ox A$ into
itself:
$$
T_1(a \ox b) := (\Dl a)(1 \ox b),  \sepword{and}
T_2(a \ox b) := (a \ox 1)(\Dl b).
$$
We say that $A$ is a \textit{multiplier Hopf algebra} 
\cite{VanDaeleMult} if $T_1$ and $T_2$ are bijective.

When $A$ is a (unital) Hopf algebra, one finds that 
$T_1^{-1}(a \ox b) = ((\id \ox S)\Dl a)(1 \ox b)$ and 
$T_2^{-1}(a \ox b) = (a \ox 1)((S \ox \id)\Dl b)$. In fact,
\begin{align*}
T_1(((\id \ox S)\Dl a)(1 \ox b)) 
&= \tsum T_1(a_{:1} \ox S(a_{:2})b) = \tsum a_{:1}\ox a_{:2}S(a_{:3})b
\\
&= \tsum a_{:1} \ox \eps(a_{:2})b = a \ox b,
\end{align*}
and $T_2((a \ox 1)((S \ox \id)\Dl b)) = a \ox b$ by a similar
argument. The bijectivity of $T_1$ and $T_2$ is thus a proxy for the
existence of an antipode. It is shown in \cite{VanDaeleMult} that from
the stated properties of $\Dl$, $T_1$ and $T_2$, one can construct
both a counit $\eps\: A \to \FF$ and an antipode $S$, though the 
latter need only be an antihomomorphism from $A$ to $M(A)$. 

The motivating example is the case where $A$ is an algebra of
functions on a locally compact group $G$ (with finite support, say, to 
keep the context algebraic), and $\Dl f(x,y) := f(xy)$ as before.
Then $T_1(f \ox g) : (x,y) \mapsto f(xy)g(y)$ also has finite support 
and the formula $(T_1^{-1}F)(x,y) := F(xy^{-1},y)$ shows that $T_1$ is
bijective; similarly for~$T_2$. A fully topological theory,
generalizing Hopf algebras to include $C_0(G)$ for any locally compact
group $G$ and satisfying Pontryagin duality, is now available: the
basic paper on that is~\cite{KustermansV}.

\marker
\textit{Duality} is an important aspect of Hopf algebras. If
$(C,\Dl,\eps)$ is a coalgebra, the linear dual space
$C^* := \Hom(C,\FF)$ is an algebra, as we have already seen, where the
product $f \ox g \mapsto (f \ox g)\Dl$ is just the restriction of
$\Dl^{\!t}$ to $C^* \ox C^*$; the unit is~$\eps^t$, where $^t$ denotes
transpose. (By convention, we do not write the multiplication
in~$\FF$, implicit in the identification $\FF \ox \FF \simeq \FF$.)
However, if $(A,m,u)$ is an algebra, then $(A^*,m^t,u^t)$ need not
be a coalgebra because $m^t$ takes $A^*$ to $(A \ox A)^*$ which
is generally much larger than $A^* \ox A^*$. Given a Hopf algebra
$(H,m,u,\Dl,\eps,S)$, we can replace $H^*$ by the subspace
$H^\circ := \set{f \in H^* : m^t(f) \in H^* \ox H^*}$; one can check
that $(H^\circ,\Dl^{\!t},\eps^t,m^t,u^t,S^t)$ is again a Hopf algebra,
called the \textit{finite dual} (or ``Sweedler dual'') of~$H$.

To see why $H^\circ$ is a coalgebra, we must check that
$m^t(H^\circ) \subseteq H^\circ \ox H^\circ$. So suppose that
$f \in H^*$ satisfies $m^t(f) = \sum_{j=1}^m g_j \ox h_j$, a
\textit{finite} sum with $g_j,h_j \in H^*$. We may suppose that the
$g_j$ are linearly independent, so we can find elements
$\row a1m \in H$ such that $g_j(a_k) = \dl_{jk}$. Now
$$
h_k(ab) = \sum_{j=1}^m g_j(a_k) h_j(ab) = f(a_k ab)
 = \sum_{j=1}^m g_j(a_k a) h_j(b),
$$
so $m^t(h_k) = \sum_{j=1}^m f_{jk} \ox h_j$, where
$f_{jk}(a) := g_j(a_k a)$; thus $h_k \in H^\circ$. A similar argument
shows that each $g_j \in H^\circ$, too.

However, $H^\circ$ is often too small to be useful: in practice, one
works with two Hopf algebras $H$ and $H'$, where each may be regarded
as included in the dual of the other. That is to say, we can write
down a bilinear form $\dst{a}{f} := f(a)$ for $a \in H$ and
$f \in H'$ with an implicit inclusion $H' \hookto H^*$. The
transposing of operations between the two Hopf algebras boils down to
the following five relations, for $a,b \in H$ and $f,g \in H'$:
\begin{gather*}
\dst{ab}{f} = \dst{a \ox b}{\Dl'f},  \qquad
\dst{a}{fg} = \dst{\Dl a}{f \ox g},  \qquad
\dst{S(a)}{f} = \dst{a}{S'(f)},
\\
\eps(a) = \dst{a}{1_{H'}},  \sepword{and}  \eps'(f) = \dst{1_H}{f}.
\end{gather*}
The nondegeneracy conditions which allow us to assume that
$H' \subseteq H^*$ and $H \subseteq {H'}^*$ are:
(i) $\dst{a}{f} = 0$ for all $f \in H'$ implies $a = 0$, and
(ii) $\dst{a}{f} = 0$ for all $a \in H$ implies $f = 0$.

Let $G$ be a compact connected Lie group whose Lie algebra is~$\g$.
The function algebra $\Rr(G)$ is a commutative Hopf algebra, whereas
$\U(\g)$ is a cocommutative Hopf algebra. On identifying $\g$ with the
space of left-invariant vector fields on the group manifold~$G$, we
can realize $\U(\g)$ as the algebra of left-invariant differential
operators on~$G$. If $X \in \g$, and $f \in \Rr(G)$, we define
$$
\dst{X}{f} := Xf(1) = \ddto{t} f(\exp tX),
$$
and more generally, $\dst{X_1\dots X_n}{f} := X_1(\cdots(X_nf))(1)$;
we also set $\dst{1}{f} := f(1)$. This yields a duality between
$\Rr(G)$ and $\U(\g)$. Indeed, the Leibniz rule for vector fields,
namely $X(fh) = (Xf)h + f(Xh)$, gives
\begin{align}
\dst{X}{fh} &= Xf(1) h(1) + f(1) Xh(1)
 = (X \ox 1 + 1 \ox X)(f \ox h)(1 \ox 1)
\notag \\
&= \Dl X(f \ox h)(1 \ox 1) = \dst{\Dl X}{f \ox h},
\label{eq:Leibniz-Ug} 
\end{align}
while
\begin{align*}
\dst{X \ox Y}{\Dl f} &= \ddto{t} \ddto{s} (\Dl f)(\exp tX \ox \exp sY)
 = \ddto{t} \ddto{s} f(\exp tX \exp sY)
\\
&= \ddto{t} (Yf)(\exp tX) = X(Yf)(1) = \dst{XY}{f}.
\end{align*}
If $\dst{D}{f} = 0$ for all $D \in U(\g)$, then $f$ has a vanishing
Taylor series at the identity of~$G$. Since representative functions
are real-analytic~\cite{Knapp}, this forces $f = 0$. On the other
hand, if $\dst{D}{f} = 0$ for all $f$, the left-invariant differential
operator determined by~$D$ is null, so $D = 0$ in~$\U(\g)$. The
remaining properties are easily checked.

\begin{defn}
\label{df:prim-elt} 
The relation \eqref{eq:Leibniz-Ug} shows that
$\Dl X = X \ox 1 + 1 \ox X$ encodes the Leibniz rule for vector
fields. In any Hopf algebra $H$, an element $h \in H$ for which
$\Dl h = h \ox 1 + 1 \ox h$ is called \textbf{primitive}. It follows
that $\eps(h) = 0$ and that $S(h) = -h$. In the enveloping algebra
$\U(\g)$, elements of~$\g$ are obviously primitive. If $a$ and $b$ are
primitive, then so is $ab - ba$, so the vector space $\Prim(H)$ of
primitive elements of~$H$ is actually a Lie algebra.
\end{defn}

Indeed, since the field of scalars $\FF$ has characteristic zero, the
\textit{only} primitive elements of $\U(\g)$ are those in $\g$, i.e.,
$\Prim(\U(\g)) = \g$: see \cite{Bourbaki}, \cite[Lemma~1.21]{Polaris}
or \cite[Prop.~5.5.3]{Montgomery}. (Over fields of prime
characteristic, there are other primitive elements in~$\U(\g)$
\cite{Montgomery}.)

\marker
If $H$ is a bialgebra and $A$ is an algebra, and if
$\phi,\psi\: H \to A$ are algebra homomorphisms, their convolution 
$\phi * \psi \in \Hom(H,A)$ is a linear map, and will be also a
homomorphism provided that $A$ is \textit{commutative}. Indeed,
$\phi * \psi = m(\phi \ox \psi)\Dl$ is a composition of three
homomorphisms in this case; the commutativity of $A$ is needed to
ensure that $m\: A \ox A \to A$ is multiplicative. A particularly 
important case arises when $A = \FF$.

\begin{defn}
\label{df:Hopf-char} 
A {\bf character} of an algebra is a nonzero linear functional which
is also multiplicative, that is,
$$
\mu(ab) = \mu(a)\,\mu(b) \sepword{for all} a,b;
$$
notice that $\mu(1) = 1$. The counit $\eps$ of a bialgebra is a
character. Characters of a bialgebra can be convolved, since
$\mu * \nu = (\mu \ox \nu)\Dl$ is a composition of homomorphisms. The
characters of a Hopf algebra $H$ form a \textbf{group} $\G(H)$ under
convolution, whose neutral element is~$\eps$; the inverse of $\mu$ is
$\mu S$.

A {\bf derivation} or ``infinitesimal character'' of a Hopf algebra
$H$ is a linear map $\dl: H \to \FF$ satisfying
$$
\dl(ab) = \dl(a) \eps(b) + \eps(a) \dl(b) \sepword{for all} a,b \in H.
$$
This entails $\dl(1_H) = 0$. The previous relation can also be
written as $m^t(\dl) = \dl \ox \eps + \eps \ox \dl$, which shows
that $\dl$ belongs to $H^\circ$ and is primitive there; in particular,
the bracket $[\dl,\del] := \dl * \del - \del * \dl$ of two derivations
is again a derivation. Thus the vector space $\Der_\eps(H)$ of
derivations is actually a Lie algebra.
\end{defn}

In the commutative case, there is another kind of duality to consider:
one that matches a Hopf algebra with its character group. A compact
topological group $G$ admits a normalized left-invariant integral
(the Haar integral): this can be thought of as a functional
$J\: \Rr(G) \to \R$, where the left-invariance translates as
$(\id \ox J)\Dl = \eta J$. (We leave it as an exercise to show that
this corresponds to the usual definition of an invariant integral.)
The evaluations at points of $G$ supply all the characters of this
Hopf algebra: $\G(\Rr(G)) \simeq G$. Conversely, if $H$ is a
commutative Hopf algebra possessing such a left-invariant
functional~$J$, then its character group is compact, and
$H \simeq \Rr(\G(H))$. These results make up the Tannaka--Kre\u{\i}n
duality theorem ---for the proofs, see \cite{Polaris} or
\cite{HochschildStruct}--- and it is important either to use real
scalars, or to consider only hermitian characters if complex scalars
are used. The totality of all $\C$-valued characters of $\Rr(G)$,
hermitian or not, is a complex group $G^\C$ called the
complexification of $G$ \cite[III.8]{BroeckerD}; for instance, if
$G = SU(n)$, then $G^\C \simeq SL(n,\C)$.

\marker
The action of vector fields in~$\g$ and differential operators
in~$\U(\g)$ on the space of smooth functions on~$G$, and more
generally on any manifold carrying a transitive action of the
group~$G$, leads to the notion of a \textit{Hopf action} of a Hopf
algebra $H$ on an algebra~$A$.

\begin{defn}
\label{df:Hopf-mod-alg} 
Let $H$ be a Hopf algebra. A (left) \textbf{Hopf $H$-module algebra}
$A$ is an algebra which is a (left) module for the algebra $H$ such
that $h\.1_A = \eps(h)\,1_A$ and
\begin{equation}
h\.(ab) = \tsum (h_{:1}\.a)(h_{:2}\.b)
\label{eq:Hopf-Leib} 
\end{equation}
whenever $a,b \in A$ and $h \in H$.
\end{defn}

Grouplike elements act by endomorphisms of $A$, since
$g\.(ab) = (g\.a)(g\.b)$ and $g\.1 = 1$ if $g$ is grouplike. On the
other hand, primitive elements of $H$ act by the usual Leibniz rule:
$h\.(ab) = (h\.a)b + a(h\.b)$ and $h\.1 = 0$ if
$\Dl h = h \ox 1 + 1 \ox h$. Thus \eqref{eq:Hopf-Leib} is a sort of
generalized Leibniz rule.

\marker
Duality suggests that an \textit{action} of $\U(\g)$ should manifest
itself as a \textit{coaction} of $\Rr(G)$.

\begin{defn}
\label{df:coact-right} 
A vector space $V$ is called a right \textit{comodule} for a Hopf
algebra $H$ if there is a linear map $\Phi\: V \to V \ox H$ (the
right \textit{coaction}) satisfying
\begin{equation}
(\Phi \ox \id) \Phi = (\id \ox \Dl) \Phi : V \to V \ox H \ox H, \qquad
(\id \ox \eps)\Phi = \id : V \to V.
\label{eq:coact-right} 
\end{equation}
In Sweedler notation, we may write the coaction as
$\Phi(v) =: \tsum v_{:0} \ox v_{:1}$, so
$\tsum v_{:0}\,\eps(v_{:1}) = v$ and
$\tsum v_{:0:0} \ox v_{:0:1} \ox v_{:1}
 = \tsum v_{:0} \ox v_{:1:1} \ox v_{:1:2}$; we can rewrite both sides
of the last equality as $\tsum v_{:0} \ox v_{:1} \ox v_{:2}$, where,
by convention, $v_{:r} \in H$ for $r \neq 0$ while $v_{:0} \in V$.

Left $H$-comodules are similarly defined; a linear map
$\Phi\: V \to H \ox V$ is a left coaction if
$$
(\id \ox \Phi) \Phi = (\Dl \ox \id) \Phi  \sepword{and}
(\eps \ox \id)\Phi = \id;
$$
it is convenient to write $\Phi(v) =: \tsum v_{:-1} \ox v_{:0}$
in this case.

If a $H$-comodule $A$ is also an algebra and if the coaction
$\Phi\: A \to A \ox H$ is an algebra homomorphism, we say that $A$ is
a (right) \textit{$H$-comodule algebra}. In this case,
$\tsum (ab)_{:0} \ox (ab)_{:1} = \tsum a_{:0}b_{:0} \ox a_{:1}b_{:1}$.
\end{defn}

If $H$ and $U$ are two Hopf algebras in duality, then any right
$H$-comodule algebra $A$ becomes a left $U$-module algebra, under
$$
X\.a := \tsum a_{:0}\, \dst{X}{a_{:1}},
$$
for $X \in U$ and $a \in A$. In symbols: $X$ acts as the operator
$(\id \ox \bra{X})\Phi$ on~$A$. Indeed, it is enough to note that
\begin{align*}
X\.(ab) &= \tsum a_{:0}b_{:0}\, \dst{X}{a_{:1}b_{:1}}
 = \tsum a_{:0}b_{:0}\, \dst{\Dl X}{a_{:1} \ox b_{:1}}
\\
& = \tsum a_{:0}b_{:0}\, \dst{X_{:1} \ox X_{:2}}{a_{:1} \ox b_{:1}}
 = \tsum a_{:0}\, \dst{X_{:1}}{a_{:1}}\, b_{:0}\, \dst{X_{:2}}{b_{:1}}
\\
&= \tsum (X_{:1} \. a)\, (X_{:2} \. b).
\end{align*}

The language of coactions is used to formulate what one obtains by
applying the Gelfand cofunctor (loosely speaking) to the concept of a
\textit{homogeneous space under a group action}. If a compact group
$G$ acts transitively on a space $M$, one can write
$M \approx G/K$, where $K$ is the closed subgroup fixing a basepoint
$z_0 \in M$ (i.e., $K$ is the ``isotropy subgroup'' of~$z_0$). Then
any function on~$M$ is obtained from a function on~$G$ which is
constant on right cosets of~$K$. If $\F(G)$ and $\F(M)$ denote
suitable algebras of functions on $G$ and~$M$ (we shall be more
precise about these algebras in a moment), then there is a
corresponding algebra of right $K$-invariant functions
$$
\F(G)^K :=
\set{f \in \F(G) : f(xw) = f(x) \text{ whenever } w \in K,\ x \in G}.
$$
If $\bar x \in M$ corresponds to the right coset $xK$ in $G/K$, then
$$
\zeta f(\bar x) := f(x)
$$
defines an algebra isomorphism $\zeta\: \F(G)^K \to \F(M)$. [For
aesthetic reasons, one may prefer to work with left $K$-invariant
functions; for that, one should instead identify $M$ with the space
$K\backslash G$ of left cosets of~$K$.]

Suppose now that the chosen spaces of functions satisfy
\begin{equation}
\F(G) \ox \F(M) \simeq \F(G \x M),
\label{eq:tens-Cartes} 
\end{equation}
where $\ox$ denotes, as before, the \textit{algebraic} tensor product.
Then we can define $\rho\: \F(M) \to \F(G) \ox \F(M)$ by
$\rho f(x,\bar y) := f(\overline{xy})$. It follows that
\begin{equation}
[\rho\zeta f](x,\bar y) = \zeta f(\overline{xy}) = f(xy) = \Dl f(x,y)
 = [(\id \ox \zeta)\Dl f](x,\bar y),
\label{eq:coact-twine} 
\end{equation}
so that $\rho\zeta = (\id \ox \zeta)\Dl: \F(G)^K \to \F(G) \ox \F(M)$.
Notice, in passing, that the coproduct $\Dl$ maps $\F(G)^K$ into
$\F(G) \ox \F(G)^K$, which consists of functions $h$ on $G \x G$ such
that $h(x,yw) = h(x,y)$ when $w \in K$. [Had we used left cosets and
left-invariant functions, the corresponding relations would be
$\Dl(\F(G)^K) \subseteq \F(G)^K \ox \F(G)$,
$\rho\:\F(M) \to \F(M) \ox \F(G)$, and
$\rho\zeta = (\zeta \ox \id)\Dl$.] In Hopf algebra language, $\rho$
defines a left [or right] coaction of $\F(G)$ on the algebra $\F(M)$,
implementing the left [or right] action of the group $G$ on~$M$,
and $\zeta$ intertwines this with left [or right] regular coaction on
$K$-invariant functions induced by the coproduct~$\Dl$. We get an
instance of the following definition.

\begin{defn}
\label{df:homog-space} 
In the lore of quantum groups ---see, for instance,
\cite[\S 11.6]{KlimykS}--- a (left) \textbf{embedded homogeneous
space} for a Hopf algebra $H$ is a left $H$-comodule algebra $A$ with
coaction $\rho\: A \to H \ox A$, for which there exists a
\textit{subalgebra} $B \subseteq H$ and an algebra isomorphism
$\zeta\: B \to A$ such that
$\rho\zeta = (\id \ox \zeta)\Dl: B \to H \ox A$.

A right embedded homogeneous space is defined, \textit{mutatis
mutandis}, in the same way.
\end{defn}

There are two ways to ensure that the relation \eqref{eq:tens-Cartes}
holds. One way is to choose $\F(G) := \Rr(G)$, which is a bona-fide
Hopf algebra, and then to \textit{define} $\Rr(M)$ as the image
$\zeta(\Rr(G)^K)$ of the $K$-invariant representative functions. For
instance, if $G = SU(2)$ and $K = U(1)$, so that $M \approx \Sf^2$ is
the usual 2-sphere of spin directions, then $\Rr(G)$ is spanned by
the matrix elements $\D_{mn}^j$ of the $(2j+1)$-dimensional unitary
irreducible representations of $SU(2)$: see \cite{BiedenharnL}, for
example. Now $\D_{mn}^j$ is right $U(1)$-invariant if and only if $j$
is an integer (not a half-integer) and $n = 0$; moreover, the
functions $Y_{lm} := \sqrt{(2l + 1)/4\pi}\, \D_{m0}^{l*}$ are the
usual spherical harmonics on the 2-sphere. In other words:
$\Rr(\Sf^2)$ is the algebra of spherical harmonics on~$\Sf^2$.

\marker
To move closer to noncommutative geometry, it would be better to use
either \textit{continuous} functions (at the $C^*$-algebra level) or
\textit{smooth} functions on $G$ and $M$; that is, one should work
with $\F = C$ or with $\F = \Coo$. Notice that formulas
like~\eqref{eq:coact-twine} make perfect sense in those cases; but the
tensor product relation~\eqref{eq:tens-Cartes} is false in the
continuous or smooth categories, unless the algebraic $\ox$ is
replaced by a more suitable \textit{completed} tensor product.

In the continuous case, for compact $G$ and~$M$, the relation
$$
C(G) \ox C(M) \simeq C(G \x M)
$$
is valid, where $\ox$ denotes the ``minimal'' tensor product of
$C^*$-algebras. (There may be several compatible $C^*$-norms on a
tensor product of two $C^*$-algebras; but they all coincide if the
algebras are commutative.) In the smooth case, we may fall back on a
theorem of Grothendieck~\cite{Grothendieck}, which says that
$$
\Coo(G) \hatox \Coo(M) \simeq \Coo(G \x M),
$$
where $\hatox$ denotes the projective tensor product of Fr\'echet
spaces. But then, it is necessary to go back and reexamine our
definitions: for instance, the coproduct need only satisfy
$\Dl(A) \subseteq A \ox A$ for a completed tensor product, which is a
much weaker statement than the original one --- the formula
$\Dl a = \tsum a_{:1} \ox a_{:2}$ need no longer be a finite sum, but
only some kind of convergent series. The bad news is that, in the
$C^*$-algebra case, the product map $m\: A \ox A \to A$ is usually not
continuous; the counit $\eps$ and antipode $S$ become unbounded linear
maps and one must worry about their domains; and so on. We shall meet
examples of these generalized Hopf algebras in
subsection~\ref{sec:Moyal-cpt-qgroup}.

\subsection{Hopf actions of differential operators: an example}
\label{sec:Hopf-diff} 

The Hopf algebras which are currently of interest are typically
neither commutative, like $\Rr(G)$, nor cocommutative, like $\U(\g)$.
The enormous profusion of ``quantum groups'' which have emerged in the
last twenty years provide many examples of such noncommutative,
noncocommutative Hopf algebras: see
\cite{ChariP,KasselQGroups,KlimykS,Majid} for catalogues of these. A
class of Hopf algebras which are commutative but are not cocommutative
were introduced a few years ago, first by Kreimer in a quantum field
theory context~\cite{KreimerHopf}, and independently by Connes and
Moscovici~\cite{ConnesMHopf} in connection with a local index formula
for foliations; in both cases, the Hopf algebra becomes a device to
organize complicated calculations. We shall discuss the QFT version at
length in the next section; here we look at the geometric example
first.

If one wishes to deal with gravity in a noncommutative geometric
framework~\cite{ConnesBrisure}, one must be able to handle the
geometrical invariants of spacetime under the action of local
diffeomorphisms. We consider an oriented $n$-dimensional manifold $M$,
without boundary. By local diffeomorphisms on~$M$ we mean
diffeomorphisms $\psi\: \Dom\psi \to \Ran\psi$, where both the domain
$\Dom\psi$ and range $\Ran\psi$ are open subsets of~$M$; and we shall
always assume that $\psi$ preserves the given orientation on~$M$. Two
such local diffeomorphisms can be composed if and only if the range of
the first lies within the domain of the second, and any local
diffeomorphism can be inverted: taken all together, they form what is
called a pseudogroup. We let $\Ga$ be a subpseudogroup (with the
discrete topology), and consider the pair $(M,\Ga)$.

The orbit space $M/\Ga$ has in most cases a very poor topology. The
noncommutative geometry approach is to replace this singular space by
an \textit{algebra} which captures the action of $\Ga$ on~$M$. The
initial candidate, a ``crossed product'' algebra $C(M) \semi \Ga$,
still has a very complicated structure; but much progress can be made
\cite{ConnesFoli} by replacing $M$ by the bundle $F \to M$ of
oriented frames on $M$. This is a principal fibre bundle whose
structure group is $GL^+(n,\R)$, the $n \x n$ matrices with positive
determinant.

Any $\psi \in \Ga$ admits a \textit{prolongation} to the frame bundle
described as follows. Let $x = (\row x1n)$ be local coordinates on $M$
and let $y = (y_1^1,y_1^2,\dots,y_n^n)$ be local coordinates for the
frame at~$x$. To avoid a ``debauch of indices'', we mainly consider
the $1$-dimensional case, where $M \approx \Sf^1$ is a circle and
$F$ is a cylinder (but we use a matrix notation to indicate how to
proceed for higher dimensions; the details for the general case are
carefully laid out in~\cite{WulkenhaarDiff}). Then $\psi$ acts locally
on~$F$ through $\Onda\psi$, given by
$$
\Onda\psi(x,y) := (\psi(x), \psi'(x)y).
$$
The point is that, while $M$ need not carry any $\Ga$-invariant
measure, the top-degree differential form $\nu = y^{-2}\,dy \w dx$
on $F$ is $\Ga$-invariant:
$$
\Onda\psi^* \nu
 = y^{-2} \psi'(x)^{-2} \,\psi'(x)\,dy \w \psi'(x)\,dx = \nu,
$$
so we can build a Hilbert space $L^2(F,\nu)$ and represent the action
of each $\psi \in \Ga$ by the \textit{unitary} operator $U_\psi$
defined by $U_\psi\xi(x,y) := \xi(\Onda\psi^{-1}(x,y))$. It is
slightly more convenient to work with the adjoint unitary operators
$U_\psi^\7\xi(x,y) := \xi(\Onda\psi(x,y))$. These unitaries intertwine
multiplication operators coming from functions on~$F$ (specifically,
smooth functions with compact support) as follows:
\begin{equation}
U_\psi f U_\psi^\7  = f^\psi,  \sepword{where}
 f^\psi(x,y) := f(\Onda\psi^{-1}(x,y)).
\label{eq:covt-repn} 
\end{equation}

The local action of $\Ga$ on~$F$ can be described in the language of
smooth groupoids~\cite{ConnesMDiff}, or alternatively by introducing a
``crossed product'' algebra which incorporates the groupoid
convolution. This is a pre-$C^*$-algebra $\A$ obtained by suitably
completing the algebra
$$
\lin\set{f U_\psi^\7 : \psi \in \Ga,\ f \in \Coo_c(\Dom\Onda\psi)}.
$$
The relation \eqref{eq:covt-repn} gives the multiplication rule
\begin{equation}
(f U_\psi^\7)(g U_\phi^\7) = f(U_\psi^\7 g U_\psi) U_\psi^\7 U_\phi^\7
 = f(g \circ \Onda\psi) U_{\phi\psi}^\7,
\label{eq:smash-prod} 
\end{equation}
Any two such elements are composable, since the support of
$f(g \circ \Onda\psi)$ is a compact subset of
$\Dom\Onda\psi \cap \Onda\psi^{-1}(\Dom\Onda\phi)
 \subseteq \Dom(\Onda\phi \Onda\psi)$.

This construction is called the \textit{smash product} in the Hopf
algebra books: if  $H$ is a Hopf algebra and $A$ is a left Hopf
$H$-module algebra, the smash product is the algebra $A \aspa H$ which
is defined as the vector space $A \ox H$ with the product rule
$$
(a \ox h)(b \ox k) := \tsum a (h_{:1}\.b) \ox h_{:2} k.
$$
If $h$ is a grouplike element of~$H$, this reduces to
$(a \ox h)(b \ox k) := a (h\.b) \ox hk$, of which
\eqref{eq:smash-prod} is an instance.

A local basis $\{X,Y\}$ of vector fields on the bundle $F$ is
defined by the ``vertical'' vector field $Y := y\,\del/\del y$,
generating translations along the fibres, and the ``horizontal''
vector field $X := y\,\del/\del x$, generating displacements
transverse to the fibres. In higher dimensions, the basis contains
$n^2$ vertical vector fields $Y_j^i$ and $n$ horizontal vector fields
$X_k$ \cite{WulkenhaarDiff}. Under the lifted action of~$\Ga$, $Y$ is
invariant:
$$
\Onda\psi_* Y = \psi'(x)y\,\dd{\psi'(x)y} = y\,\dd{y} = Y,
$$
but $X$ is not. To see that, consider the $1$-forms
$\a := y^{-1}\,dx$ and $\om := y^{-1}\,dy$. The form $\a$ is the
so-called canonical $1$-form on~$F$, which is invariant since
$\Onda\psi^* \a = y^{-1}\psi'(x)^{-1}\,d\psi(x) = y^{-1}\,dx = \a$,
whereas $\om$ is not invariant:
$$
\Onda\psi^* \om = y^{-1}\,dy + \psi'(x)^{-1} \,d\psi'(x)
 = y^{-1}\,dy + \frac{\psi''(x)}{\psi'(x)} \,dx.
$$
This transformation rule shows that $\om$ is a connection $1$-form on
the principal bundle $F \to M$; and the horizontality of~$X$ means,
precisely, that $\om(X) = 0$. Notice also that $\a(X) = 1$. Now the
vector field $\Onda\psi^{-1}_*X$ can be computed from the two
equations
$\a(\Onda\psi^{-1}_*X) = \Onda\psi^*\a(\Onda\psi^{-1}_*X) = \a(X) = 1$
and $\Onda\psi^*\om(\Onda\psi^{-1}_*X) = \om(X) = 0$; we get
\begin{subequations}
\label{eq:x-change} 
\begin{equation}
\Onda\psi^{-1}_*X = y \dd{x} - y^2 \frac{\psi''(x)}{\psi'(x)} \dd{y}
 = X - h_\psi Y,
\label{eq:x-change-psi} 
\end{equation}
where
\begin{equation}
h_\psi(x,y) := y \,\frac{\psi''(x)}{\psi'(x)}
 = y\,\dd{x} \bigl( \log\psi'(x) \bigr).
\label{eq:x-change-cocyc} 
\end{equation}
\end{subequations}

Any vector field $Z$ on~$F$ determines a linear operator on~$\A$, also
denoted by~$Z$, by
\begin{equation}
Z(f U_\psi^\7) := (Zf)\, U_\psi^\7,
\label{eq:lifted-vecfld} 
\end{equation}
which makes sense since
$\supp(Zf) \subseteq \supp f \subset \Dom\Onda\psi$. When applied to
products, this operator gives
\begin{align}
Z(f U_\psi^\7\,g U_\phi^\7) &= Z(f(g \circ \Onda\psi)) U_{\phi\psi}^\7
 = (Zf)(g \circ \Onda\psi) U_{\phi\psi}^\7
   + f Z(g \circ \Onda\psi) U_{\phi\psi}^\7
\notag \\
&= (Zf) U_\psi^\7 \,g U_\phi^\7
   + f U_\psi^\7\,(Z(g\circ \Onda\psi) \circ \Onda\psi^{-1}) U_\phi^\7
\notag \\
&= (Zf) U_\psi^\7 \,g U_\phi^\7
   + f U_\psi^\7 \,\Onda\psi_*Z(g) U_\phi^\7.
\label{eq:Hopf-actn-Z} 
\end{align}
Since the vector field $Y$ is invariant, $\Onda\psi_* Y = Y$, so the
lifted operator $Y$ is a derivation on the algebra~$\A$:
$$
Y(f U_\psi^\7 \,g U_\phi^\7) =
(Yf) U_\psi^\7 \,g U_\phi^\7 + f U_\psi^\7 \,(Yg) U_\phi^\7,
$$

\begin{prop}
\label{pr:coprod-Diff} 
The operator $X$ on~$\A$ is \emph{not} a derivation; however, there
is a derivation $\la_1$ on~$\A$ such that $X$ obeys the generalized
Leibniz rule
\begin{equation}
X(ab) = X(a) b + a X(b) + \la_1(a) Y(b) \sepword{for all} a,b \in \A.
\label{eq:Hopf-actn-X} 
\end{equation}
\end{prop}

\begin{proof}
Using the invariance of $Y$ and~\eqref{eq:x-change-psi}, we get
$$
\Onda\psi_* X - X = \Onda\psi_* (X - \Onda\psi^{-1}_* X)
 = \Onda\psi_* (h_\psi Y) = (h_\psi \circ \Onda\psi^{-1}) Y,
$$
and it follows that
$$
f U_\psi^\7 \,(\Onda\psi_* X(g) - Xg) U_\phi^\7
 = f U_\psi^\7 \,(h_\psi \circ \Onda\psi^{-1}) (Yg) U_\phi^\7
 = f\,h_\psi U_\psi^\7 \,(Yg) U_\phi^\7.
$$
If we define
\begin{equation}
\la_1(f U_\psi^\7) := h_\psi f U_\psi^\7,
\label{eq:la-mult} 
\end{equation}
then \eqref{eq:Hopf-actn-Z} for $Z = X$ now reads
$$
X(f U_\psi^\7 \,g U_\phi^\7)
 = X(f U_\psi^\7) \,g U_\phi^\7 + f U_\psi^\7 \,X(g U_\phi^\7)
   + \la_1(f U_\psi^\7) \, Y(g U_\phi^\7).
$$
Thus, \eqref{eq:Hopf-actn-X} holds on generators. We leave the reader
to check that the formula extends to finite products of generators,
provided that $\la_1$ is indeed a derivation. Now
\eqref{eq:x-change-cocyc} implies
$$
h_{\phi\psi}(x,y)
 = y\,\dd{x} \bigl( \log\phi'(\psi(x)) + \log\psi'(x) \bigr)
 = h_\phi(\Onda\psi(x,y)) + h_\psi(x,y),
$$
so that $h_{\phi\psi} = \Onda\psi^* h_\phi + h_\psi$, and the
derivation property of $\la_1$ follows:
\begin{align*}
\la_1(f U_\psi^\7 \,g U_\phi^\7)
&= (\Onda\psi^* h_\phi + h_\psi)\,f(g\circ\Onda\psi) U_{\phi\psi}^\7
\\
&= f\,((h_\phi g) \circ \Onda\psi) U_{\phi\psi}^\7
   + h_\psi f U_\psi^\7 \,g U_\phi^\7
\\
&= (f U_\psi^\7)(h_\phi g U_\phi^\7)
   + (h_\psi f U_\psi^\7)(g U_\phi^\7).
\tag* \qed
\end{align*}
\hideqed
\end{proof}

Consider now the Lie algebra obtained from the operators $X$, $Y$ and
$\la_1$. The vector fields $X$, $Y$ have the commutator
$[y\,\del/\del y, y\,\del/\del x] = y\,\del/\del x$ and the
corresponding operators on~$\A$ satisfy $[Y,X] = X$. Next,
$[Y,\la_1](f U_\psi^\7) = f (Yh_\psi) U_\psi^\7$,
and from $Yh_\psi = h_\psi$ we get $[Y,\la_1] = \la_1$. Similarly,
$[X,\la_1](f U_\psi^\7) = f (Xh_\psi) U_\psi^\7$, where
$Xh_\psi = y\,\del/\del x \bigl(y \,\psi''(x)/\psi'(x)\bigr)
 = y^2\,\del^2/\del x^2 \bigl( \log\psi'(x) \bigr)$. Introduce
$$
h_\psi^n = y^n \,\frac{d^n}{dx^n} \log\psi'(x),
$$
for $n = 1,2,\dots$, and define
$\la_n(f U_\psi^\7) := f h_\psi^n U_\psi^\7$, then $\la_2 = [X,\la_1]$
and by induction we obtain $\la_{n+1} = [X,\la_n]$ for all~$n$.
Clearly $Yh_\psi^n = nh_\psi^n$, which implies $[Y,\la_n] = n\la_n$.
The operators $\la_n$ commute among themselves. We have constructed a
Lie algebra, linearly generated by $X$, $Y$, and all the~$\la_n$.

We can make the associative algebra with these same generators into a
Hopf algebra~\cite{ConnesMHopf} by defining their coproducts as
follows. Since $Y$ and $\la_1$ act as derivations, they must be
primitive:
\begin{subequations}
\label{eq:CM-coprod} 
\begin{align}
\Dl Y &:= Y \ox 1 + 1 \ox Y,
\label{eq:CM-coprod-Y}  
\\
\Dl\la_1 & := \la_1 \ox 1 + 1 \ox \la_1.
\label{eq:CM-coprod-la} 
\end{align}
The coproduct of~$X$ can be read off from \eqref{eq:Hopf-actn-X}:
\begin{equation}
\Dl X := X \ox 1 + 1 \ox X + \la_1 \ox Y.
\label{eq:CM-coprod-X}  
\end{equation}
\end{subequations}
Moreover, $\eps(Y) = \eps(\la_1) = 0$ since $Y$ and $\la_1$ are
primitive, and $\eps(X) = 0$ since $X = [Y,X]$ is a commutator;
moreover, $\eps(\la_n) = 0$ for all $n \geq 2$ for the same reason.
The commutation relations yield the remaining coproducts; for
instance,
$$
\Dl\la_2 := [\Dl X, \Dl\la_1]
 = \la_2 \ox 1 + 1 \ox \la_2 + \la_1 \ox \la_1.
$$
The antipode is likewise determined: $S(Y) = -Y$ and
$S(\la_1) = -\la_1$ since $Y$ and $\la_1$ are primitive, and
$(\id * S)(X) = \eps(X)1 = 0$ gives $X + S(X) + \la_1 Y = 0$, so
$S(X) = -X + \la_1 Y$. The relation $S(\la_{n+1}) = [S(\la_n),S(X)]$
yields all $S(\la_n)$ by induction.

\begin{defn}
\label{df:Hopf-CM} 
The Hopf algebra $H_{CM}$ generated as an algebra by $X$, $Y$
and~$\la_1$, with the coproduct determined by~\eqref{eq:CM-coprod} and
the indicated counit and antipode, will be called the
\textit{Connes--Moscovici Hopf algebra}.
\end{defn}

\begin{exer}
\label{xr:CMHopf-subalg} 
Show that the commutative subalgebra generated by
$\set{\la_n : n = 1,2,3,\dots}$ is indeed a Hopf subalgebra which is
not cocommutative.
\end{exer}

The example $H_{CM}$ arose in connection with a local index formula
computation, which is already very involved when the base space $M$
has dimension~$1$ (the case treated above). In higher dimensions, one
may start~\cite{WulkenhaarDiff} with the vertical vector fields
$Y_j^i = y_j^\mu \,\del/\del y_i^\mu$ and a matrix-valued connection
$1$-form
$\om_j^i = (y^{-1})_\mu^i (dy_j^\mu + \Ga_{\a\b}^\mu y_j^\a \,dx^\b)$,
which may be chosen torsion-free, with Christoffel symbols
$\Ga_{\a\b}^\mu = \Ga_{\b\a}^\mu$. With respect to this connection
form, there are horizontal vector fields
$X_k = y_k^\mu(\del/\del x^\mu
 - \Gamma_{\a\mu}^\nu y_j^\a \,\del/\del y_\nu^j)$. One obtains the
Lie algebra relations $[Y_i^j, Y_k^l] = \dl_k^j Y_i^l - \dl_i^l Y_k^j$
and $[Y_i^j, X_k] = \dl_k^j X_i$, involving ``structure constants'';
however, $[X_k, X_l] = R_{jkl}^i \,Y_i^j$ where $R_{jkl}^i$ are the
components of the curvature of the connection~$\om$, and these
coefficients are in general not constant, for $n > 1$.

At first, Connes and Moscovici decided to use flat connections
only~\cite{ConnesMHopf}, which entails $[X_k, X_l] = 0$; then, on
lifting the $Y_j^i$ and the $X_k$ using~\eqref{eq:lifted-vecfld}, a
higher-dimensional analogue of $H_{CM}$ is obtained. For instance, one
gets~\cite{WulkenhaarDiff}:
$$
\Dl X_k = X_k \ox 1 + 1 \ox X_k + \la_{kj}^i \ox Y_i^j,
$$
where the $\la_{kj}^i$ are derivations of the form~\eqref{eq:la-mult}.

A better solution was later found~\cite{ConnesMDiff}: one can allow
commutation relations like $[X_k, X_l] = R_{jkl}^i \,Y_i^j$ if one
modifies the original setup to allow for ``transverse differential
operators with nonconstant coefficients''. The algebra $\A$ remains
the same as before, but the base field $\C$ is replaced by the
algebra $\Rr = \Coo(F)$ of smooth functions on~$F$. Now $\A$ is an
$\Rr$-bimodule under the commuting left and right actions
\begin{subequations}
\label{eq:mult-op} 
\begin{align}
\a(b) : f U_\psi^\7 \mapsto b\.(f U_\psi^\7) &:= (bf) U_\psi^\7,
\label{eq:left-mult-op}  
\\
\b(b) : f U_\psi^\7 \mapsto (f U_\psi^\7)\.b
&:= (b \circ \Onda\psi)\. (f U_\psi^\7)
  = (f(b \circ \Onda\psi)) U_\psi^\7.
\label{eq:right-mult-op} 
\end{align}
\end{subequations}
Letting $H$ now denote the algebra of operators on~$\A$ generated by
these operators \eqref{eq:mult-op} and the previous ones
\eqref{eq:lifted-vecfld}, then we no longer have a Hopf algebra
over~$\C$, but $(H,\Rr,\a,\b)$ gives an instance of a more general
structure called a \textit{Hopf algebroid} over~$\Rr$~\cite{Lu}. For
instance, the coproduct is an $\Rr$-bimodule map from $H$ into
$H \ox_\Rr H$, where elements of this range space satisfy
$(h\.b) \ox_\Rr k = h \ox_\Rr (b\.k)$ by construction, for any
$b \in \Rr$. Just as Hopf algebras are the noncommutative counterparts
of groups, Hopf algebroids are the noncommutative counterparts of
groupoids: see \cite{Lu,Xu} for instance. For the details of these
recent developments, we refer to~\cite{ConnesMDiff}.

\newpage


\section{The Hopf Algebras of Connes and Kreimer}

\subsection{The Connes--Kreimer algebra of rooted trees}
\label{sec:CKr-rooted-tree} 

A very important Hopf algebra structure is the one found by Kreimer
\cite{KreimerHopf} to underlie the combinatorics of subdivergences in
the computation of perturbative expansions in quantum field theory.
Such calculations involve several layers of complication, and it is no
small feat to remove one such layer by organizing them in terms of a
certain coproduct: indeed, the corresponding antipode provides a
method to obtain suitable counterterms. Instead of addressing this
matter from the physical side, the approach taken here is algebraic,
in order first to understand why the Hopf algebras which emerge are in
the nature of things.

A given Feynman graph represents a multiple integral (say, over
momentum space) where the integrand is assembled from a definite
collection of Rules, and before renormalization will often be
superficially divergent, as determined by power counting. Even if not
itself divergent, it may well contain one or several subgraphs which
yield divergent partial integrations: the first order of business is
to catalogue and organize the various graphs according to this nesting
of subdivergences. Kreimer's coproduct separates out the divergences
of subgraphs from those of the overall graph. In consequence, when
expressed in terms of suitable generators of a Hopf algebra, the
coproduct turns out to be polynomial in its first tensor factor, but
merely linear in the second factor, and is therefore highly
noncocommutative. Our starting point is to find a source of Hopf
algebras with this kind of noncocommutativity.

\marker
We start with an apparently unrelated digression into the homological
classification of (associative) algebras.

There is a natural homology theory for associative algebras, linked
with the name of Hochschild. Given an algebra $\A$ over any field
$\FF$ of scalars, one forms a \textit{complex} by setting
$C_n(\A) := \A^{\ox(n+1)}$, and defining the boundary operator
$b\: C_n(\A) \to C_{n-1}(\A)$ by
$$
b(a_0 \ox a_1 \oxyox a_n)
 := \sum_{j=0}^{n-1} (-1)^j a_0 \oxyox a_ja_{j+1} \oxyox a_n
    + (-1)^n a_na_0 \ox a_1 \oxyox a_{n-1},
$$
where the last term ``turns the corner''. By convention, $b = 0$
on~$C_0(\A) = \A$. One checks that $b^2 = 0$ by cancellation. For
instance, $b(a_0 \ox a_1) := [a_0, a_1]$, while
$$
b(a_0 \ox a_1 \ox a_2)
 := a_0a_1 \ox a_2 - a_0 \ox a_1a_2 + a_2a_0 \ox a_1.
$$

There are two important variants of this definition. One comes from
the presence of a ``degenerate subcomplex'' $D_\8(\A)$ where, for each
$n = 0,1,2,\dots$, the elements of $D_n(\A)$ are finite sums of terms
of the form $a_0 \oxyox a_j \oxyox a_n$, with $a_j = 1$ for some
$j = 1,2,\dots,n$; elements of the quotient
$\Om^n\A := C_n(\A)/D_n(\A) = \A \ox \Bar\A^{\ox n}$, where
$\Bar\A = \A/\FF$, are sums of expressions $a_0 \,da_1 \dots da_n$
where $d(ab) = da\,b + a\,db$. The direct sum
$\Om^\8\A = \bigoplus_{n\geq 0} \Om^n\A$ is the universal graded
differential algebra generated by $\A$ in degree zero; using it, $b$
can be rewritten as
\begin{align}
b(a_0 \,da_1 \dots da_n)
&:= a_0a_1 \,da_2 \dots da_n
    + \sum_{j=1}^{n-1} (-1)^j a_0\,da_1\dots d(a_ja_{j+1})\dots da_n
\notag \\
&\qquad + (-1)^n a_na_0 \,da_1 \dots da_{n-1}.
\label{eq:Oma-bdry} 
\end{align}
The second variant involves replacing the algebra $\A$ in degree~0 by
any $\A$-bimodule $\E$, and taking $C_n(\A,\E) := \E \ox \A^{\ox n}$;
in the formulas, the products $a_na_0$ and $a_0a_1$ make sense even
when $a_0 \in \E$. We write its homology as $H_\8(\A,\E)$ and
abbreviate $HH_n(\A) := H_n(\A,\A)$.

Hochschild \textit{cohomology}, with values in an $\A$-bimodule $\E$,
is defined using cochains in $C^n = C^n(\A,\E)$, the vector space of
$n$-linear maps $\psi\: \A^n \to \E$; this itself becomes an
$\A$-bimodule by writing
$(a'\.\psi\.a'')(\row a1n) := a'\.\psi(\row a1n)\.a''$. The coboundary
map $b\: C^n \to C^{n+1}$ is given by
\begin{align}
b\psi(\row a1{n+1})
&:= a_1\.\psi(\row a2{n+1})
     + \sum_{j=1}^n (-1)^j \psi(\row a1j \row a{j+1}{n+1})
\notag \\
&\qquad + (-1)^{n+1} \psi(\row a1n)\.a_{n+1}.
\label{eq:CnAE-cobdry} 
\end{align}
The standard case is $\E = \A^*$ as an $\A$-bimodule, where for
$\psi\in \A^*$ we put $(a'\.\psi\.a'')(c) := \psi(a''ca')$. Here, we 
identify $\psi \in C^n(\A,\E)$ with the $(n+1)$-linear map 
$\vf\: \A^{n+1} \to \C$ given by 
$\vf(a_0,\row a1n) := \psi(\row a1n)(a_0)$; then, from the first
summand in \eqref{eq:CnAE-cobdry} we get $a_1\.\psi(\row a2{n+1})(a_0)
 = \psi(\row a2{n+1})(a_0a_1) = \vf(a_0\row a1{n+1})$, while the last 
sum\-mand gives $\psi(\row a1n)\.a_{n+1}(a_0)
 = \psi(\row a1n)(a_{n+1}a_0) = \vf(a_{n+1}\row a0n)$. In this case, 
\eqref{eq:CnAE-cobdry} reduces to
\begin{equation}
b\vf(\row a0{n+1})
 := \sum_{j=0}^n (-1)^j \vf(\row a0j \row a{j+1}{n+1})
     + (-1)^{n+1} \vf(a_{n+1}\row a0n).
\label{eq:Hoch-cobrdy} 
\end{equation}
The $n$-th Hochschild cohomology group is denoted $H^n(\A,\E)$ in the
general case, and we also write $HH^n(\A) := H^n(\A,\A^*)$.

Suppose that $\mu\: \A \to \FF$ is a character of~$\A$. We denote by
$\A_\mu$ the bimodule obtained by letting $\A$ act on itself on the
left by the usual multiplication, but on the right through~$\mu$:
$$
a'\.c\.a'' := a'c\,\mu(a'')  \sepword{for all} a',a'',c \in \A.
$$
In \eqref{eq:CnAE-cobdry}, the last term on the right must be replaced
by $(-1)^{n+1} \vf(\row a1n)\mu(a_{n+1})$.

\marker
We return now to the Hopf algebra setting, by considering a dual kind
of Hochschild cohomology for coalgebras. Actually, we now consider a
\textit{bialgebra} $B$; the dual of the coalgebra $(B,\Dl,\eps)$ is an
algebra $B^*$, and the unit map $\eta$ for~$B$ transposes to a
character $\eta^t$ of~$B^*$. Thus we may define the Hochschild
cohomology groups $H^n(B^*,B^*_{\eta^t})$. An ``$n$-cochain'' now
means a linear map $\ell\: B \to B^{\ox n}$ which transposes to an
$n$-linear map $\vf = (B^*)^n \to B^*$ by writing
$\vf(\row a1n) := \ell^t(a_1 \oxyox a_n)$. Its coboundary is defined
by
$$
\dst{a_1 \oxyox a_{n+1}}{b\ell(x)} := \dst{b\vf(\row a1{n+1})}{x},
 \qquad  x \in B.
$$
We compute $b\ell$ using \eqref{eq:CnAE-cobdry}. First,
$$
\dst{a_1\.\vf(\row a2{n+1})}{x}
 = \dst{a_1 \ox \vf(\row a2{n+1})}{\Dl x}
 = \dst{a_1 \ox a_2 \oxyox a_{n+1}}{(\id \ox \ell)\Dl x}.
$$
Next, if $\Dl_j\: B^{\ox n} \to B^{\ox(n+1)}$ is the homomorphism
which applies the coproduct on the $j$th factor only, then
$\dst{\vf(\row a1j \row a{j+1}{n+1})}{x}
 = \dst{a_1 \oxyox a_{n+1}}{\Dl_j(\ell(x))}$. Finally, notice that
$\dst{\vf(\row a1n)\eta^t(a_{n+1})}{x}
 = \dst{a_1 \oxyox a_{n+1}}{\ell(x) \ox 1}$. Thus the Hochschild
coboundary operator simplifies to
\begin{equation}
b\ell(x) := (\id \ox \ell) \Dl(x)
 + \sum_{j=1}^n (-1)^j \Dl_j(\ell(x)) + (-1)^{n+1} \ell(x) \ox 1.
\label{eq:ell-cobdry} 
\end{equation}
In particular, a linear form $\la\: B \to \FF$ is a $0$-cochain, and
$b\la = (\id \ox \la)\Dl - \la \ox 1$ is its coboundary; and a
\textit{$1$-cocycle} is a linear map $\ell\: B \to B$ satisfying
\begin{equation}
\Dl \ell = \ell \ox 1 + (\id \ox \ell) \Dl.
\label{eq:coHoch-cocyc} 
\end{equation}

The simplest example of a nontrivial $1$-cocycle obeying
\eqref{eq:coHoch-cocyc} come from integration of polynomials in the
algebra $B = \FF[X]$; we make $\FF[X]$ a cocommutative coalgebra by
declaring the indeterminate $X$ to be \textit{primitive}, so that
$\Dl(X) = X \ox 1 + 1 \ox X$ and $\eps(X) = 0$. We immediately get
the binomial expansion
$\Dl(X^k) = (\Dl X)^k = \sum_{j=0}^k \binom{k}{j}\, X^{k-j} \ox X^j$.
If $\la$ is any linear form on $\FF[X]$, then
$$
b\la(X^k) = (\id \ox \la)\Dl(X^k) - \la(X^k) \ox 1
 = \sum_{j=1}^k \binom{k}{j}\, \la(X^{k-j}) \,X^j,
$$
so $b\la$ is a linear transformation of polynomials which does not
raise the degree. Therefore, the integration map $\ell(X^k) :=
X^{k+1}/(k+1)$ is \textit{not} a $1$-coboundary, but it is a
$1$-cocycle:
\begin{align*}
\Dl(\ell(X^k))
&= \frac{1}{k+1} \sum_{j=0}^{k+1} \binom{k+1}{j}\, X^{k+1-j} \ox X^j
 = \frac{X^{k+1}}{k+1} \ox 1
   + \sum_{j=1}^{k+1} \frac{1}{j} \binom{k}{j-1}\, X^{k+1-j} \ox X^j
\\
&= \ell(X^k) \ox 1
   + \sum_{r=0}^k \frac{1}{r+1} \binom{k}{r}\, X^{k-r} \ox X^{r+1}
 = \ell(X^k) \ox 1 + (\id \ox \ell)(\Dl(X^k)).
\end{align*}

This simple example already shows what the ``Hochschild equation''
\eqref{eq:coHoch-cocyc} is good for: it allows \textit{a recursive
definition of the coproduct} $\Dl$, with the assistance of a
degree-raising operation~$\ell$. Indeed, $\FF[X]$ is a simple example
of a \textit{connected, graded bialgebra}.

\begin{defn}
\label{df:graded-bialg} 
A bialgebra $H = \bigoplus_{n=0}^\infty H^{(n)}$ is a \textbf{graded
bialgebra} if it is graded both as an algebra and as a coalgebra:
\begin{equation}
H^{(m)} H^{(n)} \subseteq H^{(m+n)}  \sepword{and}
 \Dl(H^{(n)}) \subseteq \bigoplus_{p+q=n} H^{(p)} \ox H^{(q)}.
\label{eq:graded-bialg} 
\end{equation}
It is called \textit{connected} if the degree-zero piece consists of
scalars only: $H^{(0)} = \FF\,1 = \im\eta$.
\end{defn}

In a connected graded bialgebra, we can write the coproduct with a
modified Sweedler notation: if $a \in H^{(n)}$, then
\begin{equation}
\Dl a = a \ox 1 + 1 \ox a + \tsum a'_{:1} \ox a'_{:2},
\label{eq:gr-coprod} 
\end{equation}
where the terms $a'_{:1}$ and $a'_{:2}$ all have degrees between $1$
and $n - 1$. Indeed, for the counit equations
\eqref{eq:Swee-notn-counit} to be satisfied, $\Dl a$ must contain the
terms $a \ox 1$ in $H^{(n)} \ox H^{(0)}$ and $1 \ox a$ in
$H^{(0)} \ox H^{(n)}$; the remaining terms have intermediate
bidegrees. On applying $\eps \ox \id$, we get
$a = (\eps \ox \id)(\Dl a)
 = \eps(a)1 + a + \tsum \eps(a'_{:1})\,a'_{:2}$, so that
$\eps(a) = 0$ when $n \geq 1$: in a connected graded bialgebra, the
``augmentation ideal'' $\ker\eps$ is $\bigoplus_{n=1}^\infty H^{(n)}$,
so that $H = \FF\,1 \op \ker\eps$.

In fact, $H$ is a \textit{Hopf} algebra, since the grading allows us
to define the antipode recursively \cite[\S 8]{MilnorM}. Indeed, the
equation $m(S \ox \id)\Dl = \eta\eps$ may be solved thus: if
$a \in H^{(n)}$, we can obtain
$0 = \eps(a)\,1 = S(a) + a + \tsum S(a'_{:1})\, a'_{:2}$, where each
term $a'_{:1}$ has degree less than~$n$, just by setting
\begin{equation}
S(a) := - a - \tsum S(a'_{:1})\, a'_{:2}.
\label{eq:graded-antp} 
\end{equation}
Likewise, $m(\id \ox T)\Dl = \eta\eps$ is solved by setting
$T(1) := 1$ and recursively defining
$T(a) := - a - \tsum T(a'_{:2})\, a'_{:1}$. It follows that
$T = S * \id * T = S$, so we have indeed constructed a convolution
inverse for~$\id$.

In the same way, if there is a $1$-cocycle $\ell$ which raises the
degree, then \eqref{eq:coHoch-cocyc} gives a recursive recipe for the
coproduct: start with $\Dl(1) := 1 \ox 1$ in degree zero (since $H$
is connected, that will suffice), and use
$$
\Dl(\ell(a)) := \ell(a) \ox 1 + (\id \ox \ell) \Dl(a)
$$
as often as necessary. The point is that, at each level,
coassociativity is maintained:
\begin{align*}
(\id \ox \Dl) \Dl(\ell(a))
&= (\id \ox \Dl)(\ell(a) \ox 1 + (\id \ox \ell)(\Dl a))
 = \ell(a) \ox 1 \ox 1 + (\id \ox \Dl\ell)(\Dl a)
\\
&= \ell(a) \ox 1 \ox 1 + (\id \ox \ell)(\Dl a) \ox 1
   + (\id \ox \id \ox \ell)(\id \ox \Dl)(\Dl a),
\end{align*}
whereas
\begin{align*}
(\Dl \ox \id) \Dl(\ell(a))
&= (\Dl \ox \id)(\ell(a) \ox 1 + (\id \ox \ell)(\Dl a))
\\
&= \ell(a) \ox 1 \ox 1 + (\id \ox \ell)(\Dl a) \ox 1
   + (\id \ox \id \ox \ell)(\Dl \ox \id)(\Dl a),
\end{align*}
where we have used the trivial relation
$(\Dl \ox \id)(\id \ox \ell) = (\id \ox \id \ox \ell)(\Dl \ox \id)$.
The only remaining issues are (i) whether such a $1$-cocycle $\ell$
exists; and (ii) whether any $c \in H^{(n+1)}$ is a sum of products of
elements of the form $\ell(a)$ with $a$ of degree at most~$n$.

\marker
Both questions are answered by producing a \textit{universal} example
of a pair $(H,\ell)$ consisting of a connected graded Hopf algebra and
a $1$-cocycle $\ell$. It was pointed out by Connes and
Kreimer~\cite{ConnesKrHopf} that their Hopf algebra of \textit{rooted
trees} gives precisely this universal example. (Kreimer had first
introduced a Hopf algebra of ``parenthesized words''
\cite{KreimerHopf}, where the nesting of subdivergences was indicated
by parentheses, but rooted trees are nicer, and both Hopf algebras are
isomorphic by the same universality.)

\begin{defn}
\label{df:rooted-tree} 
A \textbf{rooted tree} is a tree (a finite, connected graph without
loops) with oriented edges, in which all the vertices but one have
exactly one incoming edge, and the remaining vertex, the
\textit{root}, has only outgoing edges.
\end{defn}

Here are the rooted trees with at most four vertices (up to
isomorphism). To draw them, we place the root at the top with a
$\circ$ symbol, and denote the other vertices with $\bullet$ symbols:
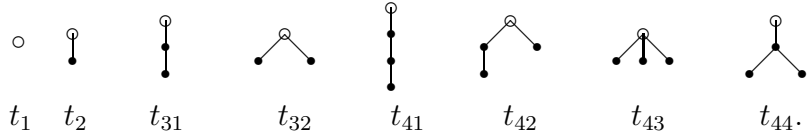
\begin{align*}
\punto  &\quad \match  \qquad  \baton  \qquad  \legs  \qquad
\stick  \qquad  \crook  \qquad  \claw   \qquad  \biped
\\[4\jot]
t_1   &\quad  t_2    \qquad  t_{31} \qquad\quad  t_{32} \qquad\enspace
t_{41} \qquad\enspace  t_{42} \qquad\quad  t_{43} \qquad\quad  t_{44}.
\end{align*}

The \textbf{algebra of rooted trees} $H_R$ is the commutative algebra
generated by symbols $T$, one for each isomorphism class of rooted
trees, plus a unit~$1$ corresponding to the empty tree. We shall write
the product of trees as the juxtaposition of their symbols. There is
an obvious grading making $H_R$ a graded algebra, by assigning to each
tree $T$ the number of its vertices $\#T$. The counit
$\eps: H_R \to \FF$ is the linear map defined by $\eps(1) := 1$ and
$\eps(T_1T_2\dots T_n) = 0$ if $\row T1n$ are trees; this ensures that
$H_R = \FF\,1 \op \ker\eps$. To get a coproduct satisfying
\eqref{eq:gr-coprod}, we must give a rule which shows how a tree may
be cut into subtrees with complementary sets of vertices. A
\textit{simple cut} $c$ of a tree $T$ is the removal of some of its
edges, in such a way that along the path from the root to any vertex,
at most one edge is removed. Here, for instance, are the possible
simple cuts of~$t_{44}$:

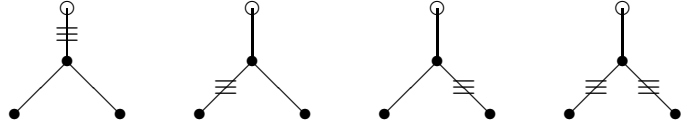
\begin{picture}(330,60)(-75,15)
\put(30,60){\circle{5}}
\put(30,60){\line(0,-1){20}}
\put(30,50){\makebox(0,0){$\equiv$}}
\put(30,40){\circle*{4}}
\put(30,40){\line(1,-1){20}}
\put(30,40){\line(-1,-1){20}}
\put(10,20){\circle*{4}}
\put(50,20){\circle*{4}}
\put(100,60){\circle{5}}
\put(100,60){\line(0,-1){20}}
\put(100,40){\circle*{4}}
\put(100,40){\line(1,-1){20}}
\put(100,40){\line(-1,-1){20}}
\put(90,30){\makebox(0,0){$\equiv$}}
\put(80,20){\circle*{4}}
\put(120,20){\circle*{4}}
\put(170,60){\circle{5}}
\put(170,60){\line(0,-1){20}}
\put(170,40){\circle*{4}}
\put(170,40){\line(1,-1){20}}
\put(180,30){\makebox(0,0){$\equiv$}}
\put(170,40){\line(-1,-1){20}}
\put(150,20){\circle*{4}}
\put(190,20){\circle*{4}}
\put(240,60){\circle{5}}
\put(240,60){\line(0,-1){20}}
\put(240,40){\circle*{4}}
\put(240,40){\line(1,-1){20}}
\put(250,30){\makebox(0,0){$\equiv$}}
\put(240,40){\line(-1,-1){20}}
\put(230,30){\makebox(0,0){$\equiv$}}
\put(220,20){\circle*{4}}
\put(260,20){\circle*{4}}
\end{picture}

Among the subtrees of~$T$ produced by a simple cut, exactly one, the
``trunk'' $R_c(T)$, contains the root of~$T$. The remaining ``pruned''
branches also form one or more rooted trees, whose product is denoted
by $P_c(T)$. The formula for the coproduct can now be given, on the
algebra generators, as
\begin{equation}
\Dl T := T \ox 1 + 1 \ox T + \sum_c P_c(T) \ox R_c(T),
\label{eq:coprod-tree} 
\end{equation}
where the sum extends over all simple cuts of the tree~$T$; as well as
$\Dl 1 := 1 \ox 1$, of course. Here are the coproducts of the trees
listed above:
\begin{align}
\Dl t_1 &= t_1 \ox 1 + 1 \ox t_1,
\notag \\
\Dl t_2 &= t_2 \ox 1 + 1 \ox t_2 + t_1 \ox t_1,
\notag \\
\Dl t_{31} &= t_{31} \ox 1 + 1 \ox t_{31} + t_2 \ox t_1 + t_1 \ox t_2,
\notag \\
\Dl t_{32} &= t_{32} \ox 1 + 1 \ox t_{32} + 2t_1 \ox t_2
    + t_1^2 \ox t_1,
\notag \\
\Dl t_{41} &= t_{41} \ox 1 + 1 \ox t_{41} + t_{31} \ox t_1
    + t_2 \ox t_2 + t_1 \ox t_{31},
\notag \\
\Dl t_{42} &= t_{42} \ox 1 + 1 \ox t_{42} + t_1 \ox t_{32}
 + t_2 \ox t_2 + t_1 \ox t_{31} + t_2 t_1 \ox t_1 + t_1^2 \ox t_2.
\notag \\
\Dl t_{43} &= t_{43} \ox 1 + 1 \ox t_{43} + 3t_1 \ox t_{32}
    + 3t_1^2 \ox t_2 + t_1^3 \ox t_1,
\notag \\
\Dl t_{44} &= t_{44} \ox 1 + 1 \ox t_{44} + t_{32} \ox t_1
    + 2 t_1 \ox t_{31} + t_1^2 \ox t_2.
\label{eq:coprod-list} 
\end{align}

In this way, $H_R$ becomes a connected graded commutative Hopf
algebra; clearly, it is not cocommutative. In order to prove that this
$\Dl$ is coassociative, we need only produce the appropriate
$1$-cocycle $L$ which raises the degree by~$1$. The linear operator
$L$ ---also known as $B^+$~\cite{ConnesKrHopf}--- is defined, on each
product of trees, by sprouting a new common root.

\begin{defn}
\label{df:new-root} 
Let $L\: H_R \to H_R$ be the linear map given by $L(1) := t_1$ and
\begin{equation}
L(T_1 \dots T_k) := T,
\label{eq:new-root} 
\end{equation}
where $T$ is the rooted tree obtained by conjuring up a new vertex as
its root and extending edges from this vertex to each root of
$\row T1k$. Notice, in passing, that \textit{any} tree $T$ with
$n + 1$ vertices equals $L(T_1 \dots T_k)$, where $\row T1k$ are the
rooted trees, with $n$ vertices in all, formed by removing every edge
outgoing from the root of~$T$.
\end{defn}

For instance,
$$
L\Bigl( \legs \Bigr) = \biped  \qquad\mbox{and}\qquad
L\Bigl(\ \match \match \Bigr) = \arch \ .
$$
Checking the Hochschild equation \eqref{eq:coHoch-cocyc} is a matter
of bookkeeping: see \cite[p.~229]{ConnesKrHopf} or
\cite[p.~603]{Polaris}, for instance. Here, an illustration will
suffice:
\begin{align*}
\Dl\Bigl( L\Bigl( \legs \Bigl) \Bigl)
&= \Dl\biggl( \biped \biggl)
 = \biped \ox 1 + 1 \ox \biped + \legs \ox \punto
    + 2\,\punto \ox \baton + \punto\punto \ox \match
\\
&= L\Bigl( \legs \Bigl) \ox 1 + (\id \ox L)\biggl(
    1 \ox \legs + \legs \ox 1 + 2\,\punto \ox \match
      + \punto\punto \ox \punto \biggr)
\\
&= L\Bigl( \legs \Bigl) \ox 1 + (\id \ox L) \Dl\Bigl( \legs \Bigr).
\end{align*}

Finally, suppose that a pair $(H,\ell)$ is given; we want to define a
Hopf algebra morphism $\rho\: H_R \to H$ such that
\begin{equation}
\rho(L(a)) = \ell(\rho(a)),
\label{eq:ell-rho} 
\end{equation}
where $a$ is a product of trees. Since $L(a)$ may be any tree of
degree $\#a + 1$, we may regard this as a recursive definition (on
generators) of an algebra homomorphism, starting from
$\rho(1) := 1_H$. The only thing to check is that it also yields a
coalgebra homomorphism, which again reduces to an induction on the
degree of~$a$:
\begin{align*}
\Dl(\rho(L(a)))
&= \Dl(\ell(\rho(a)))
 = \ell(\rho(a)) \ox 1 + (\id \ox \ell) \Dl(\rho(a))
\\
&= \ell(\rho(a)) \ox 1 + (\id \ox \ell)(\rho \ox \rho)(\Dl a)
\\
&= \rho(L(a)) \ox 1 + (\rho \ox \rho)(\id \ox L)(\Dl a)
\\
&= (\rho \ox \rho) \bigl( L(a) \ox 1 + (\id \ox L)(\Dl a) \bigr)
 = (\rho \ox \rho)\Dl(L(a)),
\end{align*}
where in the third line, by using
$\ell(\rho(a'_{:2})) = \rho(L(a'_{:2}))$, we have implicitly relied on
the property \eqref{eq:gr-coprod} that the nontrivial components of
the coproduct $\Dl a$ have lower degree than~$a$.

\marker
Since the Hopf algebra $H_R$ is commutative, we may look for a
cocommutative Hopf algebra in duality with it. Now, there is a
structure theorem for connected graded cocommutative Hopf algebras,
arising from contributions of Hopf, Samelson, Leray, Borel, Cartier,
Milnor, Moore and Quillen,%
\footnote{The historical record is murky; this list of contributors
is due to P.~Cartier.}
commonly known as the Milnor--Moore theorem, which states that such a
Hopf algebra $H$ is necessarily isomorphic to $\U(\g)$, with $\g$
being the Lie algebra of primitive elements of~$H$.

This dual Hopf algebra is constructed as follows.
Each rooted tree $T$ gives not only an algebra generator for $H_R$,
but also a derivation $Z_T\: H_R \to \FF$
defined by
\begin{align*}
\dst{Z_T}{T_1\dots T_k} &:= 0 \quad\text{unless $k=1$ and $T_1 = T$};
\\
\dst{Z_T}{T} &:= 1.
\end{align*}
Also, $\dst{Z_T}{1} = 0$ since $Z_T \in \Der_\eps(H)$
(Definition~\ref{df:Hopf-char}). Notice that the ideal generated by
products of two or more trees is $(\ker\eps)^2$, and any
derivation~$\dl$ vanishes there, since
$\dl(ab) = \dl(a) \eps(b) + \eps(a) \dl(b) = 0$ whenever
$a,b \in \ker\eps$. Therefore, derivations are determined by their
values on the subspace $H_R^{(1)}$ spanned by single trees ---which
equals $L(H_R)$, by the way--- and reduce to linear forms on this
subspace; thus $\Der_\eps(H)$ can be identified with the (algebraic) 
dual space $H_R^{(1)*}$. We denote by $\hl$ the linear subspace
spanned by all the~$Z_T$.

Let us compute the Lie bracket
$[Z_R, Z_S] := (Z_R \ox Z_S - Z_S \ox Z_R)\Dl$ of two such
derivations. Using \eqref{eq:coprod-tree} and
$\dst{Z_R}{1} = \dst{Z_S}{1} = 0$, we get
$$
\dst{Z_R\ox Z_S}{\Dl T} = \sum_c \dst{Z_R}{P_c(T)}\,\dst{Z_S}{R_c(T)},
$$
where $\dst{Z_R}{P_c(T)} = 0$ unless $P_c(T) = R$ and
$\dst{Z_S}{R_c(T)} = 0$ unless $R_c(T) = S$; in particular, the sum
ranges only over simple cuts which remove just \textit{one} edge
of~$T$. Let $n(R,S;T)$ be the number of one-edge cuts $c$ of $T$ such
that $P_c(T) = R$ and $R_c(T) = S$; then
$$
\dst{[Z_R,Z_S]}{T} = \dst{Z_R \ox Z_S - Z_S \ox Z_R}{\Dl T}
 = n(R,S;T) - n(S,R;T),
$$
and this expression vanishes altogether except for the finite number
of trees $T$ which can be produced either by grafting $R$ on $S$ or by
grafting $S$ on $R$. Evaluation of the derivation $[Z_R, Z_S]$ on a
product $T_1 \dots T_k$ of two or more trees gives zero, since each
$T_j \in \ker\eps$. Therefore,
$$
[Z_R, Z_S] = \sum_T \bigl( n(R,S;T) - n(S,R;T) \bigr)\, Z_T,
$$
which is a finite sum. In particular, $[Z_R, Z_S] \in \hl$, and so
$\hl$ is a Lie subalgebra of $\Der_\eps(H)$. The linear duality of
$H_R^{(1)}$ with $\hl$ then extends to a duality between the graded
Hopf algebras $H_R$ and $\U(\hl)$.

It is possible to give a more concrete description of the Hopf algebra
$\U(\hl)$ in terms of \textit{another} Hopf algebra of rooted trees
$H_{GL}$, which is cocommutative rather than commutative. This
structure was introduced by Grossman and Larson \cite{GrossmanLTrees}
and is described in \cite[\S 14.2]{Polaris}; here we mention only that
the multiplicative identity is the tree $t_1$ and that the primitive
elements are spanned by those trees which have only one edge outgoing
from the root. Panaite~\cite{Panaite} has shown that $\hl$ is
isomorphic to the Lie algebra of these primitive trees ---by matching
each $Z_T$ to the tree $L(T)$--- so that $\U(\hl) \simeq H_{GL}$.

In \cite{ConnesKrHopf}, another binary operation among the $Z_T$ was
introduced by setting $Z_R \star Z_S := \sum_T n(R,S;T)\, Z_T$. This
is \textit{not} the convolution $(Z_R \ox Z_S)\Dl$, nor is it even
associative, although it is obviously true that
$Z_R \star Z_S - Z_S \star Z_R = [Z_R, Z_S]$. This nonassociative
bilinear operation satisfies the defining property of a
\textit{pre-Lie algebra} \cite{Chapoton}:
$$
(Z_R \star Z_S) \star Z_T - Z_R \star (Z_S \star Z_T)
 = (Z_R \star Z_T) \star Z_S - Z_R \star (Z_T \star Z_S).
$$
Indeed, both sides of this equation express the formation of new trees
by grafting both $S$ and $T$ onto the tree $R$. The combinatorics of
this operation are discussed in \cite{ChapotonL}, and several
computations with it are developed in \cite{ChryssomalakosQRV}
and~\cite{KastlerSiena}.

\marker
The characters of~$H_R$ form a group $\G(H_R)$ (under convolution):
see Definition~\ref{df:Hopf-char}. This group is infinite-dimensional,
and can be thought of as the set of grouplike elements in a suitable
completion of the Hopf algebra $U = \U(\hl)$. To see that, recall 
that $U$ is a graded connected Hopf algebra; denote by $e$ its counit.
Then the sets $(\ker e)^m = \sum_{k\geq m} \hl^k$, for
$m = 1,2,\dots$, form a basis of neighbourhoods of~$0$ for a vector 
space topology on~$U$, and the grading properties
\eqref{eq:graded-bialg} entail that all the Hopf operations are 
continuous for this topology. (The basic neighbourhoods of~$0$ in
$U \ox U$ are the powers of the ideal $1 \ox \ker e + \ker e \ox 1$.)
We can form the \textit{completion} $\Uhat$ of this topological vector 
space, which is again a Hopf algebra since all the Hopf operations 
extend by continuity; an element of $\Uhat$ is a series
$\sum_{k\geq 0} z_k$ with $z_k \in \hl^k$ for each $k \in \N$, since
the partial sums form a Cauchy sequence in~$U$. The closure of $\hl$
within $\Uhat$ is $\Der_\eps(H)$.

For example, consider the exponential given by
$\vf_T := \exp Z_T = \sum_{n\geq 0} (1/n!)\, Z_T^n$; in any evaluation
$$
\vf_T(T_1\dots T_k) = \sum_{n\geq 0} \frac{1}{n!}\,
  \dst{Z_T^{\ox n}}{\Dl^{n-1}(T_1\dots T_k)},
$$
the series has only finitely many nonzero terms. More generally, 
$\vf := \exp\dl \in \Uhat$ makes sense for each $\dl \in \Der_\eps(H)$;
and $\vf \in \G(H_R)$ since $\Dl\vf = \exp(\Dl\dl)
 = \exp(\eps\ox\dl + \dl\ox\eps) = \vf \ox \vf$ by continuity 
of~$\Dl$. In fact, the exponential map is a bijection between 
$\Der_\eps(H)$ and $\G(H_R)$, whose inverse is provided by the
logarithmic series $\log(1 - x) := - \sum_{k\geq 1} x^k/k$; for if
$\mu$ is a character, the equation $\mu = \exp(\log\mu)$ holds
in~$\Uhat$, and
\begin{align*}
\Dl(\log\mu) &= \Dl(\log(\eps - (\eps - \mu))
 = \log(\eps\ox\eps - \Dl(\eps - \mu)) = \log(\mu\ox\mu)
\\
&= \log(\eps\ox\mu) + \log(\mu\ox\eps)
 = \eps \ox \log\mu + \log\mu \ox \eps,
\end{align*}
so that $\log\mu \in \Der_\eps(H)$. See
\cite[Chap.~X]{HochschildStruct} or~\cite[Chap.~XVI]{HochschildBasic}
for a careful discussion of the exponential map. In view of this
bijection, we can regard the commutative Hopf algebra $H_R$ as an
algebra of affine coordinates on the group $\G(H_R)$, in the spirit of
Tannaka--Kre\u{\i}n duality.

\marker
In any Hopf algebra, whether cocommutative or not, the determination
of the \textit{primitive elements} plays an important part. If in any
tree $T$, the longest path from the root to a leaf contains $k$ edges,
then the coproduct $\Dl T$ is a sum of at least $k + 1$ terms. In the
applications to renormalization, $T$ represents a possibly divergent
integration with $k$ nested subdivergences, while the primitive tree
$t_1$ corresponds to an integration without subdivergences. A
primitive algebraic \textit{combination} of trees represents a
collection of integrations where some of these divergences may cancel.
For that reason alone, it would be desirable to describe all the
primitive elements of~$H_R$ and then, as far as possible, to rebuild
$H_R$ from its primitives. This is a work in progress
\cite{BroadhurstK,ChryssomalakosQRV,Foissy}, which deserves a few
comments here.

To begin with, since $t_1$ is primitive and
$\Dl t_2 = t_2 \ox 1 + 1 \ox t_2 + t_1 \ox t_1$, the combination
$p_2 := t_2 - \thalf t_1^2$ is also primitive. One can check that
$p_3 := t_{31} - t_1t_2 + \tthird t_1^3$ is primitive, too.

For each~$k = 1,2,\dots$, let $t_k$ denote the ``stick'' tree with
$k - 1$ edges and $k$ vertices in a vertical progression. (In
particular, $t_3$ and $t_4$ are the trees previously referred to as
$t_{31}$ and $t_{41}$, respectively.) A simple cut severs $t_k$ into
two shorter sticks, and so
\begin{equation}
\Dl t_k = \sum_{0\leq r\leq k} t_r \ox t_{k-r},
\label{eq:stick-coprod} 
\end{equation}
with $t_0 := 1$ by convention. Thus the sticks generate a
cocommutative graded Hopf subalgebra $H_l$ of~$H_R$.

To find the primitives in~$H_l$, we follow the approach of
\cite{ChryssomalakosQRV}. Consider the formal power series
$g(x) := \sum_{k\geq 0} t_k x^k$ whose coefficients are sticks. Then
the equation~\eqref{eq:stick-coprod} can be read as saying that
$g(x)$ is grouplike in $H_l\ldbrack x\rdbrack$, that is,
$\Dl g(x) = g(x) \ox g(x)$. If we can find a power series
$p(x) = \sum_{r\geq 1} p_r x^r$, where each $p_r$ is homogeneous of
degree~$r$ in the grading of~$H_l$, such that $\exp(p(x)) = g(x)$, the
corresponding equation will be $\Dl p(x) = p(x) \ox 1 + 1 \ox p(x)$;
on comparing coefficients of each~$x^r$, we see that each $p_r$ is
primitive. The equation $\exp(p(x)) = g(x)$ is solved as
$$
\sum_{r\geq 1} p_r x^r = \log\biggl(1 + \sum_{k\geq 1} t_k x^k\biggr),
$$
by developing the Taylor series of $\log(1 + x)$. Since a monomial
$t_1^{m_1} t_2^{m_2} \dots t_r^{m_r}$ has degree
$m_1 + 2m_2 +\cdots+ rm_r$, the general formula
\cite[Prop.~9.3]{Foissy} is quickly found to be
$$
p_r = \sum_{m_1+2m_2+\cdots+rm_r=r} (-1)^{m_1+\cdots+m_r+1}
       \frac{(m_1+\cdots+m_r-1)!}{m_1!\dots m_r!}
        t_1^{m_1} \dots t_r^{m_r},
$$
where the sum ranges over the partitions of the positive integer~$r$.

\marker
Nonstick primitives are more difficult to come by, but an algorithm
which provides many of them is found in \cite{ChryssomalakosQRV},
based on formal differential calculus. Indeed, this ``differential''
approach can be extended, in principle, to deal efficiently with the
more elaborate Hopf algebras of Feynman diagrams discussed in the next
subsection.

For each $a \in H_R$, the expression
\begin{equation}
\Pi_a := \tsum S(a_{:1}) \,da_{:2}
\label{eq:left-invt} 
\end{equation}
where $d$ denotes an ordinary exterior derivative, may be regarded as
a $1$-form on~$G$; it is a straightforward generalization of the
familiar (matrix-valued) $1$-form $g^{-1} \,dg$ on a group manifold,
whose matrix elements are $\sum_j (g^{-1})_{ij} \,dg_{jk}$. We can
treat such expressions algebraically, as a ``first-order differential
calculus'' on a Hopf algebra, in the sense of
Woronowicz~\cite{WoronowiczDiff}. The commutativity of $H_R$ shows
that these $1$-forms have the following derivation property:
$$
\Pi_{ab} = \tsum S(a_{:1}b_{:1}) \,d(a_{:2}b_{:2})
 = \tsum S(b_{:1}) S(a_{:1}) a_{:2} \,db_{:2}
     + S(b_{:1}) b_{:2} S(a_{:1}) \,da_{:2}
 = \eps(a)\, \Pi_b + \Pi_a \,\eps(b).
$$
In particular, $\Pi_a = 0$ for $a \in (\ker\eps)^2$, so we need only
consider $\Pi_a$ for $a \in H_R^{(1)}$. Each $\Pi_a$ can be thought
of as a ``left-invariant'' $1$-form, as follows.

\begin{exer}
\label{xr:left-invt} 
Let $G$ be a compact Lie group and let $\Rr(G)$ be its Hopf algebra of
representative functions. If $L_t$ denotes left translation by
$t \in G$, then $L_t^* f(x) = f(t^{-1}x) = \Dl f(t^{-1},x)
 = \tsum f_{:1}(t^{-1}) \,f_{:2}(x)$, so that
$L_t^* f = \tsum f_{:1}(t^{-1}) \,f_{:2}$ for $f \in \Rr(G)$. Let
$\Pi_f$ be the smooth $1$-form on~$G$ defined by \eqref{eq:left-invt};
prove that $L_t^* \Pi_f = \Pi_f$ for all $t \in G$.
\end{exer}

Each left-invariant $1$-form \eqref{eq:left-invt} satisfies a
``Maurer--Cartan equation'':
$$
d\Pi_a = - \tsum \Pi_{a_{:1}} \w \Pi_{a_{:2}}.
$$
Indeed, since $0 = d(\eps(a)\,1) = \tsum d(S(a_{:1})\,a_{:2}) 
  = \tsum d(S(a_{:1}))\,a_{:2} + S(a_{:1})\,da_{:2}$, we find that
\begin{align*}
d(S(a)) &= \tsum d(S(a_{:1})\,\eps(a_{:2}))
 = \tsum d(S(a_{:1}))\,\eps(a_{:2})
 = \tsum d(S(a_{:1}))\,a_{:2}\,S(a_{:3})
\\
&= - \tsum S(a_{:1})\,da_{:2}\,S(a_{:3}),
\end{align*}
in analogy with $d(g^{-1}) = - g^{-1}\,dg\,g^{-1}$. Therefore,
$$
d\Pi_a = \tsum d(S(a_{:1})) \w da_{:2}
 = - \tsum S(a_{:1})\,da_{:2} \w S(a_{:3})\,da_{:4}
 = - \tsum \Pi_{a_{:1}} \w \Pi_{a_{:2}}.
$$

Suppose now that we are given some element $a \in H_R^{(1)}$ for which
$d\Pi_a = 0$. The bijectivity of the exponential map for $\G(H_R)$
suggests that this closed $1$-form should be exact: $\Pi_a = db$ for
some $b \in H_R$. It is clear from \eqref{eq:left-invt} that the
equation $\Pi_a = db$ can hold only if $b$ is primitive. Theorem~2 of
\cite{ChryssomalakosQRV} uses the Poincar\'e lemma technique to
provide a formula for~$b$, namely,
$$
b := - \Phi^{-1}(S(a)),
$$
where $\Phi$ is the operator which grades $H_R$ by the number of trees
in a product: $\Phi(T_1\dots T_k) := k\,T_1\dots T_k$. Notice that
$b = a + c$, where $c \in (\ker\eps)^2$ is a sum of higher-degree
terms.

\begin{exer}
\label{xr:tricky-prim} 
Show that $a = \claw + \biped - 2\,\crook$ satisfies $d\Pi_a = 0$,
and compute that
$$
b = \claw + \biped - 2\,\crook - \punto\legs + \match\match.
$$
Verify directly that $b$ is indeed primitive.
\end{exer}

It is still not a trivial matter to find linear combinations of trees
satisfying $d\Pi_a = 0$, but it clearly is much easier to verify this
property than to check primitivity directly on a case-by-case basis.

\marker
Finally, we comment on the link between $H_R$ and the Hopf algebra
$H_{CM}$ of differential operators, developed in~\cite{ConnesKrHopf}.
This is found by extending $H_R$ to a larger (but no longer
commutative) Hopf algebra $\Onda H_R$. Since $H_R$ is graded by the
number of vertices per tree, we regard the subspace $H_R^{(1)}$ of
single trees as an abelian Lie algebra, and introduce an extra
generator $Y$ with the commutation rule
$$
[Y,T] := (\#T)\, T.
$$
For each simple cut $c$ of~$T$, it is clear that
$\#P_c(T) + \#R_c(T) = \#T$; a glance at \eqref{eq:coprod-tree} then
shows that $\Dl[Y,T] = (\#T)\,\Dl T = [Y\ox 1 + 1\ox Y, \Dl T]$. This
forces $Y$ to be primitive:
\begin{equation}
\Dl Y := Y \ox 1 + 1 \ox Y,
\label{eq:coprod-Y} 
\end{equation}
in order to get $\Dl[Y,T] = [\Dl Y, \Dl T]$ for consistency.

Another important operator on $\H_R$ is the so-called \textit{natural
growth} of trees. We define $N(T)$, for each tree $T$ with vertices
$\row v1n$, by setting $N(T) := T_1 + T_2 +\cdots+ T_n$, where each
$T_j$ is obtained from $T$ by adding a leaf to~$v_j$. For example,
\begin{gather*}
N\bigl( \punto \bigr) := \match,  \qquad
N\bigl(\, \match \bigr) := \baton + \legs,
\\[2\jot]
N\Bigl( \baton + \legs \Bigr)
 := \stick + 3\,\crook + \claw + \biped \ .
\end{gather*}
In symbols, we write these relations as
\begin{align*}
N(t_1) &= t_2,  \qquad  N^2(t_1) = N(t_2) = t_{31} + t_{32},
\\
N^3(t_1) &= N(t_{31} + t_{32}) = t_{41} + 3 t_{42} + t_{43} + t_{44}.
\end{align*}
We rename these $\dl_1 := t_1$, $\dl_2 := N(\dl_1)$,
$\dl_3 := N^2(\dl_1)$, $\dl_4 := N^3(\dl_1)$, and in general
$\dl_{n+1} := N^n(\dl_1)$ for any~$n$. Notice that $\dl_{n+1}$ is a
sum of $n!$~trees.

$N$, defined on the algebra generators, extends uniquely to a
\textit{derivation} $N\: H_R \to H_R$. Now, we can add one more
generator $X$ with the commutation rule
$$
[X,T] := N(T).
$$
The Jacobi identity forces $[Y,X] = X$, as follows:
\begin{align*}
[[Y,X],T] &= [[Y,T],X] + [Y,[X,T]] = (\#T)\,[T,X] + [Y,N(T)]
\\
&= -(\#T)\,N(T) + (\#T + 1)\,N(T) = N(T) = [X,T].
\end{align*}

What must the coproduct $\Dl X$ be? Proposition~3.6 of
\cite{ConnesKrHopf} ---see also Proposition~14.6 of \cite{Polaris}---
proves that
\begin{equation}
\Dl N(T) = (N\ox\id) \Dl T + (\id\ox N) \Dl T + [\dl_1 \ox Y, \Dl T]
\label{eq:new-leaf} 
\end{equation}
for each rooted tree~$T$. The argument is as follows: to get
$\Dl N(T)$, we grow an extra leaf on~$T$ and then cut the resulting
trees in every allowable way. If the new edge is not cut, then it
belongs either to a pruned branch or to the trunk which remains after
a cut has been made on the original tree~$T$; this amounts to
$(N \ox \id)\Dl T + (\id \ox N)\Dl T$. On the other hand, if the new
edge is cut, the new leaf contributes a solitary vertex $\dl_1$ to
$P_c$; the new leaf must have been attached to the trunk $R_c(T)$ at
any one of the latter's vertices. Since $(\#R_c)R_c = [Y, R_c]$, the
terms wherein the new leaf is cut amount to $[\dl_1 \ox Y, \Dl T]$.
The equation~\eqref{eq:new-leaf} accounts for both possibilities.
Then, since $\Dl[X,T] = [\Dl X, \Dl T]$ must hold, we get
\begin{equation}
\Dl X = X \ox 1 + 1 \ox X + \dl_1 \ox Y.
\label{eq:coprod-X} 
\end{equation}

Let $\Onda H_R$ be the algebra generated by $X$, $Y$ and $H_R$.
We can extend the counit and antipode to it as follows.
Since $Y$ is primitive, we must take $\eps(Y) := 0$ and
$S(Y) := -Y$. Then, on applying $(\id \ox \eps)$ to
\eqref{eq:coprod-X}, $\eps(X) := 0$ follows; and by applying
$m(\id \ox S)$ or $m(S \ox \id)$ to it, we also get
$0 = X + S(X) - \dl_1 Y$, which forces $S(X) := - X + \dl_1 Y$.

Now \eqref{eq:coprod-Y} and \eqref{eq:coprod-X} reproduce exactly the
coproducts \eqref{eq:CM-coprod} for the differential operators $Y$
and~$X$ of the Hopf algebra $H_{CM}$. Indeed, since $\dl_1$, like
$\la_1 \in H_{CM}$, is primitive and since
$\dl_{n+1} = N(\dl_n) = [X,\dl_n]$, the correspondence $X \mapsto X$,
$Y \mapsto Y$, $\la_n \mapsto \dl_n$ maps $H_{CM}$ isomorphically into
$\Onda H_R$.

\subsection{Hopf algebras of Feynman graphs and renormalization}
\label{sec:Hopf-Feyn} 

In this subsection, we shall describe briefly some other
Hopf algebras which underlie the structure of a renormalizable
quantum field theory. Rather than going into the details of
perturbative renormalization, we shall merely indicate how such Hopf
algebras are involved.

In a given QFT, one is faced with the problem of computing
correlations (Green functions) from a perturbative expansion whose
terms are labelled by Feynman graphs $\Ga$, and consist of multiple
integrals where the integrand is completely specified by the
combinatorial structure of~$\Ga$ (its vertices, external and internal
lines, and loops) according to a small number of Feynman rules.
Typically, one works in momentum space of $D$ dimensions, and a
preliminary count of the powers of the momenta in the integrand
indicates, in many cases, a superficially divergent integral; even if
the graph $\Ga$ itself passes this test, it may contain subgraphs
corresponding to superficially divergent integrals. The main idea of
renormalization theory is to associate a ``counterterm'' to each
superficially divergent subgraph, in order to obtain a finite result
by subtraction.

The first step in approaching such calculations is to realize that all
superficially divergent subgraphs must be dealt with, in a recursive
fashion, before finally assigning a finite value to the full
graph~$\Ga$. Thus, each graph $\Ga$ determines a nesting of divergent
subgraphs: this nesting is codified by a \textit{rooted tree}, where
the root represents the full graph, provided that the $\Ga$ does not
contain overlapping divergences. (Even if overlapping divergences do
occur, one can replace the single rooted tree by a sum over rooted
trees after disentangling the overlaps: see \cite{KreimerOverlap} for
a detailed analysis.) A ``leaf'' is a divergent subgraph
which itself contains no further subdivergences.

The combinatorial algebra is worked out in considerable detail in a
recent article of Connes and Kreimer~\cite{ConnesKrRHI}: the following
remarks can be taken as an incentive for a closer look at that paper.
See also the survey of Kreimer~\cite{KreimerRev} for a detailed
discussion of the conceptual framework. The authors of
\cite{ConnesKrRHI} consider $\phi^3$ theory in $D = 6$ dimensions; but
one could equally well start with $\phi^4$ theory for $D = 4$
\cite{Etoile}, or QED, or any other well-known theory.

\begin{defn}
\label{df:graph-Hopf} 
Let $\Phi$ stand for any particular QFT. The Hopf algebra $H_\Phi$ is
a commutative algebra generated by one-particle irreducible (1PI)
graphs: that is, connected graphs with at least two vertices which
cannot be disconnected by removing a single line. The product is given
by disjoint union of graphs: $\Ga_1 \Ga_2$ means $\Ga_1 \uplus \Ga_2$.
The counit is given by $\eps(\Ga) := 0$ on any generator, with
$\eps(\emptyset) := 1$ (we assign the empty graph to the identity
element). The \textit{coproduct} $\Dl$ is given, on any 1PI graph
$\Ga$, by
\begin{equation}
\Dl\Ga := \Ga \ox 1 + 1 \ox \Ga
  + \sum_{\emptyset\subsetneq\ga\subsetneq\Ga} \ga \ox \Ga/\ga,
\label{eq:graph-coprod} 
\end{equation}
where the sum ranges over all subgraphs which are divergent and
proper (in the sense that removing one internal line cannot increase
the number of its connected components); $\ga$ may be either
connected or a disjoint union of several connected pieces. The
notation $\Ga/\ga$ denotes the (connected, 1PI) graph obtained
from~$\Ga$ by replacing each component of $\ga$ by a single vertex.
\end{defn}

To see that $\Dl$ is coassociative, we may reason as follows. We may
replace the right hand side of \eqref{eq:graph-coprod} by a single
sum over $\emptyset\subseteq\ga\subseteq\Ga$, allowing
$\ga = \emptyset$ or $\ga = \Ga$ and setting $\Ga/\Ga := 1$. We
observe that if $\ga \subseteq \ga' \subseteq \Ga$, then $\ga'/\ga$
can be regarded as a subgraph of $\Ga/\ga$; moreover, it is obvious
that
\begin{equation}
(\Ga/\ga)/(\ga'/\ga) \simeq \Ga/\ga'.
\label{eq:third-isom} 
\end{equation}
The desired relation $(\Dl \ox \id)(\Dl\Ga) = (\id \ox \Dl)(\Dl\Ga)$
can now be expressed as
$$
\sum_{\emptyset\subseteq\ga\subseteq\ga'\subseteq\Ga}
 \ga \ox \ga'/\ga \ox \Ga/\ga'
 = \sum_{\emptyset\subseteq\ga\subseteq\Ga,\;
         \emptyset\subseteq\ga''\subseteq\Ga/\ga}
    \ga \ox \ga'' \ox (\Ga/\ga)/\ga'',
$$
so coassociativity reduces to proving, for each subgraph $\ga$
of~$\Ga$, that
$$
\sum_{\ga\subseteq\ga'\subseteq\Ga} \ga'/\ga \ox \Ga/\ga'
 = \sum_{\emptyset\subseteq\ga''\subseteq\Ga/\ga}
    \ga'' \ox (\Ga/\ga)/\ga''.
$$
Choose $\ga'$ so that $\ga \subseteq \ga' \subseteq \Ga$; then
$\emptyset \subseteq \ga'/\ga \subseteq \Ga/\ga$. Reciprocally, to
every $\ga'' \subseteq \Ga/\ga$ there corresponds a unique $\ga'$ such
that $\ga \subseteq \ga' \subseteq \Ga$ and $\ga'/\ga = \ga''$; the
previous equality now follows from the 
identification~\eqref{eq:third-isom}.

We have now defined $H_\Phi$ as a bialgebra. To make sure that it is a
\textit{Hopf} algebra, it suffices to show that it is graded and
connected, whereby the antipode comes for free. Several grading
operators $\Ups$ are available, which satisfy the two
conditions~\eqref{eq:graded-bialg}:
$$
\Ups(\Ga_1 \Ga_2) = \Ups(\Ga_1) + \Ups(\Ga_2)  \sepword{and}
 \Ups(\ga) + \Ups(\Ga/\ga) = \Ups(\Ga)
$$
whenever $\ga$ is a divergent proper subgraph of~$\Ga$. One such
grading is the loop number $\ell(\Ga) := I(\Ga) - V(\Ga) + 1$, if
$\Ga$ has $I(\Ga)$ internal lines and $V(\Ga)$ vertices. If
$\ell(\Ga) = 0$, then $\Ga$ would be a tree graph, which is
never~1PI; thus $\ker\ell$ consists of scalars only, so $H_\Phi$ is
connected. The antipode is now given recursively
by~\eqref{eq:graded-antp}:
\begin{equation}
S(\Ga)
 = - \Ga + \sum_{\emptyset\subsetneq\ga\subsetneq\Ga} S(\ga)\,\Ga/\ga.
\label{eq:graph-antp} 
\end{equation}

As it stands, the Hopf algebra $H_\Phi$ corresponds to a formal
manipulation of graphs. It remains to understand how to match these
formulas to expressions for numerical values, whereby the antipode $S$
delivers the counterterms. This is done in two steps. First of all,
the Feynman rules for the unrenormalized theory can be thought of as
prescribing a linear map
$$
f : H_\Phi \to \A,
$$
into some commutative algebra $\A$, which is multiplicative on disjoint
unions: $f(\Ga_1 \Ga_2) = f(\Ga_1)\,f(\Ga_2)$. In other words, $f$ is
actually a homomorphism of algebras. For instance, $\A$ is often an
algebra of Laurent series in some (complex) regularization
parameter~$\eps$: in dimensional regularization, after adjustment by a
mass unit~$\mu$ so that each $f(\Ga)$ is dimensionless, one computes
the corresponding integral in dimension $d = D + \eps$, for
$\eps \neq 0$. We shall also suppose that $\A$ is the direct sum of
two \textit{subalgebras}:
$$
\A = \A_+ \op \A_-.
$$
Let $T\: \A \to \A_-$ be the projection on the second subalgebra,
with $\ker T = \A_+$. When $\A$ is a Laurent-series algebra, one
takes $\A_+$ to be the holomorphic subalgebra of Taylor series and
$\A_-$ to be the subalgebra of polynomials in $1/\eps$ without
constant term; the projection $T$ picks out the pole part, as in a
minimal subtraction scheme. Now $T$ is not a homomorphism, but the
property that both its kernel and image are subalgebras is reflected
in a ``multiplicativity constraint'':
\begin{equation}
T(ab) + T(a)\,T(b) = T(T(a)\,b) + T(a\,T(b))
 \sepword{for all} a,b \in \A.
\label{eq:mult-contr} 
\end{equation}

\begin{exer}
\label{xr:mult-ctrt} 
Check \eqref{eq:mult-contr} by examining the four cases $a \in \A_\pm$,
$b \in \A_\pm$ separately.
\end{exer}

The second step is to invoke the renormalization scheme. It can now
be summarized as follows. If $\Ga$ is 1PI and is \textit{primitive}
(i.e., it has no subdivergences), we set
$$
C(\Ga) := - T(f(\Ga)),  \sepword{and then}  R(\Ga) := f(\Ga) + C(\Ga),
$$
where $C(\Ga)$ is the \textit{counterterm} and $R(\Ga)$ is the
desired finite value: in other words, for primitive graphs one simply
removes the pole part. Next, we may recursively define Bogoliubov's
$\Rbar$-operation by setting
$$
\Rbar(\Ga) = f(\Ga)
 + \sum_{\emptyset\subsetneq\ga\subsetneq\Ga} C(\ga)\,f(\Ga/\ga),
$$
with the proviso that
\begin{equation}
C(\ga_1\dots\ga_r) := C(\ga_1) \dots C(\ga_r),
\label{eq:mult-ctrtm} 
\end{equation}
whenever $\ga = \ga_1\dots\ga_r$ is a disjoint union of several
components. The final result is obtained by removing the pole part of
the previous expression: $C(\Ga) := - T(\Rbar(\Ga))$ and
$R(\Ga) := \Rbar(\Ga) + C(\Ga)$. In summary,
\begin{subequations}
\label{eq:graph-renorm} 
\begin{align}
C(\Ga) &:= - T\biggl[ f(\Ga)
+ \sum_{\emptyset\subsetneq\ga\subsetneq\Ga}C(\ga)\,f(\Ga/\ga)\biggr],
\label{eq:graph-renorm-C} 
\\
R(\Ga) &:= f(\Ga) + C(\Ga)
 + \sum_{\emptyset\subsetneq\ga\subsetneq\Ga} C(\ga)\,f(\Ga/\ga).
\label{eq:graph-renorm-R} 
\end{align}
\end{subequations}

The equation \eqref{eq:graph-renorm-C} is what is meant by saying that
``the antipode delivers the counterterm'': one replaces $S$ in the
calculation \eqref{eq:graph-antp} by~$C$ to obtain the right hand
side, before projection with~$T$. From the definition of the coproduct
in $H_\Phi$, \eqref{eq:graph-renorm-R} is a \textit{convolution} in
$\Hom(H_\Phi,\A)$, namely, $R = C * f$. To show that $R$ is
multiplicative, it is enough to verify that the counterterm map $C$ is
multiplicative, since the convolution of homomorphisms is a
homomorphism because $\A$ is commutative. In other words, we must
check that \eqref{eq:mult-ctrtm} and \eqref{eq:graph-renorm-C} are
compatible.

This is easy to do by induction on the degree of the grading
of~$H_\Phi$. We shall use the modified Sweedler notation
of~\eqref{eq:gr-coprod}, to simplify the calculation. Starting from
$C(1) := 1_\A$, we define, for $a \in \ker\eps$,
\begin{equation}
C(a) := - T\bigl[ f(a) + \tsum C(a'_{:1})\,f(a'_{:2})\bigr],
\label{eq:defn-ctrtm} 
\end{equation}
assuming $C(b)$ to be already defined, and multiplicative, whenever
$b$ has smaller degree than~$a$. By comparing the expansions of
$\Dl(ab)$ and $(\Dl a)(\Dl b)$, we see that
\begin{align*}
\tsum (ab)'_{:1} \ox (ab)'_{:2}
&= a \ox b + b \ox a + \tsum ab'_{:1} \ox b'_{:2}
   + b'_{:1} \ox ab'_{:2}
\\
&\qquad  + a'_{:1}b \ox a'_{:2} + a'_{:1} \ox a'_{:2}b
  + a'_{:1}b'_{:1} \ox a'_{:2}b'_{:2}.
\end{align*}
Using the multiplicativity constraint \eqref{eq:mult-contr} and the
definition $C(a) := - T(\Rbar(a))$, we get
\begin{align*}
C(a) C(b)
&= T\bigl[ \Rbar(a) \bigr] \, T\bigl[ \Rbar(b) \bigr]
 = -T\bigl[\Rbar(a)\,\Rbar(b) + C(a)\,\Rbar(b) + \Rbar(a)\,C(b)\bigr]
\\
&= -T\bigl[f(a) f(b) + C(a) f(b) + f(a) C(b)
     + \tsum C(a) C(b'_{:1}) f(b'_{:2}) + f(a) C(b'_{:1}) f(b'_{:2})
\\
&\qquad\quad + \tsum C(a'_{:1})f(a'_{:2})C(b)
     + C(a'_{:1})f(a'_{:2})f(b)
     + C(a'_{:1}) f(a'_{:2}) C(b'_{:1}) f(b'_{:2}) \bigr]
\\
&= -T\bigl[f(a) f(b) + C(a) f(b) + C(b) f(a)
     + \tsum C(ab'_{:1}) f(b'_{:2}) + C(b'_{:1}) f(ab'_{:2})
\\
&\qquad\quad + \tsum C(a'_{:1}b) f(a'_{:2}) + C(a'_{:1}) f(a'_{:2}b)
     + C(a'_{:1}b'_{:1}) f(a'_{:2}b'_{:2}) \bigr]
\\
&= -T\bigl[f(ab) + \tsum C((ab)'_{:1}) f((ab)'_{:2}) \bigr]
 = C(ab),
\end{align*}
where, in the penultimate line, we have used the assumed
multiplicativity of~$C$ in lower degrees.

\marker
The decomposition $R = C * f$ has a further consequence. Assume that
the unrenormalized integrals, although divergent at $\eps = 0$, make
sense on the circle $S$ in the complex plane where
$|\eps| = |d - D| = r_0$, say. Evaluation at any $d = z$ defines a
character $\chi_z \: \A \to \C$ of the Laurent-series algebra.
Composing this character with $f\: H_\Phi \to \A$ gives a loop of
characters of~$H_\Phi$:
$$
\ga(z) := \chi_z \circ f,  \sepword{for any} z \in S.
$$
Likewise, $\ga_-(z) := \chi_z\circ C$ and $\ga_+(z) := \chi_z\circ R$
define characters of~$H_\Phi$ ---here is where we use the
multiplicativity of~$C$ and~$R$--- and $R = C * f$ entails
$\ga_+(z) = \ga_-(z) \ga(z)$, or equivalently,
\begin{equation}
\ga(z) = \ga_-(z)^{-1} \,\ga_+(z),  \sepword{for all} z \in S.
\label{eq:Birk-decomp} 
\end{equation}
The properties of the subalgebras $\A_+$ and $\A_-$ show that
$\ga_+(z)$ extends holomorphically to the disc $|z - D| < r_0$, while
$\ga_-(z)$ extends holomorphically to the outer region
$|z - D| > r_0$ with $\ga_-(\infty)$ being finite. Since a function
holomorphic on both regions must be constant (Liouville's theorem), we
can normalize the factorization \eqref{eq:Birk-decomp} just by setting
$\ga_-(\infty) := 1$. The renormalization procedure thus corresponds
to replacing the loop $\set{\ga(z) : z \in S}$ by the finite
evaluation $\ga_+(D)$.

The decomposition \eqref{eq:Birk-decomp} of a group-valued loop is
known as the \textit{Birkhoff factorization}, and arises in the study
of linear systems of differential equations
$$
y'(z) = A(z) \,y(z),
$$
where $A(z)$ is a meromorphic $n \x n$ matrix-valued function with
simple poles. The solution involves constructing a loop around one of
these poles $z_0$ with values in the Lie group $GL(n,\C)$. We refer to
\cite[Chap.~8]{PressleyS} for an instructive discussion of this
problem. Any such loop factorizes as follows:
$$
\ga(z) = \ga_-(z)^{-1} \,\la(z) \,\ga_+(z),
$$
where $\ga_+(z)$ is holomorphic for $|z - z_0| < r_0$,
$\ga_-(z)$ is holomorphic for $|z - z_0| > r_0$ with
$\ga_-(\infty) = 1$, and $\set{\la(z) : |z - z_0| = r_0}$ is a loop
with values in the $n$-torus of diagonal matrices. The loop $\la$
provides clutching functions for $n$~line bundles over the Riemann
sphere, and these are obstructions to the solvability of the
differential system. However, in our context, the Lie group $GL(n,\C)$
is replaced by the topologically trivial group $\G(H_\Phi)$, so that
the loop $\la$ becomes trivial and the decomposition
\eqref{eq:Birk-decomp} goes through as stated, thereby providing a
general recipe for computing finite values in renormalizable theories.

\newpage


\section{Cyclic Cohomology}

\subsection{Hochschild and cyclic cohomology of algebras}
\label{sec:Hoch-cyc} 

We have already discussed briefly, in
subsection~\ref{sec:CKr-rooted-tree}, the Hochschild cohomology of
associative algebras. Recall that a Hochschild $n$-cochain, for an
algebra over the complex field, is a multilinear map
$\vf\: \A^{n+1} \to \C$, with the coboundary map given
by~\eqref{eq:Hoch-cobrdy}. These $n$-cochains make up an $\A$-bimodule
$C^n = C^n(\A,\A^*)$; the $n$-cocycles
$Z^n = \set{\vf \in C^n : b\vf = 0}$ and the $n$-coboundaries
$B^n = \set{b\psi : \psi \in C^{n-1}}$ conspire to form the Hochschild
cohomology module $HH^n(\A) := Z^n/B^n$. A $0$-cocycle $\tau$ is a
trace on~$\A$, since
$\tau(a_0a_1) - \tau(a_1a_0) = b\tau(a_0,a_1) = 0$.

In the commutative case, when $\A = \Coo(M)$ is an algebra of smooth
functions on a manifold~$M$ (we take $\A$ unital and $M$ compact, as
before), there is a theorem of Connes~\cite{ConnesNCDiffG}, which
dualizes an older result in algebraic geometry due to Hochschild,
Kostant and Rosenberg~\cite{HochschildKR}, to the effect that
Hochschild classes for $\Coo(M)$ correspond exactly to de~Rham
currents on~$M$. (Currents are the objects which are dual to
differential forms, and can be thought of as formal linear
combinations of domains for line and surface integrals within~$M$.)
The correspondence $[\vf] \mapsto C_\vf$ is given by
skewsymmetrization of $\vf$ in all arguments but the first:
$$
\int_{C_\vf} a_0 \,da_1 \wyw da_k
 := \frac{1}{k!} \sum_{\pi\in S_k} (-1)^\pi\,
     \vf(a_0,\row a{\pi(1)}{\pi(k)}).
$$
Dually, Hochschild \textit{homology} classes on~$\Coo(M)$ correspond
to differential \textit{forms} on~$M$; that is,
$HH_k(\Coo(M)) \simeq \A^k(M)$ for $k = 0,1,\dots,\dim M$.

On the de~Rham side, the vector spaces $\D_k(M)$ of currents of
dimension~$k$ form a complex, but with zero maps between them, so that
each Hochschild class $[\vf]$ matches with a single current $C_\vf$
rather than with its homology class. To deal with the homology
\textit{classes}, we must bring in an algebraic expression for the
de~Rham boundary. This turns out to be a degree-lowering operation on
Hochschild cochains: if $\psi \in C^k$, then $B\psi \in C^{k-1}$,
given by
\begin{align}
B\psi(\row a0{k-1})
&:= \sum_{j=0}^{k-1} (-1)^{j(k-1)} \psi(1,\row aj{k-1},\row a0{j-1})
\notag \\
&\qquad\quad + (-1)^{(j-1)(k-1)} \psi(\row aj{k-1},\row a0{j-1},1),
\label{eq:cohoml-B} 
\end{align}
does the job. Indeed, if $C$ is a $k$-current and $\vf_C$ is the
(already skewsymmetric) cochain
$$
\vf_C(a_0,\row a1k) := \int_C a_0 \,da_1 \wyw da_k,
$$
then $\vf_C(\row a0{k-1},1) = 0$, and therefore
\begin{align*}
B\vf_C(\row a0{k-1})
&= \sum_{j=0}^{k-1} (-1)^{j(k-1)}
    \int_C da_j \wyw da_{k-1} \w da_0 \wyw da_{j-1}
\\
&= \sum_{j=0}^{k-1} \int_C da_0 \wyw da_{k-1}
 = k \int_{\del C} a_0 \,da_1 \wyw da_{k-1},
\end{align*}
by using Stokes' theorem; thus $B\vf_C = k\,\vf_{\del C}$. Up to the
normalization factor~$k = \deg C$, the algebraic operator $B$ delivers
the de~Rham boundary. Thus, the algebraic picture for de~Rham
homology involves a cohomology of algebras which uses both $b$ and~$B$.

\marker
Dually, the Hochschild homology of algebras supports a degree-raising
operator, also called~$B$, which is closely related related to the
de~Rham coboundary (that is, the exterior derivative). Indeed, if we
use the version of Hochschild homology where the chains belong to the
universal graded differential algebra $\Om^\8\A$, with $b$ given
by~\eqref{eq:Oma-bdry}, then $B\: \Om^k\A \to \Om^{k+1}\A$ is given
simply by
\begin{equation}
B(a_0 \,da_1 \dots da_k)
 := \sum_{j=0}^k (-1)^{kj} da_j\dots da_k\,da_0\dots da_{j-1}.
\label{eq:homl-B} 
\end{equation}
which mimics the operation $\om \mapsto k\,d\om$ on differential
$k$-forms. In the manifold case, the various $da_j$ anticommute, but
for more general algebras they do not, so the cyclic summation in
\eqref{eq:homl-B} is unavoidable. From the formula, it is obvious
that $B^2 = 0$. One checks easily that $bB + Bb = 0$, too.

\begin{exer}
\label{xr:baby-Chern} 
If $e \in \A$ is an idempotent element, that is, $e^2 = e$, and $k$
is even, check that
\begin{align*}
b(e\,(de)^k) &= e\,(de)^{k-1}, &  b((de)^k) &= (2e-1)\,(de)^{k-1},
\\
B(e\,(de)^k) &= (k + 1)\,(de)^{k+1}, &  B((de)^k) &= 0.
\end{align*}
If $k$ is odd, show that instead,
$$
b(e(de)^k) = b((de)^k) = 0  \sepword{and}  B(e(de)^k) = B((de)^k) = 0.
\eqno \qef
$$
\hideqef
\end{exer}

Moving back to cohomology, one can check that $b^2 = 0$, $B^2 = 0$,
and $bB + Bb = 0$ hold there, too. This gives rise to a
\textit{bicomplex}:
$$
\begin{CD}
\vdots @.  \vdots @.  \vdots @. \vdots   \\
@A{b}AA    @A{b}AA    @A{b}AA   @A{b}AA  \\
C^3 @>B>>  C^2 @>B>>  C^1 @>B>> C^0      \\
@A{b}AA    @A{b}AA    @A{b}AA            \\
C^2 @>B>>  C^1 @>B>>  C^0                \\
@A{b}AA    @A{b}AA                       \\
C^1 @>B>>  C^0                           \\
@A{b}AA                                  \\
C^0
\end{CD}
$$
Folding this up along the diagonals, we get a ``total complex'' whose
coboundary operator is $b + B$, and whose module in degree~$n$ is
$$
C^n \op C^{n-2} \op C^{n-4} \opyop C^{\#n},
$$
where $\#n = 0$ or~$1$ according as $n$ is even or odd. The cohomology
of this total complex is, by definition, the \textbf{cyclic
cohomology} $HC^\8(\A)$ of the algebra~$\A$. (The letters $HC$ stand
for ``homologie cyclique'': on replacing $C^k$ by $\Om^k(\A)$ and
running all the arrows backwards, we get a dual bicomplex; the
homology $HC_\8(\A)$ of its total complex is the \textit{cyclic
homology} of~$\A$.)

\marker
There is an alternative description of cyclic cohomology, which in
some ways is simpler. Let $\tau$ be the operation of cyclic
permutation of the arguments of a Hochschild cochain:
\begin{equation}
\tau\vf(\row a0n) := \vf(a_n,\row a0{n-1}).
\label{eq:cyclic-tau} 
\end{equation}
We say that $\vf$ is \textit{cyclic} if $\tau\vf = (-1)^n \vf$
---notice that $(-1)^n$ is the sign of this cyclic permutation--- and
denote the subspace of cyclic $n$-cochains by $C_\la^n = C_\la^n(\A)$
(the notation $\la = (-1)^n \tau$ is often used). If $Z_\la^n(\A)$ and
$B_\la^n(\A)$ are the respective cyclic $n$-cocycles and cyclic
$n$-coboundaries, an exercise in homological algebra shows that
$HC^n(\A) \simeq Z_\la^n(\A)/B_\la^n(\A)$.

Let us compute $HC^\8(\A)$ for a simple example: the algebra
$\A = \C$, which is the coordinate algebra of a single point. The
module $\C^n$ is one-dimensional, since
$\vf(\row a0n) = a_0\dots a_n \vf(1,1,\dots,1)$; it has a basis
element $\vf^n$ determined by $\vf^n(1,1,\dots,1) := 1$. Clearly,
$b\vf^n = \sum_{j=0}^{n+1} (-1)^j \vf^{n+1} = 0$ or~$\vf^{n+1}$,
according as $n$ is even or odd. We also find that $B\vf^n = 0$
or~$2n\vf^{n-1}$, according as $n$ is even or odd. The total complex
is of the form
$$
\C \longto^0 \C \longto^{d_1} \C^2 \longto^0 \C^2
   \longto^{d_2} \C^3 \longto^0 \C^3 \longto^{d_3} \cdots
$$
each $d_j$ being injective with range of codimension~$1$; for
instance, $d_2(\vf^3, \vf^1) = (\vf^4, 7\vf^2, 2\vf^0)$. The
alternative approach, using cyclic $n$-cocycles, argues more simply
that $\tau\vf^n = \vf^n$, so that $Z_\la^n(\C) = \C$ or~$0$ according
as $n$ is even or odd, while $B_\la^n(\C) = 0$ for all~$n$. Either
way, $HC^n(\C) = \C$ if $n$ is even, and $HC^n(\C) = 0$ if $n$ is odd.

This periodicity might seem surprising: the de~Rham cohomology of a
one-point space is $\C$ in degree zero, and $0$ in all higher
degrees. Now we may notice that there is an obvious ``shifting
operation'' $S$ on the bicomplex, moving all modules right and up by
one step (and pushing the total complex along by two steps); it leaves
behind the first column, which is just the Hochschild complex of~$\A$.
At the level of cohomology, we get a pair of maps
$$
HC^{n-2}(\A) \longto^S HC^n(\A) \longto^I HH^n(\A),
$$
which actually splice together into a long exact sequence:
$$
\cdots \longto HC^n(\A) \longto^I HH^n(\A) \longto^B HC^{n-1}(\A)
  \longto^S HC^{n+1}(\A) \longto^I HH^{n+1}(\A) \longto \cdots
$$
whose connecting homomorphism comes from the aforementioned $B$ at the
level of cochains. The detailed calculations which back up these
plausible statements are long and tedious; they are given in
\cite[Chap.~2]{Loday} for cyclic homology, and in
\cite[\S 10.1]{Polaris} is the cohomological setting. The upshot is
that, by iterating the periodicity operator~$S$, one can compute two
direct limits, which capture the main algebraic invariants of~$\A$.

\begin{defn}
\label{df:HP-gps} 
The periodicity maps $S\: HC^n \to HC^{n+2}$ define two directed
systems of abelian groups; their inductive limits
$$
HP^0(\A) := \injlim HC^{2k}(\A),  \qquad
HP^1(\A) := \injlim HC^{2k+1}(\A),
$$
are called the even and odd \textit{periodic cyclic cohomology} groups
of the algebra~$\A$. In particular, $HP^0(\C) = \C$ and
$HP^1(\C) = 0$.
\end{defn}

In the commutative case $\A = \Coo(M)$, it turns out that $HC^\8(\A)$
does not quite capture the de~Rham homology of $M$. The exact result
---see \cite[Thm.~III.2.2]{ConnesBook} or \cite[Thm.~10.5]{Polaris}---
is
$$
HC^k(\Coo(M)) \simeq Z_k^\dR(M) \op H_{k-2}^\dR(M) \op H_{k-4}^\dR(M)
  \opyop H_{\#k}^\dR(M),
$$
where $Z_k^\dR(M)$ is the vector space of closed $k$-currents on~$M$,
$H_r^\dR(M)$ is the $r$th de~Rham homology group of~$M$, and $\#k = 0$
or~$1$ according as $k$ is even or~odd. However, one may use $S$ to
promote the closed $k$-currents, two degrees at a time, until the
full de~Rham homology is obtained, since $Z_k^\dR(M) = 0$ for
$k > \dim M$; then we get de~Rham homology exactly, albeit rolled up
into even and odd degrees:
$$
HP^0(\Coo(M)) \simeq H_\even^\dR(M),  \qquad
HP^1(\Coo(M)) \simeq H_\odd^\dR(M).
$$
There is also a dual result, which matches a periodic variant of the
cyclic homology of $\Coo(M)$ with the even/odd de~Rham cohomology
of~$M$.

\marker
The importance of this algebraic scheme for de~Rham co/homology is
that it provides many \textit{Chern characters}, even for highly
noncommutative algebras. Generally speaking, Chern characters are
tools to compute algebraic invariants from the more formidable
$K$-theory and $K$-homology of algebras. The idea is to associate, to
any pair of classes $[x] \in K_\8(\A)$ and $[D] \in K^\8(\A)$ another
pair of classes $\ch_\8 x \in HC_\8(\A)$ and $\ch^\8 D \in HC^\8(\A)$,
given by explicit and manageable formulas, so that the \textit{index
pairing} $\dst{[x]}{[D]}$ can be computed from a cyclic co/homology
pairing $\dst{\ch_\8 x}{\ch^\8 D}$, which is usually more tractable.
We look at the $K$-theory version first, and distinguish the even and
odd cases.

Suppose first that $e = e^2$ is an idempotent in $\A$, representing a
class $[e] \in K_0(\A)$; we define
$\ch e := \sum_{k=0}^\infty \ch_k e \in \Om^\even\A$, where the
component chains are
$$
\ch_k e
 := (-1)^k \frac{(2k)!}{k!} (e - \thalf)\,(de)^{2k} \in \Om^{2k}\A,
$$
It follows from Exercise~\ref{xr:baby-Chern} that $(b+B)(\ch e) = 0$.
Next, if $u \in \A$ is invertible, representing a class
$[u] \in K_1(\A)$; we define
$\ch u := \sum_{k=0}^\infty \ch_{k+\shalf} u \in \Om^\odd\A$, with
components
$$
\ch_{k+\shalf} u := (-1)^k k!\, u^{-1}\,du\,(d(u^{-1})\,du)^k
  = k!\,(u^{-1}\,du)^{2k+1} \in \Om^{2k+1}\A.
$$
Again, one checks that $(b + B)(\ch u) = 0$. Actually, it is fairly
rare that $K$-theory classes arise from idempotents or invertibles in
the original algebra~$\A$; more often, $e$ and $u$ belong to
$M_r(\A)$, the algebra of $r \x r$ matrices with entries in~$\A$, for
some $r = 1,2,3,\dots$; so in the definitions we must insert a trace
over these matrix elements; the previous equations must be modified to
\begin{subequations}
\label{eq:Chern-chain} 
\begin{align}
\ch_k e &:= (-1)^k \frac{(2k)!}{k!}
   \tr\bigl( (e - \thalf)\,(de)^{2k} \bigr) \in \Om^{2k}\A,
\label{eq:Chern-chain-even} 
\\[1\jot]
\ch_{k+\shalf} u &:= k!\,\tr(u^{-1}\,du)^{2k+1} \in \Om^{2k+1}\A.
\label{eq:Chern-chain-odd} 
\end{align}
\end{subequations}
For instance, $\tr(e\,de\,de) = \tsum e_{ij} \,de_{jk} \,de_{ki}$.
The pairing of, say, the $2$-chain $\ch_1 e$ and a $2$-cochain $\vf$
is given by
$$
\dst{\vf}{\ch_1 e}
 = -2 \tsum \vf(e_{ij} - \thalf\dl_{ij}, e_{jk}, e_{ki}).
$$

\marker
The Chern character from $K$-homology to cyclic cohomology is trickier
to define. First of all, what is a $K$-cycle over the algebra~$\A$? It
turns out that it is just a spectral triple $(\A,\H,D)$, of
Definition~\ref{df:spec-tri}: an even spectral triple is a
$K^0$-cycle, an odd spectral triple is a $K^1$-cycle. The
unboundedness of the selfadjoint operator~$D$ may cause trouble, but
one can always replace $D$ (using the homotopy
$D \mapsto D\,|D|^{-t}$ for $0 \leq t \leq 1$) with its \textit{sign
operator} $F := D\,|D|^{-1}$, which is a symmetry, that is, a bounded
selfadjoint operator such that $F^2 = 1$. The compactness of
$|D|^{-1}$ translates to the condition that $[F,a]$ be compact for
each $a \in \A$; in the even case, $F$ anticommutes with the grading
operator $\chi$, just like $D$ does. The triple $(\A,\H,F)$,
satisfying these conditions, is called a \textit{Fredholm module}; it
represents the same $K$-homology class as the spectral triple
$(\A,\H,D)$.

Although $F$ is bounded, it is analytically a much more singular
object than~$D$, as a general rule. For instance, if
$D = (2\pi i)^{-1} \,d/d\th$ is the Dirac operator on the unit
circle $\Sf^1$, one finds that $F$ is given by a principal-value
integral:
$$
Fh(\a) = \PVint_0^1 i\, h(\a - \th)\, \cot\pi\th \,d\th,
$$
which is a trigonometric version of the \textit{Hilbert transform} on
$L^2(\R)$,
$$
Fh(x) = \frac{i}{\pi} \PVint \frac{h(x-t)}{t} \,dt
 := \frac{i}{\pi} \lim_{\eps\downto 0} \int_{|t|>\eps}
     \frac{h(x-t)}{t} \,dt.
$$
This can be seen by writing both operators in a Fourier basis for
$\H = L^2(\Sf^1)$:
$$
D(e^{2\pi ik\th}) = k\, e^{2\pi ik\th},  \qquad
F(e^{2\pi ik\th}) = (\sign k)\, e^{2\pi ik\th},
$$
with the convention that $\sign 0 = 1$. This analytic intricacy of $F$
must be borne in mind when regarding the formula for the Chern
character of its $K$-homology class, which is given by the cyclic
$n$-cocycle
\begin{equation}
\tau_F^n(\row a0n) := \frac{\Ga(\tfrac{n}{2} + 1)}{2\,n!}
     \Tr\bigl(\chi F [F,a_0] \dots [F,a_n] \bigr),
\label{eq:Chern-char-F} 
\end{equation}
provided $n$ is large enough that the operator in parentheses is
trace-class. (The Fredholm module is said to be ``finitely summable''
if this is true for a large enough~$n$.) One can always replace $n$ by
$n + 2$, because it turns out that $S\tau_F^n$ and $\tau_F^{n+2}$ are
cohomologous, so that the Chern character is well-defined as a
periodic class. Much effort has gone into finding more tractable
``local index formulas'' for this Chern character, in terms of more
easily computable cocycles: see \cite{ConnesMIndex}
or~\cite{Benameur}.

\marker
An important example of a cyclic $1$-cocycle ---historically one of
the first to appear in the literature \cite{ArakiCAR,ArakiCyclic}---
is the \textit{Schwinger term} of a $1+1$-dimensional QFT. In that
context, there is a fairly straightforward ``second quantization'' in
Fock space: we recall here only a few aspects of the formalism. In
``first quantization'', one starts with a real vector space $V$ of
solutions of a Dirac-type equation $(i\,\del/\del t - D)\psi = 0$,
together with a symmetric bilinear form $g$ making it a \textit{real}
Hilbert space. If $E_+$ and $E_-$ denote the orthogonal projectors on
the subspaces of positive- and negative-frequency solutions,
respectively, the sign operator is $F := E_+ - E_-$; moreover,
$J := iF = iE_+ - iE_-$ is an orthogonal complex structure on~$V$ (in
other words, $J^2 = -1$), which can be used to make $V$ into a
\textit{complex} Hilbert space $V_J$ with the scalar product
$$
\<u,v>_\ssJ := g(u,v) + ig(Ju,v).
$$
(In examples representing charged fields, $V$ is already a complex
Hilbert space with an ``original'' complex structure $Q = i$; the
construction of the new Hilbert space with complex structure $J$ is
equivalent to ``filling up the Dirac sea'', and $Q$ is the charge, a
generator of global gauge transformations.)

The fermion Fock space $\F_J(V)$ is simply the exterior algebra over
$V_J$; the scalars in $\La^0 V$ are the multiples of the vacuum vector
$\ket{0}$. If $\{u_j\}$ is an orthonormal basis for $V_J$, there are
corresponding creation and annihiliation operators on $\F_J(V)$:
\begin{align*}
a_i^\7(u_1 \wyw u_k) &:= u_i \w u_1 \wyw u_k,
\\
a_i(u_1 \wyw u_k) &:= \sum_{j=1}^k (-1)^{j-1}
  \<u_i,u_j>_\ssJ\, u_1 \wyw \miss{u_j} \wyw u_k.
\end{align*}
Any real-linear operator $B$ on~$V$ can be written as $B = B_+ + B_-$
where $B_+ := \thalf(B - JBJ)$ gives a complex-linear operator
on~$V_J$ because it commutes with~$J$, but $B_- := \thalf(B + JBJ)$
is antilinear: $JB_- = - B_-J$. A skewsymmetric operator $B$ is
quantizable, by a result of Shale and Stinespring~\cite{ShaleS},
if and only if $[J,B] = 2JB_-$ is Hilbert--Schmidt operator, and the
second-quantization rule is $B \mapsto \dm(B)$, where $\dm(B)$ is the
following operator on Fock space:
\begin{equation}
\dm(B) := \frac{1}{2} \sum_{k,l} \<u_k, B_- u_l>_\ssJ \,a_k^\7 a_l^\7
 + 2\,\<u_k, B_+ u_l>_\ssJ \,a_k^\7 a_l
 - \<B_- u_l, u_k>_\ssJ \,a_l a_k.
\label{eq:second-quantn} 
\end{equation}
The rule complies with normal ordering, because
$\braCket{0}{\dm(B)}{0} = 0$, i.e., the vacuum expectation value is
zero. However, this implies that \eqref{eq:second-quantn} is not quite
a representation of the Lie algebra
$\set{B = -B^t : B_-\text{ is Hilbert--Schmidt}}$. The anomalous
commutator, or Schwinger term, is given by
$$
[\dm(A), \dm(B)] - \dm([A,B]) = -\thalf \Tr[A_-, B_-].
$$
This is a well-known result: see \cite{Rhea} or
\cite[Thm.~6.7]{Polaris} for a proof. The trace here is taken on the
Hilbert space $V_J$; notice that, although $[A_-,B_-]$ is a traceclass
commutator, its trace need not vanish, because it is the commutator of
\textit{antilinear} operators.

The claim is that $\a(A,B) := -\thalf \Tr[A_-,B_-]$ defines a cyclic
$1$-cocycle on the algebra generated by such $A$ and~$B$. For that, we
rewrite it in terms of a trace of operators on the complexified space
$V^\C := V \op iV$; any real-linear operator $B$ on~$V$ extends to a
$\C$-linear operator on~$V^\C$ in the obvious way:
$B(u + iv) := B(u) + iB(v)$. For instance, $F := E_+ - E_-$ where
$E_+$ and $E_-$ now denote complementary orthogonal projectors
on~$V^\C$. Taking now the trace over $V^\C$, too, we find that
\begin{equation}
\a(A,B) = \teighth \Tr(F[F,A][F,B]).
\label{eq:Schwng-term} 
\end{equation}
To see that, first notice that $F[F,B] = B - FBF = -[F,B]F$, and so
$\Tr(F[F,A][F,B]) = \Tr([F,B]F[F,A]) = - \Tr(F[F,B][F,A])$. The right
hand side of~\eqref{eq:Schwng-term} is unchanged under
skewsymmetrization:
$\frac{1}{8} \Tr(F[F,A][F,B]) = \thalf \Tr(A_- F B_-)
 = -\tquarter \Tr(F[A_-,B_-])$. Thus, in turn, equals
\begin{align*}
-\tquarter \Tr(F[A_-,B_-])
&= -\tquarter \Tr(E_+[A_-,B_-]E_+) +\tquarter \Tr(E_-[A_-,B_-]E_-)
\\
&= -\thalf \Tr(E_+ A_- E_- B_- E_+ - E_+ B_- E_- A_- E_+) = \a(A,B).
\end{align*}
This is a cyclic cochain, since $\a(A,B) = -\a(B,A)$; and it is a
cocycle because
\begin{align*}
b\a(A,B,C)
&= \teighth \Tr(F[F,AB][F,C] - F[F,A][F,BC] + F[F,CA][F,B])
\\
&= \teighth
    \Tr(FA[F,B][F,C] - F[F,A][F,B]C + FC[F,A][F,B] + F[F,C]A[F,B])
\\
&= \teighth
    \Tr(FA[F,B][F,C] - [F,A][F,B]FC + FC[F,A][F,B] - [F,C]FA[F,B])
\\
&= 0.
\end{align*}
The Schwinger term is actually just a multiple of the Chern character
$\tau_F^1$, as specified by~\eqref{eq:Chern-char-F}, of the Fredholm
module defined by~$F$. The Shale--Stinespring condition shows that
$F[F,A][F,B]$ is trace-class, so that, in this case, the character
formula makes sense already for $n = 1$.

\subsection{Cyclic cohomology of Hopf algebras}
\label{sec:cyc-Hopf} 

We now take a closer look at the algebraic operators $b$ and~$B$, in
the cohomological setting. They can be built up from simpler
constituents. First of all, the coboundary $b\: C^{n-1} \to C^n$ may
be written as $b = \sum_{i=0}^n (-1)^i \dl_i$, where
\begin{align*}
\dl_i\vf(\row a0n) &:= \vf(\row a0i\row a{i+1}n),
  \qquad i = 0,1,\dots,n-1,
\\
\dl_n\vf(\row a0n) &:= \vf(a_n\row a0{n-1}).
\end{align*}
We also introduce maps $\sg_j\: C^{n+1} \to C^n$, for
$j = 0,1,\dots,n$, given by
$$
\sg_j\vf(\row a0n) := \vf(\row a0j,1,\row a{j+1}n),
$$
and recall the ``cyclic permuter'' $\tau\: C^n \to C^n$
of~\eqref{eq:cyclic-tau}:
$$
\tau\vf(\row a0n) := \vf(a_n,\row a0{n-1}).
$$
Notice that $\tau^{n+1} = 1$ on $C^n$. The operator $B$ is built from
the $\sg_j$ and $\tau$, as follows. The ``cyclic skewsymmetrizer''
$N := \sum_{k=0}^n (-1)^{nk} \tau^k$ acts on $C^n$ as
$$
N\vf(\row a0n) = \vf(\row a0n)
 + \sum_{k=1}^n (-1)^{nk} \vf(\row a{n-k+1}n,\row a0{n-k}).
$$
The formula \eqref{eq:cohoml-B} now reduces to
$$
B = (-1)^n N(\sg_0 \tau^{-1} + \sg_n) : C^{n+1} \to C^n.
$$

The algebraic structure of cyclic cohomology is essentially determined
by the relations between the elementary maps $\dl_i$, $\sg_j$
and~$\tau$. For instance, the associativity of the algebra $\A$ is
captured by the rule $\dl_{i+1} \dl_i = \dl_i^2$ as maps from
$C^{n-1}$ to $C^{n+1}$. Here is the full catalogue of these
composition rules:
\begin{align}
\dl_j \dl_i &= \dl_i \dl_{j-1} \sepword{if} i < j;
\notag \\
\sg_j \sg_i &= \sg_i \sg_{j+1} \sepword{if} i \leq j;
\notag \\
\sg_j \dl_i &= \begin{cases}
                \dl_i \sg_{j-1} &\text{if } i < j, \\
                            \id &\text{if } i = j \text{ or } j+1, \\
                \dl_{i-1} \sg_j &\text{if } i > j+1; \end{cases}
\notag \\
\tau \dl_i &= \dl_{i-1} \tau : C^{n-1} \to C^n
 \text{ for } i = 1,\dots,n,  \sepword{and} \tau \dl_0 = \dl_n,
\notag \\
\tau \sg_j &= \sg_{j-1} \tau : C^{n+1} \to C^n
 \text{ for } j = 1,\dots,n, \sepword{and} \tau \sg_0 = \sg_n \tau^2,
\notag \\
\tau^{n+1} &= \id  \text{ on } C^n.
\label{eq:cycl-relns} 
\end{align}
The first three rules, not involving~$\tau$, arise when working with
simplices of different dimensions, where the ``face maps'' $\dl_i$
identify an $(n-1)$-simplex with the $i$th face of an $n$-simplex,
while the ``degeneracy maps'' $\sg_j$ reduce an $(n+1)$-simplex to an
$n$-simplex by collapsing the edge from the $j$th to the $(j+1)$st
vertex into a point. A set of simplices, one in each dimension,
together with maps $\dl_i$ and $\sg_j$ complying with the above rules,
forms the so-called ``simplicial category'' $\Dl$ ---see \cite{Loday},
for instance--- and any other instance of those rules defines a
functor from $\Dl$ to another category: in other words, $\Dl$ is a
universal model for those rules.

By bringing in the next three rules involving $\tau$ also, Connes
defined a ``cyclic category'' $\La$ which serves as a universal model
for cyclic cohomology \cite{ConnesLambda}. Essentially, one
supplements $\Dl$ with the maps which cyclically permute the vertices
of each simplex (an ordering of the vertices is given). The point of
this exercise is its universality, so that any system of maps
complying with \eqref{eq:cycl-relns} gives a bona-fide cyclic
cohomology theory, complete with periodicity properties and so on.
Indeed, one can show \cite[Lemma~10.4]{Polaris} that if
$\ga\: C^{n-1} \to C^n$ is defined by
$\ga := \sum_{k=1}^n (-1)^k k\,\dl_k$, then $S := (n^2+n)^{-1} b\ga$
defines the periodicity operator on cyclic $(n-1)$-cocycles.

\marker
Important cyclic cocycles, such as the characteristic classes for the
algebras which typically arise in noncommutative geometry, can be
quite difficult to compute. This is especially true for crossed
product algebras, such as those of subsection~\ref{sec:Hopf-diff}.
It is time to discuss how this problem may be addressed by transfer
from cyclic cocycles of an associated Hopf algebra which acts on the
algebra in question.

We recall from subsection~\ref{sec:Hopf-diff} that such a crossed
product algebra $\A$, obtained from the action of local diffeomorphisms
on the frame bundle over a manifold, carries an action of a certain
Hopf algebra $H$ of differential (and multiplication) operators, where
the Hopf action itself codifies the generalized Leibniz rules for
these operators. To define characteristic classes in $HC^\8(\A)$, we
introduce a new cyclic cohomology for $H$ and then show how to map
$H$-classes to $\A$-classes.

This cyclic cohomology for $H$ was introduced in \cite{ConnesMHopf}
and developed further in \cite{ConnesMModl,ConnesMSymm,ConnesMDiff}
and also in \cite{CrainicWeil,CrainicThesis}. Its definition will make
full use of the Hopf algebra structure, so we proceed in a
``categorical'' fashion. We shall first assume that the antipode $S$
is \textit{involutive}, that is, $S^2 = \id_H$. As indicated earlier,
this holds true for commutative or cocommutative Hopf algebras,
although not for the Hopf algebra $H_{CM}$ of
subsection~\ref{sec:Hopf-diff}; but that case can be handled by making
a suitable adjustment later on.

To set up the cyclic cohomology  of $H$, we start with the algebras
$C^n(H) := H^{\ox n}$ for $n = 1,2,3,\dots$ and $C^0(H) := \C$
(or~$\FF$, if one prefers other kinds of scalars). This looks
superficially like the \textit{chain} complex for associative
algebras, but we shall make it a cochain complex by (once again)
taking advantage of duality to replace products by coproducts, and so
on. The ``simplicial'' operations are defined by
\begin{align}
\dl_0(h^1 \oxyox h^{n-1}) &:= 1 \ox h^1 
\oxyox h^{n-1},
\notag \\
\dl_i(h^1 \oxyox h^{n-1}) &:= h^1 \oxyox \Dl(h^i) \oxyox h^{n-1},
  \quad i = 1,\dots,n-1,
\notag \\
\dl_n(h^1 \oxyox h^{n-1}) &:= h^1 \oxyox h^{n-1}
\ox 1,
\notag \\
\sg_j(h^1 \oxyox h^{n+1})
&:= \eps(h^{j+1})\, h^1 \oxyox h^j \ox h^{j+2} \oxyox h^{n+1}.
\label{eq:elem-maps} 
\end{align}
For $n = 0$, these reduce to $\dl_0(1) := 1$, $\dl_1(1) := 1$, and
$\sg_0(h) := \eps(h)$. The relation $\dl_{i+1} \dl_i = \dl_i^2$ of
\eqref{eq:cycl-relns} expresses the coassociativity of~$\Dl$ and the
equation $\Dl(1) = 1 \ox 1$; the relations
$\sg_j \dl_j = \sg_j \dl_{j+1} = \id$ are equivalent to
$(\id \ox \eps)\Dl = (\eps \ox \id)\Dl = \id$; and the remaining
relations involving the $\dl_i$ and the $\sg_j$ only are trivial.

To define the cyclic permuter $\tau$, we first note that $H^{\ox n}$
is itself an $H$-module algebra under the ``diagonal'' action of~$H$:
$$
h\.(k^1 \oxyox k^n) := (\Dl^{n-1}h)\,(k^1 \oxyox k^n)
 = \tsum h_{:1}k^1 \ox h_{:2}k^2 \oxyox h_{:n}k^n.
$$
We then define
\begin{align}
\tau(h^1 \oxyox h^n) &:= S(h^1)\.(h^2 \oxyox h^n \ox 1)
 = \Dl^{n-1}(S(h^1))\,(h^2 \oxyox h^n \ox 1)
\notag \\
&= \tsum S(h_{:n}^1)h^2 \ox S(h_{:n-1}^1)h^3 \oxyox
         S(h_{:2}^1)h^n \ox S(h_{:1}^1).
\label{eq:Hopf-tau} 
\end{align}
The cyclicity property of $\tau$ is a consequence of the following
calculation.

\begin{prop}
\label{pr:cyclic-Ssq} 
The map $\tau\: H^{\ox n} \to H^{\ox n}$ satisfies
\begin{equation}
\tau^{n+1}(h^1 \ox h^2 \oxyox h^n)
 = S^2(h^1) \ox S^2(h^2) \oxyox S^2(h^n).
\label{eq:cyclic-Ssq} 
\end{equation}
\end{prop}

\begin{proof}
First we compute $\tau^2(h^1 \ox h^2 \oxyox h^n)$. The diagonal action
of $S(S(h_{:n}^1)\,h^2) = S(h^2)\,S^2(h_{:n}^1)$ gives
\begin{align*}
\tau^2(h^1 \ox h^2 \oxyox h^n)
&= \tsum S(h_{:n}^2)\,S^2(h_{:n}^1)\,S(h_{:n-1}^1)\,h^3
     \ox S(h_{:n-1}^2)\,S^2(h_{:n+1}^1)\,S(h_{:n-2}^1)\,h^4
\\
&\qquad\quad \oxyox S(h_{:2}^2)\,S^2(h_{:2n-2}^1)\,S(h_{:1}^1)
     \ox S(h_{:1}^2)\,S^2(h_{:2n-1}^1).
\end{align*}
Observe that $\tsum S^2(h_{:2})\,S(h_{:1})
 = S\bigl(\tsum h_{:1}\,S(h_{:2})\bigr) = S(\eps(h)\,1) = \eps(h)\,1$.
A further simplification is $\tsum \eps(h_{:2}) S^2(h_{:3})\,S(h_{:1})
 = \tsum S^2(h_{:2})\,S(h_{:1}) = \eps(h)\,1$, so the terms
$S^2(h_{:n+k}^1)\,S(h_{:n-k-1}^1)$ telescope from left to right,
leaving
$$
\tau^2(h^1 \ox h^2 \oxyox h^n)
 = \tsum S(h_{:n}^2)\,h^3 \ox S(h_{:n-1}^2)\,h^4
         \oxyox S(h_{:2}^2) \ox S(h_{:1}^2)\,S^2(h^1),
$$
where the sum runs over the terms in $\Dl^{n-1} S(h^2)$. After $n-1$
iterations of this process, we obtain
$$
\tau^n(h^1 \ox h^2 \oxyox h^n)
 = \tsum S(h_{:n}^n) \ox S(h_{:n-1}^n)\,S^2(h^1) \oxyox
         S(h_{:2}^n)\,S^2(h^{n-2}) \ox S(h_{:1}^n)\,S^2(h^{n-1}),
$$
and, since $\Dl^{n-1}(S(1)) = 1 \oxyox 1$, the final iteration gives
\eqref{eq:cyclic-Ssq}.
\end{proof}

This shows that the condition $S^2 = \id_H$ is necessary and
sufficient to give $\tau^{n+1} = \id$ on $C^n(H)$. We leave the
remaining relations in \eqref{eq:cycl-relns} to the reader.

\marker
However, it turns out that $S^2$ is \textit{not} the identity in the
Hopf algebra $H_{CM}$. For instance,
$$
S^2(X) = S(-X + \la_1 Y) = (X - \la_1 Y) + S(Y)\,S(\la_1)
 = X + [Y,\la_1] = X + \la_1.
$$
The day is saved by the existence of a character $\dl$ of~$H_{CM}$
such that the ``twisted antipode'' $S_\dl := \eta\dl * S$ is
involutive. Indeed, since $X$ and $\la_1$ are commutators, any
character satisfies $\dl(X) = \dl(\la_1) = 0$, so any character is
determined by its value on the other algebra generator, $Y$. We set
$\dl(Y) := 1$. (Recall that $\eps(Y) = 0$.) Now
$$
S_\dl(h) := (\eta\dl * S)(h) = \tsum \dl(h_{:1})\,S(h_{:2}),
$$
so the twisted antipode does satisfy $S_\dl^2 = \id_H$.

\begin{exer}
\label{xr:twist-antp} 
Show this by verifying $S_\dl^2(X) = X$, $S_\dl^2(Y) = Y$, and
$S_\dl^2(\la_1) = \la_1$ directly.
\end{exer}

The relation with the coproduct is given by
$$
\Dl(S_\dl(h)) = \tsum S(h_{:2}) \ox S_\dl(h_{:1}),  \qquad
\Dl^2(S_\dl(h)) = \tsum S(h_{:3}) \ox S(h_{:2}) \ox S_\dl(h_{:1}),
$$
and more generally, $\Dl^{n-1}(S_\dl(h))
 = \tsum S(h_{:n}) \oxyox S(h_{:2}) \ox S_\dl(h_{:1})$. It is also
worth noting that
$$
\tsum S_\dl(h_{:1})\,h_{:2} = \tsum \dl(h_{:1})\,S(h_{:2})\,h_{:3}
 = \tsum \dl(h_{:1})\,\eps(h_{:2})\,1 = \dl(h)\,1.
$$

The crossed product algebra $\A$ on which $H_{CM}$ acts carries a
distinguished faithful \textit{trace}, given by integration over the
frame bundle~$F$ with the $\Ga$-invariant volume form~$\nu$:
\begin{equation}
\vf(f U_\psi^\7) := 0  \hbox{ if } \psi \neq \id,  \qquad
\vf(f) := \int_F f \,d\nu.
\label{eq:invt-trace} 
\end{equation}
It follows from~\eqref{eq:smash-prod} that, for $a = f U_\psi^\7$
and $b = g U_\psi$, the equality $\vf(ab) = \vf(ba)$ reduces to
$\int_F f(g \circ \Onda\psi) \,d\nu
 = \int_F (f \circ \Onda\psi^{-1})g \,d\nu$, so that the
$\Ga$-invariance of~$\nu$ yields the tracial property of~$\vf$.

If $f \in \Coo_c(F)$, it is easily checked that
$\int_F (Xf)\,d\nu = 0$ and that $\int_F (Yf)\,d\nu = \int_F f\,d\nu$,
using integration by parts. Moreover, since
$\la_1(f) := h_\id f$ from \eqref{eq:la-mult} and $h_\id = 0$, we also
get $\int_F (\la_1 f)\,d\nu = 0$. These identities are enough to
confirm that
$$
\vf(h\.a) = \dl(h)\,\vf(a),  \sepword{for all} h \in H_{CM},\ a \in \A.
$$
It is standard to call a functional $\mu$ on~$\A$ ``invariant'' under
a Hopf action if the relation $\mu(h\.a) = \eps(h)\,\mu(a)$ holds.
Since the character $\dl$ takes the place of the counit here, we may
say that the trace $\vf$ is a \textit{$\dl$-invariant} functional.

This $\dl$-invariance may be reformulated as a rule for
\textit{integration by parts}, as pointed out in~\cite{ConnesMSymm}:
\begin{equation}
\vf((h\.a)\,b) = \vf(a\,(S_\dl(h)\.b)).
\label{eq:int-by-parts} 
\end{equation}
Indeed, one only needs to observe that
\begin{align*}
\vf((h\.a)\,b) &= \tsum \vf((h_{:1}\.a)\,\eps(h_{:2})\,b)
 = \tsum \vf((h_{:1}\.a)\,(h_{:2}\,S(h_{:3})\.b))
\\
&= \tsum \vf(h_{:1}\.(a\,(S(h_{:2})\.b)))
 = \tsum \dl(h_{:1})\, \vf(a\,(S(h_{:2})\.b)) = \vf(a\,(S_\dl(h)\.b)).
\end{align*}

The cyclic permuter $\tau$ must be redefined to take account of the
twisted antipode~$S_\dl$, as follows:
\begin{align*}
\tau(h^1 \oxyox h^n) &:= S_\dl(h^1)\.(h^2 \oxyox h^n \ox 1)
 = \Dl^{n-1}(S_\dl(h^1))\,(h^2 \oxyox h^n \ox 1)
\\
&= \tsum S(h_{:n}^1)h^2 \ox S(h_{:n-1}^1)h^3 \oxyox
         S(h_{:2}^1)h^n \ox S_\dl(h_{:1}^1).
\end{align*}
A straightforward modification of the proof of
Proposition~\ref{pr:cyclic-Ssq} yields the following
identity~\cite[Prop.~4.4]{CrainicWeil}:
$$
\tau^{n+1}(h^1 \ox h^2 \oxyox h^n)
 = S_\dl^2(h^1) \ox S_\dl^2(h^2) \oxyox S_\dl^2(h^n).
$$
Thus, $S_\dl^2 = \id_H$ entails $\tau^{n+1} = \id$ on $C^n(H)$.

\marker
The cyclic cohomology $HC_\dl^\8(H)$ is now easily defined. The maps
$b\: C^{n-1}(H) \to C^n(H)$ and $B\: C^{n+1}(H) \to C^n(H)$ are given
by the very same formulae as before:
$$
b := \sum_{i=0}^n (-1)^i \dl_i, \qquad
B := (-1)^n N(\sg_0\tau^{-1} + \sg_n),
$$
where $N := \sum_{k=0}^n (-1)^{nk} \tau^k$ on~$C^n(H)$.

\begin{exer}
\label{xr:Hopf-cocyc} 
Show that $h \in H$ is a cyclic $1$-cocycle if and only if $h$ is
primitive and $\dl(h) = 0$.
\end{exer}

It remains to show how $HC_\dl^\8(H)$ and $HC^\8(\A)$ are related; the
trace $\vf$ provides the link. For each $n = 0,1,2,\dots$, we define a
linear map $\ga_\vf\: C^n(H) \to C^n(\A,\A^*)$ by setting
$\ga_\vf(1) := \vf$ and
$$
\ga_\vf(h^1 \oxyox h^n)
 : (\row a0n) \mapsto \vf(a_0\,(h^1\.a_1) \dots (h^n\.a_n)).
$$
Following~\cite{ConnesMSymm}, we call $\ga_\vf$ the
\textit{characteristic map} associated to~$\vf$.

It is easy to check that $\ga_\vf$ intertwines the maps $\dl_i$,
$\sg_j$ and $\tau$ defined on the two cochain complexes. For instance,
if $i = 1,2,\dots,n-1$, then
\begin{align*}
\ga_\vf \dl_i(h^1 \oxyox h^n) &: (\row a0{n+1}) \mapsto
  \ga_\vf(h^1 \oxyox \Dl(h^i) \oxyox h^{n-1})\,(\row a0{n+1})
\\
&= \vf(a_0\,(h^1\.a_1) \dots (h_{:1}^i\.a_i)\,((h_{:2}^i\.a_{i+1})
       \dots (h^n\.a_{n+1}))
\\
&= \vf(a_0\,(h^1\.a_1) \dots (h^i\.(a_ia_{i+1})) \dots (h^n\.a_{n+1}))
\\
&= \dl_i\ga_\vf(h^1 \oxyox h^n)\,(\row a0{n+1}).
\end{align*}
To match the cyclic actions, we first recall that
$$
\tau(h^1 \ox h^2 \oxyox h^n) = S_\dl(h^1)\.(h^2 \oxyox h^n \ox 1)
 = \tsum S(h_{:2}^1)\.(h^2 \oxyox h^n) \ox S_\dl(h_{:1}^1).
$$
Write $b := (h^2\.a_1) \dots (h^n\.a_{n-1})$; then, using the
``integration by parts'' formula, we get
\begin{align*}
\ga_\vf \tau(h^1 &\ox h^2 \oxyox h^n)\,(\row a0n)
 = \tsum \vf\bigl(a_0\,(S(h_{:2}^1)\.b)\, S_\dl(h_{:1}^1)\.a_n\bigr)
\\
&= \tsum \vf\bigl( h_{:1}^1\.(a_0\, S(h_{:2}^1)\.b) \,a_n \bigr)
 = \tsum \vf\bigl( a_n\, (h_{:1}^1\.a_0)\,
                   (h_{:2}^1\,S(h_{:3}^1)\.b) \bigr)
\\
&= \tsum \vf\bigl( a_n\, (h_{:1}^1\.a_0)\, \eps(h_{:2}^1) b \bigr)
 = \vf(a_n\, (h^1\.a_0)\, b)
\\
&= \vf\bigl( a_n\, (h^1\.a_0)\, (h^2\.a_1) \dots (h^n\.a_{n-1}) \bigr)
 = \tau \ga_\vf(h^1 \ox h^2 \oxyox h^n)\,(\row a0n).
\end{align*}
In retrospect, we can see what lies behind the definition of~$\tau$ on
$C^n(H)$: on reading the last calculation backwards, we see that the
formula for~$\tau$ is predetermined in order to fulfil
$\ga_\vf \tau = \tau \ga_\vf$ for any $\dl$-invariant trace~$\vf$.

\marker
We conclude with two variations on this algorithm for characteristic
classes. The first concerns algebras which support a Hopf action but
have no natural $\dl$-invariant trace. In the theory of locally
compact quantum groups~\cite{KustermansV}, another possibility arises,
namely that instead of a trace the algebra supports a linear
functional $\vf$ such that $\vf(ab) = \vf(b(\sg\.a))$ where $\sg$ is a
grouplike ``modular element'' of the Hopf algebra. If $\vf$ is also
$\dl$-invariant for a character $\dl$ such that $\dl(\sg) = 1$, only
two further modifications of the elementary maps~\eqref{eq:elem-maps}
and~\eqref{eq:Hopf-tau} are needed:
\begin{align*}
\dl_n(h^1 \oxyox h^{n-1}) &:= h^1 \oxyox h^{n-1} \ox \sg,
\\
\tau(h^1 \oxyox h^n) &:= S_\dl(h^1)\.(h^2 \oxyox h^n \ox \sg).
\end{align*}
This time, the computation in Proposition~\ref{pr:cyclic-Ssq} leads to
$$
\tau^{n+1}(h^1 \oxyox h^n)
 = \sg^{-1} S_\dl^2(h^1)\,\sg \oxyox \sg^{-1} S_\dl^2(h^n)\,\sg.
$$
Thus, the necessary and sufficient condition for $\tau^{n+1} = \id$ is
$S_\dl^2(h) = \sg h \sg^{-1}$ for all~$h$. See~\cite{ConnesMModl} and
\cite[\S 14.7]{Polaris} for the detailed construction of the
characteristic map in this ``modular'' case.

The other variant concerns the application to the original problem of
finding characteristic classes for foliations, in the
higher-dimensional cases, as discussed at the end of
subsection~\ref{sec:Hopf-diff}. What is needed is a cohomology theory
which takes account of the Hopf algebroid structure, when the
coefficient is $\Rr = \Coo(F)$ instead of~$\C$. The
formula~\eqref{eq:invt-trace} continues to define a $\Ga$-invariant
faithful trace on the algebra~$\A$. Now, however, instead of seeking a
special character~$\dl$, the main role is taken by the
integration-by-parts formula~\eqref{eq:int-by-parts}. The twisted
antipode in that formula is replaced by a map $\Onda S\: H \to H$,
subject to four requirements: (a) that it be an algebra
antihomomorphism; (b) which is involutive, that is,
$\Onda S^2 = \id_H$; (c) that it exchange the algebroid actions
of~\eqref{eq:mult-op}, namely, $\Onda S\,\b = \a$; and (d) that
$m(\Onda S \ox_\Rr \id)\Dl = \b\eps\Onda S$. Connes and Moscovici show
in~\cite{ConnesMDiff} that a unique map $\Onda S$ satisfying these
properties exists, and with its help one can again build a cyclic
cohomology theory for the Hopf algebroid of transverse differential
operators, which provides the needed invariants of~$\A$.

\newpage


\section{Noncommutative Homogeneous Spaces}

\subsection{Chern characters and noncommutative spheres}
\label{sec:NC-sphere} 

A fundamental theme of noncommutative geometry is the determination of
geometric quantities from the spectra of certain operators on Hilbert
space. An early precursor is Weyl's theorem on the dimension and
volume of a compact Riemannian manifold: these are determined by the
growth of the eigenvalues of the Laplacian. For spin manifolds, one
can obtain the same data from the asymptotics of the spectra of the
Dirac operator~$\Dslash$. This phenomenon forms the background for the
study of spectral triples. We know, for instance, that a spectral
triple $(\A,\H,D)$ over the algebra $\A = \Coo(M)$, complying with the
seven requirements listed in subsection~\ref{sec:brief-NCG}, provides a
spin structure and a Riemannian metric on~$M$ for which $D$ equals
$\Dslash$ plus a torsion term.

A question raised in the paper which introduced these seven
conditions~\cite{ConnesGrav} is whether the manifold itself ---or its
algebra of smooth coordinates--- may be extracted from spectral data.
The key property here is the orientation or volume-form condition:
\begin{equation}
\pi_D(\cc) = \chi,  \sepword{with}  \cc \in C_n(\A)
  \sepword{such that} b\,\cc = 0,
\label{eq:orient-cond} 
\end{equation}
where $n$ is the classical dimension of the spin geometry. In view of
the isomorphism between $HH_n(\Coo(M)) \simeq \A^n(M)$, there is a
unique $n$-form $\nu$ matched to the class $[\cc]$ of the Hochschild
$n$-cycle. It turns out that \eqref{eq:orient-cond} entails that
$\nu$ is nonvanishing on~$M$, so that, suitably normalized, it is a
volume form; in fact, it is the Riemannian volume for the metric
associated to the Dirac-type operator~$D$.

To see how this works, recall that the standard volume form on the
$2$-sphere $\Sf^2$ is
\begin{equation}
\nu = x\,dy \w dz + y\,dz \w dx + z\,dx \w dy  \in \A^2(\Sf^2).
\label{eq:vol-sphere} 
\end{equation}
The corresponding Hochschild $2$-cycle is
\begin{equation}
\cc := \tihalf\bigl( x \,(dy\,dz - dz\,dy) + y \,(dz\,dx - dx\,dz)
         + z \,(dx\,dy - dy\,dx) \bigr) \in \Om^2(\Coo(\Sf^2)),
\label{eq:Hoch-cycle-sphere} 
\end{equation}
and \eqref{eq:orient-cond} becomes
\begin{equation}
\tihalf\bigl( x\,[[D,y], [D,z]] + y\,[[D,z], [D,x]]
 + z\,[[D,x], [D,y]] \bigr) = \chi.
\label{eq:orient-sphere} 
\end{equation}
The algebra $\A = \Coo(\Sf^2)$ is generated by the three commuting
coordinates $x,y,z$, subject to the constraint $x^2 + y^2 + z^2 = 1$.
It is important to note that one can vary the metric on~$\Sf^2$ while
keeping the volume form~$\nu$ fixed; one usually thinks of the round
metric $g = dx^2 + dy^2 + dz^2$ which is $SO(3)$-invariant, but one
can compose $g$ with any volume-preserving diffeomorphism of~$\Sf^2$
to get many another metric $g'$ whose volume form is also~$\nu$.
Therefore, the $D$ in the equation~\eqref{eq:orient-cond} is not
uniquely determined; it may be a Dirac operator $D = \Dslash_{g'}$
obtained from any such metric $g'$ (the Hilbert space $\H$ is the
vector space of square-integrable spinors on~$\Sf^2$).

On the other hand, one may think of~\eqref{eq:orient-sphere} as a
(highly nonlinear) equation for the coordinates $x,y,z$. To see how
this comes about, we collect the three coordinates for the $2$-sphere
into a single orthogonal projector (selfadjoint idempotent),
\begin{equation}
e := \frac{1}{2} \twobytwo{1 + z}{x - iy}{x + iy}{1 - z},
\label{eq:Bott-proj} 
\end{equation}
in the algebra of $2 \x 2$ matrices, $M_2(\Coo(\Sf^2))$. This is
actually the celebrated \textit{Bott projector}, whose class
$[e] \in K_0(\Coo(\Sf^2)) = K^0(\Sf^2)$ is nontrivial. It is easy to
check the following identity in exterior algebra:
$$
\tr((e - \thalf) \,de \w de) = \tihalf \nu \in \A^2(\Sf^2).
$$
Now, up to normalization and replacement of the exterior derivative by
the differential of the universal graded differential algebra
$\Om^\8(\Coo(\Sf^2))$, the left hand side is just the term $\ch_1 e$
of the cyclic-homology Chern character of~$[e]$. Notice that
$\ch_0 e = \tr(e - \thalf)$ vanishes also. The cyclic homology
computations preceding \eqref{eq:Chern-chain} show that, in full
generality,
$$
b(\ch_1 e) = - B(\ch_0 e),
$$
so that the vanishing $\ch_0 e = 0$ is enough to guarantee that
$\ch_1 e$ is a \textit{Hochschild} cycle: $b(\ch_1 e) = 0$.

\marker
We now switch to a different point of view. Suppose we wish to produce
examples of spectral triples $(\A,\H,D,C,\chi)$ satisfying the seven
conditions for a noncommutative spin geometry. We first fix the
classical dimension, which for convenience we shall suppose to be
even: $n = 2m$. Then we start from the orientation condition:
\begin{subequations}
\label{eq:orient-Chern} 
\begin{equation}
\pi_D(\ch_m e) = \chi,
\label{eq:orient-Chern-top} 
\end{equation}
subject to the constraints
\begin{equation}
\ch_0 e = 0, \quad \ch_1 e = 0, \quad\dots, \quad \ch_{m-1} e = 0,
\label{eq:orient-Chern-lower} 
\end{equation}
\end{subequations}
which guarantee that $\ch_m e$ will be a Hochschild $2m$-cycle.

Consider \eqref{eq:orient-Chern} as a system of equations for an
``unknown'' projector $e \in M_r(\A)$, $r$ being a suitable matrix
size. What does this system tell us about the coordinate algebra~$\A$?

In Connes' survey paper~\cite{ConnesSurvey}, the answer is given in
detail for the case $n = 2$, $r = 2$: it turns out that
\eqref{eq:orient-Chern-lower} forces $\A$ to be commutative, and
\eqref{eq:orient-Chern-top} ensures that its character space is the
$2$-sphere. We summarize the argument, following our
\cite[\S 11.A]{Polaris}. First of all, the selfadjointness $e^* = e$
and the equation $\ch_0 e = \tr(e - \thalf) = 0$ allow us to write
$e$ in the form~\eqref{eq:Bott-proj}, where $x,y,z$ are selfadjoint
elements of~$\A$. The positivity of the projector~$e$ implies
$-1 \leq z \leq 1$ (here we are implicitly assuming that $\A$ is a
dense subalgebra of a $C^*$-algebra). The idempotence $e^2 = e$
boils down to a pair of equations
\begin{align*}
(1 \pm z)^2 + x^2 + y^2 \pm i[x,y] &= 2(1 \pm z),
\\
(1 \mp z)(x \pm iy) + (x \pm iy)(1 \pm z) &= 2(x \pm iy),
\end{align*}
which simplify to $[x,y] = [y,z] = [z,x] = 0$ and
$x^2 + y^2 + z^2 = 1$. Thus, $x,y,z$ generate a \textit{commutative}
algebra $\A$. Moreover, by regarding them as commuting selfadjoint
operators in a faithful representation of~$\A$, the equation
$x^2 + y^2 + z^2 = 1$ tells us that their joint spectrum in~$\R^3$ is
a closed subset~$V$ of the sphere~$\Sf^2$: the $C^*$-completion of
$\A$ is~$C(V)$.

This partial description of~$\A$ has not yet used the main equation
\eqref{eq:orient-Chern-top}, whose role is to confirm that $V$ is all
of~$\Sf^2$. For convenience, we abbreviate $\d a := [D,a]$ (at this
stage, $\d$ is just an unspecified derivation on~$\A$). Since
$$
\d e = \frac{1}{2} \twobytwo{\d z}{\d x-i\,\d y}{\d x+i\,\d y}{-\d z},
$$
a short calculation gives
$$
\chi = \tr((e - \thalf)\,\d e\,\d e) = \tihalf\bigl(
  x\,[\d y, \d z] + y\,[\d z, \d x] + z\,[\d x, \d y] \bigr).
$$
This is of the form $\pi_D(\cc) = \chi$, where $\cc$ is just the
Hochschild $2$-cycle of the formula~\eqref{eq:Hoch-cycle-sphere}. The
corresponding volume form on~$V$ is precisely \eqref{eq:vol-sphere}:
but this volume is nonvanishing on all of~$\Sf^2$, so we conclude that
$V = \Sf^2$. The pre-$C^*$-algebra $\A$, generated by $x,y,z$, is none
other than $\Coo(\Sf^2)$!

\marker
The odd-dimensional case $n = 2m+1$ uses the odd Chern character
\eqref{eq:Chern-chain-odd}, and its orientation condition is
$\pi_D(\ch_{m+\shalf} u) = 1$, with constraints $\ch_{k+\shalf} u = 0$
for $k = 0,1,\dots,m-1$. The unitarity condition $u^*u = uu^* = 1$ may
be assumed. For instance, in dimension three, Connes and
Dubois-Violette~\cite{ConnesDV} have shown that, under the sole
constraint $\ch_{1/2} u = \tr(u^{-1}\,du) = 0$, all solutions of the
equation $\pi_D(\ch_{3/2} u) = 1$ form a 3-parameter family of
algebras; one of these is the commutative algebra $\Coo(\Sf^3)$, but
the others are noncommutative.

\marker
Moving on now to dimension~$4$, we take $e = e^* = e^2$ in $M_4(\A)$,
and look for solutions of \eqref{eq:orient-Chern} with $2m = 4$.
In~\cite{ConnesSurvey}, a commutative solution is again found, by
using a ``quaternionic'' prescription reminiscent of the Connes--Lott
approach to the Standard Model (see \cite[VI.5]{ConnesBook} or
\cite{Cordelia} for the story of how quaternions enter in that
approach). One writes $e$ in $2 \x 2$ blocks:
\begin{equation}
e := \frac{1}{2} \twobytwo{(1 + z)1_2}{q}{q^*}{(1 - z)1_2},
 \sepword{where}  1_2 = \twobytwoeven{1}{1},  \quad
  q = \twobytwo{\a}{\b}{-\b^*}{\a^*}.
\label{eq:quat-proj} 
\end{equation}
Here again, $z$ is a selfadjoint element of~$\A$ such that
$-1 \leq z \leq 1$, and $e^2 = e$ yields the equalities
$qq^* = (1 - z^2) = q^*q$ and $[z\,1_2, q] = 0$. Since
$qq^* = q^*q$ is diagonal, we find that $z,\a,\a^*,\b,\b^*$ are
commuting elements of~$\A$, subject to the constraint
$\a\a^* + \b\b^* = 1 - z^2$: these are coordinate relations for a
closed subset of~$\Sf^4$. Once more, the equation
\eqref{eq:orient-Chern-top} produces the standard volume form
supported on the full sphere, and the conclusion is that
$\A = \Coo(\Sf^4)$: the ordinary $4$-sphere emerges as a solution to
the cohomological equation \eqref{eq:orient-Chern} in dimension four.

Now, the particular quaternionic form of~$q$ in\eqref{eq:quat-proj} is
merely an Ansatz, and Landi soon pointed out that one could equally
well try
$$
q = \twobytwo{\a}{\b}{-\la\b^*}{\a^*},  \sepword{with}  \la \in \C.
$$
The consequences are worked out in a recent paper by Connes and
Landi~\cite{ConnesLa} ---see also \cite{ConnesY2K}. One finds that
$qq^* = (1 - z^2) = q^*q$ and $[z\,1_2, q] = 0$ still hold, but these
relations now lead to
\begin{subequations}
\label{eq:sfour-relns} 
\begin{gather}
\a\b = \bar\la^{-1}\,\b\a,  \qquad  \a^*\b = \bar\la\,\b\a^*,
\notag \\
\b\b^* = \b^*\b,  \qquad
 \a\a^* + \b\b^* = 1 - z^2 = \a^*\a + \la\bar\la\,\b^*\b. 
\label{eq:sfour-relns-DLM} 
\end{gather}
The computation of $\ch_1(e)$, carried out in~\cite{DabrowskiLM},
yields
$$
\ch_1(e) = \teighth (1 - \la\bar\la) \bigl(
 z\,[d\b, d\b^*] + \b^*\,[dz, d\b] + \b\,[d\b^*, dz] \bigr),
$$
which vanishes if and only if $\la$ is a complex number of modulus
one.

In particular, this scheme parts company with the ever-popular
deformations where $\la = q$ would be a real number other than~$\pm
1$. Foremost among these are the well-known Podle\'s spheres
$\Sf_{qc}^2$, which were originally constructed~\cite{Podles} as
homogeneous spaces of the quantum group $SU_q(2)$. Other
higher-dimensional $q$-spheres currently on the market are described
in \cite{BonechiCT,BrzezinskiG,DabrowskiLM,HongS,SitarzSphere}; the
$C^*$-algebra construction of $\Sf_q^n$ by Hong and
Szyma\'nski~\cite{HongS}, in particular, is quite far-reaching.
However, none of these arises from a Hochschild cycle in the manner
described above. On the other hand, Aschieri and
Bonechi~\cite{AschieriB} have constructed, with $R$-matrix techniques,
a multiparameter family of quantum spaces which yields the spheres
described here as limiting cases; see also~\cite{AschieriC}.

By assuming $|\la| = 1$, $\la = e^{2\pi i\th}$ from now on, the
relations \eqref{eq:sfour-relns-DLM} simplify to
\begin{gather}
\a\b = \la\,\b\a,  \qquad  \a^*\b = \bar\la\,\b\a^*,
\notag \\
\a\a^* = \a^*\a,  \qquad  \b\b^* = \b^*\b,  \qquad
\a\a^* + \b\b^* = 1 - z^2,
\label{eq:sfour-relns-CL} 
\end{gather}
\end{subequations}
which determines a noncommutative algebra $\A$, baptized
$\Coo(\Sf_\th^4)$ by Connes and Landi.

\subsection{How Moyal products yield compact quantum groups}
\label{sec:Moyal-cpt-qgroup} 

To construct a spin geometry over $\A = \Coo(\Sf_\th^4)$, we need a
representation of this algebra on a suitable Hilbert space. The key is
to notice that the relation $\a\b = e^{2\pi i\th}\,\b\a$
of~\eqref{eq:sfour-relns-CL}, for normal operators $\a$ and~$\b$
(that is, $\a\a^* = \a^*\a$ and $\b\b^* = \b^*\b$), is closely related
to the definition of the noncommutative
torus~\cite{ConnesTorus,RieffelRot}. This is a pre-$C^*$-algebra
$\Coo(\T_\th^2)$ with two generators $u$ and $v$ which are unitary:
$uu^* = u^*u = 1$, $vv^* = v^*v = 1$, subject only to the commutation
relation
\begin{equation}
uv = e^{2\pi i\th}\,vu.
\label{eq:NCtorus-reln} 
\end{equation}
One can then define ``spherical coordinates'' $(u,v,\phi,\psi)$ for
the noncommutative space $\Sf_\th^4$ by setting
\begin{equation}
\a =: u \sin\psi \cos\phi, \qquad  \b =: v \sin\psi \sin\phi,  \qquad
 z =: \cos\psi,
\label{eq:sfour-coords} 
\end{equation}
where $\phi,\psi$ are ordinary angular coordinates. It is clear that
this is equivalent to~\eqref{eq:sfour-relns}, for
$\la = e^{2\pi i\th}$.

There is a canonical action of the ordinary $2$-torus $\T^2$ on the
algebra $\Coo(\T_\th^2)$, obtained from the independent rotations
$u \mapsto e^{2\pi i\phi_1}\,u$, $v \mapsto e^{2\pi i\phi_2}\,v$ which
respect \eqref{eq:NCtorus-reln}. By substituting these rotations in
\eqref{eq:sfour-coords}, we also obtain an action of~$\T^2$ on
$\Coo(\Sf_\th^4)$.

In the commutative case $\th = 0$, this becomes an action of the
abelian Lie group $\T^2$ by rotations on the compact manifold $\Sf^4$,
and these rotations are isometries for the round metric on~$\Sf^4$.
Any smooth function on $\Sf^4$ can be decomposed as a generalized
Fourier series $f = \sum_r f_r$, indexed by $r = (r_1,r_2) \in \Z^2$,
where $f_r$ satisfies
$$
(e^{2\pi i\phi_1}, e^{2\pi i\phi_2}) \. f_r
 = e^{2\pi i(r_1\phi_1 + r_2\phi_2)} \, f_r.
$$
Indeed, each $f_r$ is of the form $u^{r_1} v^{r_2} h(\phi,\psi)$, in
terms of the coordinates \eqref{eq:sfour-coords}; all such functions
form the spectral subspace $E_r$ of~$\Coo(\Sf^4)$. The same is true
of $\Coo(\Sf_\th^4)$ when $\th \neq 0$.

If $g_s = u^{s_1} v^{s_2} k(\phi,\psi)$, then $g_s \in E_s$ and
$e^{2\pi i\th r_2s_1}\,f_r g_s = u^{r_1+s_1} v^{r_2+s_2} \,hk$ lies in
$E_{r+s}$, so we may identify the algebra $\Coo(\Sf_\th^4)$ with the
vector space $\Coo(\Sf^4)$ of smooth functions on the ordinary
$4$-sphere, gifted with the new product:
\begin{subequations}
\label{eq:sfour-Moyal} 
\begin{equation}
f_r * g_s := e^{2\pi i\th r_2s_1}\, f_r g_s,
\label{eq:sfour-Moyal-asymm} 
\end{equation}
defined on homogeneous elements $f_r \in E_r$, $g_s \in E_s$. Since
the Fourier series $f = \sum_r f_r$ converges rapidly in the Fr\'echet
topology of $\Coo(\Sf^4)$, one can show that this recipe defines a
continuous bilinear operation on that space. A more symmetric-looking
operation, which yields an isomorphic algebra, is given by
\begin{equation}
f_r \x g_s := e^{\pi i\th(r_2s_1 - r_1s_2)}\, f_r g_s.
\label{eq:sfour-Moyal-symm} 
\end{equation}
\end{subequations}
This deserves to be called a \textit{Moyal product} of functions
on~$\Sf^4$. Indeed, suppressing the coordinates $\phi,\psi$ yields
exactly the Moyal product on $\Coo(\T^2)$, which has long been
recognized to give the smooth algebras $\Coo(\T_\th^2)$ of the
noncommutative $2$-tori \cite{WeinsteinTorus}.

The only nonobvious feature of the products \eqref{eq:sfour-Moyal} is
their associativity. To check it, we generalize a little. Suppose that
$M$ is a compact Riemannian manifold on which an $l$-dimensional torus
acts by isometries (there is no shortage of examples of that). Then
one can decompose $\Coo(M)$ into spectral subspaces indexed by $\Z^l$.
A ``twisted'' product of two homogeneous functions $f_r$ and $g_s$ may
be defined by
\begin{equation}
f_r * g_s := \rho(r,s)\, f_r g_s,
\label{eq:Moyal-rho} 
\end{equation}
where the phase factors $\set{\rho(r,s) \in U(1) : r,s \in \Z^l}$ make
up a $2$-cocycle on the additive group $\Z^l$. The cocycle relation
\begin{equation}
\rho(r, s + t) \rho(s,t) = \rho(r,s) \rho(r + s, t)
\label{eq:group-cocyc} 
\end{equation}
ensures that the new product is associative. To define such a cocycle,
one could take~\cite{Atlas}:
$$
\rho(r,s) := \exp\bigl\{-2\pi i \tsum_{j<k} r_j \th_{jk} s_k \bigr\},
$$
where $\th = [\th_{jk}]$ is a real $l \x l$ matrix. Complex
conjugation of functions remains an involution for the new product
provided that the matrix $\th$ is \textit{skewsymmetric}. (When
$l = 2$, it is customary to replace the matrix $\th$ by the real
number $\th_{12} = - \th_{21}$and, rather sloppily, call this number
$\th$, too; but in higher dimensions one is forced to deal with a
matrix of parameters.) The product~\eqref{eq:Moyal-rho} defines a
$C^*$-algebra which, when $M = \T^l$, is isomorphic to that of the
noncommutative torus $C(\T_\th^l)$ with parameter matrix~$\th$, as
we shall soon see.

Moreover, we may define a ``Moyal product'':
\begin{equation}
f_r \x g_s := \sg(r,s)\, f_r g_s,
\label{eq:Moyal-sigma} 
\end{equation}
by replacing $\rho$ by its skewsymmetrized version,
\begin{equation}
\sg(r,s) := \exp\bigl\{-\pi i \tsum_{j,k=1}^l r_j \th_{jk} s_k\bigr\},
\label{eq:sigma-cocyc} 
\end{equation}
which is again a group $2$-cocycle; in fact, $\rho$ and~$\sg$ are
cohomologous as group cocycles~\cite{RieffelPMod}, therefore they
define isomorphic $C^*$-algebras.

\marker
To see why \eqref{eq:Moyal-sigma} should be called a Moyal product,
let us briefly recall the real thing. The quantum product of two
functions on the phase space $\R^{2m}$ was introduced by
Moyal~\cite{Moyal} using a series development in powers of~$\hbar$
whose first nontrivial term gives the Poisson bracket; later, it was
noticed~\cite{Pool} that it could be rewritten in an integral
form~\cite{Phobos}:
$$
(f \x_J g)(x) := (\pi\hbar)^{-2m}
 \iint f(x + s) g(x + t) \,e^{2is\.Jt/\hbar} \,ds\,dt,
$$
where $J = \twobytwoodd{1}{-1}$ is the skewsymmetric matrix giving the
standard symplectic structure on $\R^{2m}$ (and the dot is the usual
scalar product). This is in fact the Fourier transform of the
``twisted convolution'' of phase-space functions which goes back to
von~Neumann's work on the Schr\"odinger
representation~\cite{NeumannSchrod}. For suitable classes of functions
and distributions on $\R^{2m}$, it is an oscillatory integral, which
yields Moyal's series development as an \textit{asymptotic} expansion
in powers of~$\hbar$ \cite{Nereid,Voros}.

This integral form of the Moyal product is the starting point for a
general deformation theory of $C^*$-algebras, which was undertaken by
Rieffel~\cite{RieffelDefQ}. He gave it a mildly improved presentation
by rewriting it as
$$
(f \x_J g)(x) := \iint f(x + Js) g(x + t) \,e^{2\pi is\.t} \,ds\,dt,
$$
taking $\hbar = 2$ and rescaling the measure on $\R^{2m}$. He then
replaced the functions $f,g$ by elements $a,b$ of any $C^*$-algebra
$A$, and the translations $f(x) \mapsto f(x + t)$ by a strongly
continuous action $\a$ of $\R^l$ on~$A$ by automorphisms; and he
replaced the original matrix $J$ by any skewsymmetric real $l \x l$
matrix, still called~$J$, ending up with
\begin{equation}
a \x_J b
 := \iint_{\R^l\x\R^l} \a_{Js}(a) \a_t(b) \,e^{2\pi is\.t} \,ds\,dt.
\label{eq:Marc-prod} 
\end{equation}
This formula makes sense, as an oscillatory integral, for elements
$a,b$ in the subalgebra
$A^\infty := \set{a \in A : t \mapsto \a_t(a) \text{ is smooth}}$,
which is a Fr\'echet pre-$C^*$-algebra (as a subalgebra of the
original $C^*$-algebra~$A$).

We wish to complete the algebra $(A^\infty, \x_J)$ to a $C^*$-algebra
$A_J$, which in general is not isomorphic to~$A$ (for instance, $A$
may be commutative while the new product is not). The task is to find
a new norm $\|\.\|_J$ on $A^\infty$ with the $C^*$-property
$\|a^* \x_J a\|_J = \|a\|_J^2$; then $A_J$ is just the completion of
$A^\infty$ in this norm. Rieffel achieved this by considering the left
multiplication operators $L_a^J = L^J(a)$ given by
$$
L_a^J f(x)
 := \iint \a_{x+Js}(a) f(x + t) \,e^{2\pi is\.t} \,ds\,dt,
$$
where $f$ is a smooth $A$-valued function which is rapidly decreasing
at infinity. A particular ``Schwartz space'' of such functions $f$ is
identified in~\cite{RieffelDefQ}, on which the obvious $A$-valued
pairing $\(f,g) := \int_{\R^l} f(x)^* g(x) \,dx$ yields a
Hilbert-space norm by setting $\snorm{f}^2 := \|\(f,f)\|_A$. It can
then be shown that if $a \in A^\infty$, $L_a^J$ is a bounded operator
on this Hilbert space; $\|a\|_J$ is defined to be the operator norm
of~$L_a^J$. Importantly, $L^J$ is a homomorphism:
\begin{align}
L^J(a \x_J b)f(x)
&= \iint \a_{x+Js}(a \x_J b) f(x + t) \,e^{2\pi is\.t} \,ds\,dt
\notag \\
&= \iiiint \a_{x+Js+Ju}(a) \a_{x+Js+v}(b) f(x + t)
     \,e^{2\pi i(s\.t + u\.v)} \,du\,dv\,ds\,dt
\notag \\
&= \iiiint \a_{x+Ju'}(a) \a_{x+v+Js}(b) f(x + v + t')
     \,e^{2\pi i(s\.t' + u'\.v)} \,ds\,dt'\,du'\,dv
\notag \\
&= \iint \a_{x+Ju'}(a) L_b^J f(x + v) \,e^{2\pi iu'\.v} \,du'\,dv
\notag \\
&= L_a^J L_b^J f(x),
\label{eq:Moyal-lmult} 
\end{align}
so that $L^J(a \x_J b) = L^J(a) L^J(b)$. The calculation uses only the
change of variable $t' := t - v$, $u' := s + u$, for which
$s\.t + u\.v = s\.t' + u'\.v$.

Rieffel's construction provides a \textit{deformation} $A \mapsto A_J$
of $C^*$-algebras which is explicit only on the smooth subalgebra
$A^\infty$. This construction has several useful functorial
properties which we now list, referring to the
monograph~\cite{RieffelDefQ} for the proofs.
\begin{itemize}
\item
If $A$ and $B$ are two $C^*$-algebras carrying the respective actions
$\a$ and $\b$ of~$\R^l$, and if $\phi\: A \to B$ is a $*$-homomorphism
intertwining them: $\phi\,\a_t = \b_t\,\phi$ for all~$t$, then
$\phi(A^\infty) \subseteq B^\infty$ and the restriction of $\phi$ to
$A^\infty$ extends uniquely to a $*$-homomorphism
$\phi_J \: A_J \to B_J$.
\item
The map $\phi_J$ is injective if and only if $\phi$ is injective, and
$\phi_J$ is surjective if and only if $\phi$ is surjective.
\item
When $A = B$ and $\a = \b$, we may take $\phi = \a_s$ for any~$s$,
because $\a_s\a_t = \a_{s+t} = \a_t\a_s$ for all~$t$; thus
$\a_J\: s \mapsto (\a_s)_J$ is an action of $\R^l$ on~$A_J$ by
automorphisms, whose restriction to $A^\infty$ coincides with the
original action~$\a$.
\item
Deforming $(A_J,\a_J)$ with another skewsymmetric matrix $K$ gives
a $C^*$-algebra isomorphic to $A_{J+K}$. In particular, if $K = -J$,
the second deformation recovers the original algebra~$A$.
\item
The smooth subalgebra $(A_J)^\infty$ of $A_J$ under the action~$\a_J$
coincides exactly with the original smooth subalgebra $A^\infty$
(although their products are different).
\end{itemize}

When the action $\a$ of $\R^l$ is periodic, so that $\a_t = \id_A$ for
each $t$ in a subgroup~$L$, then $\a$ is effectively an action of the
abelian group $H = \R^l/L$, and $H \simeq \T^k \x \R^{l-k}$ for
some~$k$. Suppose that $H$ is compact, i.e., $k = l$ and
$H \simeq \T^l$. Then $A^\infty$ decomposes into
spectral subspaces $\set{E_p : p \in L}$ where
$\a_s(a_p) = e^{2\pi ip\.s} a_p$ for $a_p \in E_p$. If $b_q \in E_q$
also, one can check \cite[Prop.~2.21]{RieffelDefQ} that
$$
a_p \x_J b_q = e^{-2\pi ip\.Jq} a_p b_q.
$$
On comparing this with \eqref{eq:Moyal-sigma}, we see that if
$A = C(\T^l)$ and $J := \thalf\th$, then $A_J$ is none other than the
noncommutative $l$-torus $C(\T_\th^l)$. Moreover, if $A = C(\Sf^4)$
and $\th$ is a real number, then the rotation action of $\T^2$ on
$\Sf^4$ and the parameter matrix
$$
Q := \frac{1}{2} \twobytwoodd{\th}{-\th}
$$
define a deformation such that $C(\Sf^4)_Q \simeq C(\Sf_\th^4)$.

\marker
We now apply this machinery to the case of the $C^*$-algebra $C(G)$,
where $G$ is a compact connected Lie group. The dense subalgebra
$\Rr(G)$ is a Hopf algebra: we may ask how its coalgebra structure is
modified by this kind of deformation. The answer is: not at all! It
turns out that, for suitable parameter matrices $J$, the coproduct
remains an algebra homomorphism for the new product~$\x_J$. This was
seen early on by Dubois-Violette \cite{DuboisVQGp} in the context of
Woronowicz' compact quantum groups: he noticed that the matrix
corepresentations of $C(SU_q(N))$ and similar bialgebras could be seen
as different products on the same coalgebra.

There are many ways in which a torus can act on~$G$. Indeed, any
connected abelian closed subgroup $H$ of~$G$ is a torus; by the
standard theory of compact Lie groups \cite{BroeckerD,SimonRepns}, any
such $H$ is included in a maximal torus, and all maximal tori are
conjugate. Thus $H$ can act on $G$ by left translation, right
translation, or conjugation. In what follows, we shall focus on the
action of the doubled torus $H \x H$ on~$G$, given by
\begin{equation}
(h,k) \. x := h x k^{-1}.
\label{eq:dbl-mult} 
\end{equation}
The corresponding action on $C(G)$ is
$[(h,k)\.f](x) := f(h^{-1} x k)$. If $\hl$ is the Lie algebra of~$H$,
we may pull this back to a periodic action of the $\hl \op \hl$
on~$C(G)$. For notational convenience, we choose and fix a basis for
the vector space $\hl \simeq \R^l$, which allows to write the
exponential mapping as a homomorphism $e\: \R^l \to H$ whose kernel is
the integer lattice $\Z^l$. If $\la := e(1,1,\dots,1)$, we may write
$\la^s := e(s)$ for $s \in \R^l$; and the action of $\hl \op \hl$ on
$C(G)$ becomes
\begin{equation}
[\a(s,t) f](x) := f(\la^{-s} x \la^t).
\label{eq:hCGh-action} 
\end{equation}
The coefficient matrix $J$ for the Moyal product \eqref{eq:Marc-prod}
is now a skewsymmetric matrix in $M_{2l}(\R)$. It is argued in
\cite{RieffelNonComp} ---see also \cite[\S 4]{Larissa}--- that
compatibility with the coalgebra structure is to be expected only if
$J$ splits as the direct sum of two opposing $l \x l$ matrices:
\begin{equation}
J := \twobytwoeven{Q}{-Q}
\label{eq:double-Q} 
\end{equation}
where $Q \in M_l(\R)$ is evidently skewsymmetric. Here, we accept this
as an Ansatz and explore where it leads.

The Moyal product on the group manifold $G$ can now be written as
\begin{equation}
(f \x_J g)(x) := \int_{\hl^4} f(\la^{-Qs} x \la^{-Qt})
 g(\la^{-u} x \la^v) \,e^{2\pi i(s\.u + t\.v)} \,ds \,dt \,du \,dv.
\label{eq:Moyal-gpprod} 
\end{equation}
We remind ourselves that this makes sense as an oscillatory integral
provided $f,g \in \Coo(G)$, since the smooth subalgebra of $C(G)$ for
the action \eqref{eq:hCGh-action} certainly includes $\Coo(G)$; it
could, however, be larger, for instance if the torus $H$ is not
maximal.

In subsection~\ref{sec:Hopf-basics}, the coproduct, counit and
antipode for the Hopf algebra $\Rr(G)$ are defined by
\begin{equation}
\Dl f(x,y) := f(xy),  \qquad  \eps(f) := f(1),  \qquad
Sf(x) := f(x^{-1}).
\label{eq:coalg-relns}
\end{equation}
These formulas make sense in $\Coo(G)$, which includes $\Rr(G)$ since
representative functions are real-analytic, or even in $C(G)$. In
accordance with the remarks at the end of
subsection~\ref{sec:Hopf-basics}, we shall now discard the algebraic
tensor product and work in the smooth category. The coproduct may now
be regarded as a homomorphism
$$
\Dl : \Coo(G) \to \Coo(G \x G),
$$
the counit is a homomorphism $\eps\: \Coo(G) \to \C$, and the
coalgebra relations $(\Dl \ox \id) \Dl = (\id \ox \Dl) \Dl$ and
$(\eps\ox\id) \Dl = (\id\ox\eps)\Dl = \id$ continue to hold. Moreover,
the antipode $S$ is an algebra antiautomorphism of~$\Coo(G)$.

Let us check that all of those statements continue to hold when the
pointwise product of functions in $\Coo(G)$ is replaced by a Moyal
product. The following calculations are taken
from~\cite{RieffelComQGp}; they all make use of changes of variable
similar to that of~\eqref{eq:Moyal-lmult}. First of all,
\begin{align*}
(\Dl f \x_J &\Dl g)(x,y)
\\
&= \int_{\hl^8} f(\la^{-Qs} x \la^{-Qt-Qs'} y \la^{-Qt'})
     g(\la^{-u} x \la^{v-u'} y \la^{v'})
     \,e^{2\pi i(s\.u + t\.v + s'\.u' + t'\.v')} \,ds \dots dv'
\\
&= \int_{\hl^8} f(\la^{-Qs} x \la^{-Qt''} y \la^{-Qt'})
     g(\la^{-u} x \la^{-u''} y \la^{v'})
     \,e^{2\pi i(s\.u + t''\.v + s'\.u'' + t'\.v')} \,ds \dots dv'
\\
&= \int_{\hl^6} f(\la^{-Qs} x \la^{-Qt''} y \la^{-Qt'})
     g(\la^{-u} x \la^{-u''} y \la^{v'})
     \,e^{2\pi i(s\.u + t'\.v')} \,\dl(t'') \,\dl(u'') \,ds \dots dv'
\\
&= \int_{\hl^4} f(\la^{-Qs} xy \la^{-Qt'}) g(\la^{-u} xy \la^{v'})
     \,e^{2\pi i(s\.u + t'\.v')} \,ds\,dt'\,du\,dv'
\\
&= (f \x_J g)(xy) = \Dl(f \x_J g)(x,y).
\end{align*}
Integrations like $\int_\hl e^{2\pi it''\.v} \,dv = \dl(t'')$ are a
convenient shorthand for the Fourier inversion theorem. Next,
$$
(f \x_J g)(1)
 = \int_{\hl^4} f(\la^{-Q(s+t)}) g(\la^{v-u})
     \,e^{2\pi i(s\.u + t\.v)} \,ds \,dt \,du \,dv,
$$
which simplifies to
\begin{align*}
\int_{\hl^4} & f(\la^{-Qs'}) g(\la^{v'})
     \,e^{2\pi i(s'\.u + t\.v')} \,ds' \,dt \,du \,dv'
\\
&= \int_{\hl^2} f(\la^{-Qs'}) g(\la^{v'}) \,\dl(s') \,\dl(v')
     \,ds' \,dv' = f(1) \, g(1),
\end{align*}
so $\eps(f \x_J g) = \eps(f) \eps(g)$. Finally, if $Q$ is invertible,
then
\begin{align*}
(Sf \x_J Sg)(x)
&= \int_{\hl^4} f(\la^{Qt} x^{-1} \la^{Qs}) g(\la^{-v} x^{-1} \la^u)
     \,e^{2\pi i(s\.u + t\.v)} \,ds \,dt \,du \,dv
\\
&= (\det Q)^{-2} \int_{\hl^4} f(\la^{-t'} x^{-1} \la^{s'})
     g(\la^{-v} x^{-1} \la^{-u})
     \,e^{-2\pi i(Q^{-1}t'\.v + Q^{-1}s'\.u)} \,ds' \,dt' \,du \,dv
\\
&= \int_{\hl^4} f(\la^{-t'} x^{-1} \la^{s'})
     g(\la^{-Qv'} x^{-1} \la^{-Qu'})
     \,e^{2\pi i(t'\.v' + s'\.u')} \,ds' \,dt' \,du' \,dv'
\\
&= (g \x_J f)(x^{-1}) = S(g \x_J f)(x),
\end{align*}
where the skewsymmetry of $Q$ has been used. On the other hand, if
$Q = 0$, then $f \x_J g = fg$ and the calculation reduces to
$(Sf \x_J Sg)(x) = f(x^{-1}) g(x^{-1}) = S(g \x_J f)(x)$; since we may
integrate separately over the nullspace of~$Q$ and its orthogonal
complement, the relation $Sf \x_J Sg = S(g \x_J f)$ holds in general.

\begin{exer}
\label{xr:Moyal-antp}
Show, by similar calculations, that
$$
m(\id \ox S)(\Dl f) = m(S \ox \id)(\Dl f) = \eps(f)\,1
$$
whenever $f \in \Coo(G)$.
\end{exer}

The functoriality of Rieffel's construction then lifts these maps to
the $C^*$-level, without further calculation. That is: the maps $\Dl$,
$\eps$ and $S$, defined as above on smooth functions only, extend
respectively to a $*$-homomorphism
$\Dl_J\: C(G)_J \to C(G)_J \ox C(G)_J$ (using the minimal tensor
product of $C^*$-algebras), a character $\eps_J\: C(G)_J \to \C$, and
a $*$-antiautomorphism $S_J\: C(G)_J \to C(G)_J$.

However, the Moyal product itself on $\Coo(G)$ generally need not
extend to a continuous linear map from $C(G)_J \ox C(G)_J$ to
$C(G)_J$. This may happen because the product map~$m$ is generally
\textit{not} continuous for the minimal tensor product. (There is an
interesting category of ``Hopf $C^*$-algebras'', introduced by Vaes
and van~Daele \cite{VaesVD}, which does have continuous products, but
the link with Moyal deformations remains to be worked out.)

The $C^*$-algebras $C(G)_J$, arising from Moyal products whose
coefficient matrices are of the form \eqref{eq:double-Q}, are fully
deserving of the name \textit{compact quantum groups}. Indeed, they
are thus baptized in~\cite{RieffelComQGp}. They differ from the
compact quantum groups of Woronowicz~\cite{WoronowiczComp} in that
they explicitly define the algebraic operations on smooth subalgebras,
and are thus well-adapted to the needs of noncommutative geometry.

\subsection{Isospectral deformations of homogeneous spin geometries}
\label{sec:isosp-spin-geom} 

The Connes--Landi spheres $\Sf_\th^4$ can now be seen as homogeneous
spaces for compact quantum groups. The ordinary $4$-sphere is
certainly a homogeneous space; in fact, it is ---almost by
definition--- an orbit of the $5$-dimensional rotation group: thus,
$\Sf^4 \approx SO(5)/SO(4)$. Now, $SO(5)$ is a compact simple Lie
group of rank two; that is to say, its maximal torus is $\T^2$. We can
exhibit this maximal torus as the group of block-diagonal matrices
$$
h = \begin{pmatrix}
 \cos\phi_1 & \sin\phi_1 \\
-\sin\phi_1 & \cos\phi_1 \\
&& \cos\phi_2 & \sin\phi_2 \\
&&-\sin\phi_2 & \cos\phi_2 \\
&&&& 1 \end{pmatrix}.
$$
By regarding $\Sf^4$ as the orbit of $(0,0,0,0,1)$ in~$\R^5$, whose
isotropy subgroup is $SO(4)$, we see that the maximal torus of $SO(4)$
is also $\T^2$. When the $4$-sphere is identified as the right-coset
space $SO(5)/SO(4)$, and the doubled torus $\T^2 \x \T^2$ is made to
act on~$SO(5)$ by left-right multiplication as in~\eqref{eq:dbl-mult},
then the right action of the second $\T^2$ is absorbed in the cosets,
but the left action of the first $\T^2$ passes to the quotient. This
is a group-theoretical description of how the $2$-torus acts by
rotations on the $4$-sphere. The action is isometric since the left
translations preserve the invariant metric on the group, and also
preserve the induced $SO(5)$-invariant metric on the coset space.

There is an immediate generalization, proposed in~\cite{Larissa},
which highlights the nature of this torus action. Consider a tower of
subgroups
$$
H \leq K \leq G,
$$
where $G$ is a compact connected Lie group, $K$ is a closed subgroup
of~$G$, and $H$ is a closed connected \textit{abelian} subgroup
of~$K$, i.e., a torus. The example we have just seen reappears in
higher dimensions as
$$
\T^l \leq SO(2l) \leq SO(2l + 1),
 \sepword{with}  \Sf^{2l} \approx SO(2l + 1)/SO(2l).
$$
Odd-dimensional spheres yield a slightly different case:
$$
\T^l \leq SO(2l + 1) \leq SO(2l + 2),
 \sepword{with}  \Sf^{2l+1} \approx SO(2l + 2)/SO(2l + 1).
$$
This time, $H$ is a maximal torus in $K$ but not in~$G$.

Since $H \leq K$, the left-right action \eqref{eq:dbl-mult} of
$H \x H$ on both $G$ and $K$ induces a left action of $H$ on the
quotient space $M := G/K$, since the right action of~$H$ is absorbed
in the right $K$-cosets. If we deform $C(G)$, under the action of
$H \x H$, by means of a Moyal product with parameter matrix
$J = Q \op (-Q)$, the natural thing to expect is that the
$C^*$-algebra $C(G/K)$ undergoes a deformation governed by $Q$~only.
We now prove this, following~\cite{Larissa}.

It helps to recall the discussion of homogeneous spaces at the end of
subsection~\ref{sec:Hopf-basics}. We are now in a position to replace
the generic function space $\F(G)$ used there by either $\Coo(G)$ or
$C(G)$, according to need. In particular, the algebra isomorphism
$\zeta\: \Coo(G)^K \to \Coo(G/K)$ given by $\zeta f(\bar x) := f(x)$
intertwines the coproduct $\Dl$ on~$\Coo(G)$ with the coaction
$\rho\: \Coo(M) \to \Coo(G) \ox \Coo(G/K)$ defined by
$\rho f(x,\bar y) := f(\overline{xy})$.

We can distinguish three abelian group actions here. First there is
action $\a$ of $\hl \op \hl$ on $C(G)$, already given
by~\eqref{eq:hCGh-action}. Next, the formula
$(\b_t h)(\bar x) := h(\overline{\la^{-t}x})$ determines an
action~$\b$ of~$\hl$ on $C(G/K)$. Then there is action $\ga$ of $\hl$
on $C(G)^K$ where $(\ga_t f)(x) := f(\la^{-t}x)$; it can be regarded
as an action of $\hl \op \hl$ where the second factor acts trivially,
so that $\ga$ is just the restriction of $\a$ to the subspace $C(G)^K$
of $C(G)$.

Let $f,g \in \Coo(G)^K$ be smooth right $K$-invariant functions. Then
\begin{align*}
(f \x_J g)(x)
&= \int_{\hl^4} f(\la^{-Qs} x \la^{-Qt}) g(\la^{-u} x \la^v)
    \,e^{2\pi i(s\.u + t\.v)} \,ds\,dt \,du\,dv
\\
&= \int_{\hl^4} f(\la^{-Qs} x) g(\la^{-u} x)
    \,e^{2\pi i(s\.u + t\.v)} \,ds\,dt \,du\,dv
\\
&= \int_{\hl^2} f(\la^{-Qs} x) g(\la^{-u} x) \,e^{2\pi is\.u} \,ds\,du
 = (f \x_Q g)(x),
\end{align*}
where the $Q$-product comes from the action $\ga$ on $C(G)^K$. On
passing to $\Coo(G/K)$ with the isomorphism $\zeta$, which obviously
intertwines the actions $\ga$ and~$\b$, this calculation shows that
$$
\zeta(f \x_J g) = \zeta f \x_Q \zeta g
  \sepword{for all}  f,g \in \Coo(G)^K.
$$
In other words, the $J$-product on $C(G)$ induces the $Q$-product, as
claimed.

The reason for this bookkeeping with actions and isomorphisms is to be
able to lift everything to the $C^*$-level, using Rieffel's
functoriality theorems. First, since $\zeta \ga_t = \b_t \zeta$ for
each~$t \in \hl$, the isomorphism
$\zeta^{-1}\: \Coo(G/K) \to \Coo(G)^K$ extends to a $*$-isomorphism of
$C(G/K)_Q$ onto $C(G)^K_Q$. Since $\ga$ is the restriction of
$\a$ to~$C(G)^K$, the inclusion $\Coo(G)^K \hookto \Coo(G)$ is
equivariant for the actions $\ga$ and $\a$, so it extends to an
injective $*$-homomorphism from $C(G)^K_Q$ to $C(G)_J$. We may
summarize by saying that the isomorphism and inclusion
$$
C(G/K) \simeq C(G)^K \hookto C(G)
$$
restricts to the smooth subalgebras
$$
\Coo(G/K) \simeq \Coo(G)^K \hookto \Coo(G),
$$
and from there extends to an isomorphism and inclusion
$$
C(G/K)_Q \simeq C(G)^K_Q \hookto C(G)_J.
$$
This shows that the deformed $C^*$-algebra $C(G/K)_Q$ is an
\textit{embedded homogeneous space} for the compact quantum group
$C(G)_J$.

\begin{exmp}
\label{eg:CL-spheres} 
To get the noncommutative spheres of Connes and Landi, just
take $G = SO(2l+1)$, $K = SO(2l)$ and let $H = \T^l$ be the maximal
torus for both. Then let $Q = \thalf\th$, where $\th$ is any real
skewsymmetric $l \x l$ matrix. The resulting deformation of
$C(\Sf^{2l})$ is the $C^*$-algebra $C(\Sf_\th^{2l})$, and its smooth
subalgebra (for the $\T^l$-action) is just
$\Coo(\Sf_\th^{2l}) := \Coo(\Sf^{2l})$ with the Moyal product~$\x_Q$.
\end{exmp}

The odd-dimensional spheres $\Sf^{2l+1} = SO(2l+2)/SO(2l+1)$ may be
deformed in like manner, using, say, the maximal torus $\T^l$ of
$SO(2l+1)$. However, in this case, since this torus is not maximal in
the full group $SO(2l+2)$, one can regard the $\T^l$-rotations as an
action of the torus $\T^{l+1}$ that is trivial in one direction. The
algebras $C(\Sf_\th^{2l+1})$, with their $\T^l$-actions and the
corresponding deformations of $C(SO(2l+2))$, have recently been
discussed extensively by Connes and Dubois-Violette~\cite{ConnesDV}
from the cohomological standpoint.

In fact, in \cite{ConnesDV}, the noncommutative spheres are
constructed in another way, by directly obtaining generators and
relations for the corresponding algebras from twistings of Clifford 
algebras, as already outlined in \cite{ConnesLa}, before checking
that those algebras also come from $\theta$-deformations. The
advantage of this procedure is that what is obtained is manifestly
spherical, in the sense that the homology-sphere condition 
\eqref{eq:orient-Chern}, or its odd-dimensional counterpart, is
built-in~\cite{DuboisVPvt}.

The simplest \textit{nonspherical} examples in even dimensions are
$\C P^2 \simeq SU(3)/U(2)$ and the $6$-dimensional flag manifold
$F^6 = SU(3)/\T^2$. With $G = SU(3)$ and $H = K = \T^2$ and any
irrational $\th = 2Q_{12}$, one obtains a family of $6$-dimensional
quantized flag manifolds.

\marker
We have outlined a general construction of noncommutative algebras,
including all the Connes--Landi spheres, which come equipped with
dense pre-$C^*$-algebras. The final step is to build noncommutative
spin geometries based on these algebras. This was done by Connes and
Landi for their spheres~\cite{ConnesLa} by means of an
\textit{isospectral deformation}. It was observed in~\cite{Larissa},
and likewise in~\cite{ConnesDV}, that their algorithm extends directly
to any of the aforementioned quantum homogeneous spaces, with only
notational changes.

The compact homogeneous manifold $G/K$ can be regarded as a Riemannian
manifold, since it has a $G$-invariant metric. We shall assume that
$G/K$ also has a homogeneous spin structure (this is not always the
case; for instance, $\C P^2$ is only spin$^c$, while $SU(3)/SO(3)$
does not even admit a spin$^c$ structure~\cite[\S2.4]{Friedrich}), and
we let $\Dslash$ be the corresponding Dirac operator; it is a
selfadjoint operator on the Hilbert space $\H$ of square-integrable
spinors. As we shall see, in the end we only need that the metric, and
the Dirac operator, be invariant under the action of the torus~$H$
rather than the full group~$G$.

It is important to remark that the action of $H$ by isometries 
on~$G/K$ does not lift directly to the spinor space $\H$ (or, if one 
prefers, to the spinor bundle~$S$). Rather, in view of the double 
covering $\Spin(n) \to SO(n)$ where $n = \dim G/K$, there is a double 
covering $\Onda H \longto^\pi H$ and a homomorphism
$\Onda H \to \Aut(S)$ which covers the homomorphism
$H \to \Isom(G/K)$ \cite[\S 13]{ConnesDV}. This yields a group of 
unitaries $\set{V_{\tilde x} : \tilde x \in \Onda H}$ on~$\H$ which
preserve the subspace $\Gaoo(G/K,S)$ of smooth spinors and cover the 
isometries $\set{I_x : x \in H}$ of $G/K$. More precisely: if
$\phi,\psi \in \Gaoo(G/K,S)$ and $f \in \Coo(G/K)$, then
$$
V_{\tilde x}(f\psi) = I_x(f) \,V_{\tilde x}\psi,  \sepword{and}
(V_{\tilde x}\phi)^\7 V_{\tilde x}\psi = I_x(\phi^\7 \psi),
$$
where $x = \pi(\tilde x)$. Consequently, the Dirac operator $\Dslash$
on $\H$ commutes with each~$V_{\tilde x}$.

Now choose a basis $\row{X}{1}{l}$ of the Lie algebra $\hl$, and for
$j = 1,\dots,l$, let $p_j$ be the selfadjoint operator representing
$X_j$ on~$\H$; if $\exp\: \hl \to H$ and $\Exp\: \hl \to \Onda H$
denote the exponential maps, then $\pi(\Exp(tX_j)) = \exp(tX_j/2)$ and
$p_j = -i\frac{d}{dt}\bigr|_{t=0} V_{\Exp(tX_j)}$. Therefore, the
spectrum of each operator $p_j$ lies either in~$\Z$ or $\Z + \thalf$.
For each $r \in \R^l$, we may define a unitary operator
\begin{equation}
\sg(p,r) := \exp\bigl\{-2\pi i \tsum_{j,k} p_j Q_{jk} r_k\bigr\},
\label{eq:unitary-cocyc} 
\end{equation}
by formally replacing half of the arguments of the group cocycle
\eqref{eq:sigma-cocyc} with the operators~$p_j$; its inverse is the
similarly defined operator $\sg(r,p)$. These operators commute with
each other and also with $\Dslash$, but they do not commute with the
representation of $\Coo(G/K)$ on~$\H$ (multiplication of spinors by
functions).

The unitary conjugations $T \mapsto \sg(p,t) T \sg(t,p)$ define an
action of $\R^l$ on the algebra of bounded operators on~$\H$, which is
periodic on account of the half-integer spectra of the~$p_j$, and this
action gives a grading of operators into spectral subspaces, indexed
by~$\Z^l$. Therefore, any bounded operator $T$ in the common smooth
domain of these transformations has a decomposition
$T = \sum_{r\in\Z^l} T_r$, where the components satisfy the
commutation rules
$$
\sg(p,r)\,T_s = T_s\,\sg(p+s,r)  \sepword{for}  r,s \in \Z^l.
$$
For any multiplication operator $f$ obtained from the representation
of the algebra $\Coo(G/K)$ on spinors, this grading coincides with the
previous decomposition $f = \sum_{r\in\Z^l} f_r$.

The operator $\Z^l$-grading allows us to define a ``left twist''
of~$T$ by
$$
L(T) := \sum_{r\in\Z^l} T_r \,\sg(p,r).
$$
If $f,g \in \Coo(G/K)$, the group cocycle property
\eqref{eq:group-cocyc} of~$\sg$ shows that
\begin{align*}
L(f) L(g) &= \sum_{r,s} f_r \,\sg(p,r) \,g_s \,\sg(p,s)
 = \sum_{r,s} f_r \,g_s \,\sg(p+s,r) \,\sg(p,s)
\\
&= \sum_{r,s} f_r \,g_s \,\sg(r,s) \,\sg(p,r+s) = L(f \x_Q g),
\end{align*}
on account of~\eqref{eq:Moyal-sigma}. Therefore, $L$ yields a
representation of $\Coo(G/K)_Q := (\Coo(G/K), \x_Q)$ on~$\H$. In other
words, the Moyal product gives not only an abstract deformation of the
algebra $\Coo(G/K)$, but also ---more importantly--- it yields a
deformation of the spinor \textit{representation} of $\Coo(G/K)$,
without disturbing the underlying Hilbert space.

The recipe for creating new spin geometries should now be clear: one
deforms the algebra (and its representation), while keeping unchanged
all the other terms of the spectral triple: the Hilbert space $\H$
together with its grading~$\chi$ if $\dim(G/K)$ is even, the operator
$\Dslash$, and the charge conjugation~$C$. This deformation is
\textit{isospectral} \cite{ConnesLa} in the tautological sense that
the spectrum in question is that of the operator~$\Dslash$, which
remains the same.

It remains to check that the new spectral triple satisfies the
conditions governing a spin geometry. First of all, each
$[\Dslash, L(f)]$, for $f \in \Coo(G/K)$, must be a bounded operator;
this is ensured by noting that
$$
[\Dslash, L(f)] = \sum_r [\Dslash,f_r]\,\sg(p,r) = L([\Dslash,f]),
$$
since each $[\Dslash,f]$ is bounded. The grading operator $\chi$ is
unaffected by the torus action on $G/K$ since the metric is taken to
be $H$-invariant: this implies $L(\chi) = \chi$. In view of the
previous equation, the orientation equation
$\pi_\Dslash(\cc) = \chi$ survives after application of~$L$ to both
sides.

The reality condition is more interesting. The charge conjugation
operator $C$ on spinors \cite[Chap.~9]{Polaris} commutes with all
$\sg(p,r)$ (again, due to $H$-invariance of the metric). It follows
from \eqref{eq:unitary-cocyc} and the antilinearity of~$C$ that
$C p_j C^{-1} = -p_j$ for each~$j$. We can now define a ``right
twist''
$$
R(T) := C L(T)^* C^{-1} = \sum_{r\in\Z^l} \sg(r,p)\, C T_r^* C^{-1}
 = \sum_{r\in\Z^l} C T_r^* C^{-1} \,\sg(r,p).
$$
Now, $C$ intertwines multiplication operators from $\Coo(G/K)$ with
their complex conjugates: $C f^* C^{-1} = f$ for $f$. Therefore,
$R(f) = \sum_{r\in\Z^l} f_r \,\sg(r,p)$, from which one can check that
$R(f)R(g) = R(f \x_{-Q} g)$; in other words, $R$ gives an
\textit{anti}representation of $\Coo(G/K)_Q$ on~$\H$. This commutes
with the representation~$L$:
\begin{align*}
L(f) R(g)
&= \sum_{r,s} f_r \,\sg(p, r) \,g_s \,\sg(s, p)
 = \sum_{r,s} f_r g_s \,\sg(p+s, r) \,\sg(s, p)
\\
&= \sum_{r,s} g_s f_r \,\sg(s, p+r) \,\sg(p, r)
 = \sum_{r,s} g_s \,\sg(s, p) \,f_r \,\sg(p, r) = R(g) L(f).
\end{align*}

The first-order property of the spin geometry is now immediate
$$
[[\Dslash, L(f)], R(g)]
= \sum_{r,s} \sg(p,r)\, [[\Dslash, f_r], g_s] \,\sg(s,p) = 0,
$$
since $[[\Dslash, f_r], g_s] = 0$ in the commutative case (the
commutator $[\Dslash, f_r]$ is an operator of order zero which
commutes with multiplication operators). Regularity and finiteness are
straightforward, since the smooth subalgebra $\Coo(G/K)$ does not grow
or shrink under deformations. Poincar\'e duality also goes through, on
account of another theorem of Rieffel, to the effect that the
$K$-theory of the pre-$C^*$-algebras remains unaffected by
deformations~\cite{RieffelKDef}.

The construction is now complete. We sum up with the following
Proposition.

\begin{prop}
\label{pr:isosp-defm} 
Let $H \leq K \leq G$ be a tower of compact connected Lie groups
where $H$ is a torus, such that $G/K$ admits a spin structure. Let
$(\Coo(G/K), \H, \Dslash, C, \chi)$ denote any commutative spin
geometry over $\Coo(G/K)$ where $\H$ is the spinor space and $\Dslash$
is the Dirac operator for an $H$-invariant metric on~$G/K$. Then there
is a noncommutative spin geometry obtained from it by isospectral
deformation, whose algebra $\Coo(G/K)_Q$ is that of any quantum
homogeneous space obtained from a Moyal product $\x_Q$ on $\Coo(G/K)$.
\qed
\end{prop}

\newpage


\end{document}